\documentclass[10pt,twocolumn,letterpaper]{article}

\usepackage{iccv}
\usepackage{times}
\usepackage{epsfig}
\usepackage{graphicx}
\usepackage{amsmath}
\usepackage{amssymb}

\usepackage{float}
\usepackage[normalem]{ulem}
\usepackage{enumitem}
\usepackage{textcomp}
\usepackage{array}
\usepackage{colortbl}

\setitemize{noitemsep,topsep=0pt,parsep=0pt,partopsep=0pt}

\usepackage{color}

\usepackage{booktabs}
\usepackage{subfigure}
\usepackage[percent]{overpic}
\usepackage{tikz}
\usetikzlibrary{positioning, spy}

\usepackage[pagebackref=true,breaklinks=true,colorlinks,bookmarks=false]{hyperref}

\iccvfinalcopy 

\ificcvfinal\fi

\makeatletter
\def\@fnsymbol#1{\ensuremath{\ifcase#1\or \dagger\or \ddagger\or \mathsection\or \mathparagraph\or \|\or **\or \dagger\dagger \or \ddagger\ddagger \else\@ctrerr\fi}}
\makeatother

\begin{document}

\definecolor{darkyellow}{rgb}{1, 0.6, 0}
\definecolor{darkred}{rgb}{0.8, 0.0, 0.0}
\definecolor{darkgreen}{rgb}{0, 0.4, 0}

\renewcommand{\etal}{et~al.}

\newcommand{\PSF}{\mathit{PSF}}
\newcommand{\SRF}{\mathit{SRF}}
\newcommand{\level}{\ell}
\newcommand{\lossfun}{\mathcal{L}}

\newcommand{\lannotatepic}[1]{\scriptsize \color{white} \textbf{\textsf{#1}}}
\newcommand{\rannotatepic}[1]{\flushright \scriptsize \color{white} \textbf{\textsf{#1}}}
\newcommand{\imgcell}[2][1in]{%
\begin{minipage}[c]{#1}%
  \vspace{2pt}%
  \includegraphics[width=\linewidth]{#2}%
  \vspace{1pt}%
\end{minipage}%
}
\newcommand{\imgtextcell}[3][1in]{%
\begin{minipage}[c]{#1}%
  \vspace{2pt}%
  \begin{overpic}[width=\linewidth]{#2}%
    \put(5,5){\parbox[b][0.9\linewidth][b]{0.9\linewidth}{\rannotatepic{#3}}}%
  \end{overpic}%
  \vspace{1pt}%
\end{minipage}%
}

\newcommand{\grayline}{\arrayrulecolor[rgb]{0.8,0.8,0.8}\hline\arrayrulecolor[rgb]{0,0,0}}

\makeatletter
\renewcommand{\paragraph}{%
  \@startsection{paragraph}{4}%
  {\z@}{1.5ex \@plus 1ex \@minus .2ex}{-1em}%
  {\normalfont\normalsize\bfseries}%
}
\makeatother
\newcommand{\myparagraph}[1]{\paragraph{#1}}

\tikzset{
  zoomin/.style={
    node distance = 0pt,
    inner sep = 0pt,
    anchor = south west,
    spy using outlines={rectangle, green, magnification=3, every spy on node/.append style={thin}}
  }
}

\newlength\Origarrayrulewidth
\newcommand{\Cline}[1]{%
  \noalign{\global\setlength\Origarrayrulewidth{\arrayrulewidth}}%
  \noalign{\global\setlength\arrayrulewidth{1pt}}\arrayrulecolor{red}\cline{#1}%
  \noalign{\global\setlength\arrayrulewidth{\Origarrayrulewidth}}%
}

\newcommand\Thickvrule[1]{%
  \multicolumn{1}{!{\color{red}\vrule width 1pt}c!{\color{red}\vrule width 1pt}}{#1}%
}

\title{How to Train Neural Networks for Flare Removal}

\author{
  {Yicheng Wu}\textsuperscript{\rm 1 \thanks{This work was done while Yicheng Wu was an intern at Google Research. He is currently a Research Scientist at Snap Research.}}   \quad
  {Qiurui He}\textsuperscript{\rm 2}\quad
  {Tianfan Xue}\textsuperscript{\rm 2}\quad
  {Rahul Garg}\textsuperscript{\rm 2} \quad
  {Jiawen Chen}\textsuperscript{\rm 3} \\
  {Ashok Veeraraghavan}\textsuperscript{\rm 1} \quad
  {Jonathan T. Barron}\textsuperscript{\rm 2} \\
  {\textsuperscript{1}Rice University \quad
  \textsuperscript{2}Google Research \quad
  \textsuperscript{3}Adobe Inc.} \\
}

\maketitle

\ificcvfinal\thispagestyle{empty}\fi

\begin{abstract}
    When a camera is pointed at a strong light source, the resulting photograph may contain lens flare artifacts.
    Flares appear in a wide variety of patterns (halos, streaks, color bleeding, haze, etc.) and this diversity in appearance makes flare removal challenging.
    Existing analytical solutions make strong assumptions about the artifact's geometry or brightness, and therefore only work well on a small subset of flares.
    Machine learning techniques have shown success in removing other types of artifacts, like reflections, but have not been widely applied to flare removal due to the lack of training data.
    To solve this problem, we explicitly model the optical causes of flare either empirically or using wave optics, and generate semi-synthetic pairs of flare-corrupted and clean images.
    This enables us to train neural networks to remove lens flare for the first time.
    Experiments show our data synthesis approach is critical for accurate flare removal, and that models trained with our technique generalize well to real lens flares across different scenes, lighting conditions, and cameras.
    
\end{abstract}

\section{Introduction} \label{sec:intro}
Photographs of scenes with a strong light source often exhibit lens flare---a salient visual artifact caused by unintended reflections and scattering within the camera.
Flare artifacts can be distracting, reduce detail, and occlude image content.
Despite significant efforts in optical design to minimize lens flare, even small light sources can still produce substantial artifacts when imaged by consumer cameras.

\begin{figure}[t]
  \centering
  \footnotesize
  \setlength{\tabcolsep}{1pt}
  \begin{tabular}{@{}cccc@{}}
     \imgcell[57pt]{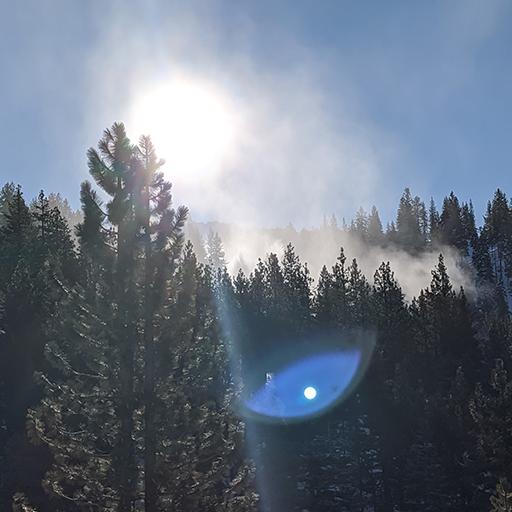} &
     \imgcell[57pt]{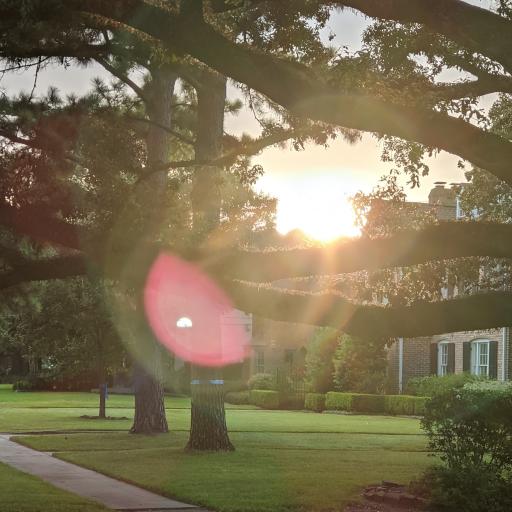} &
     \imgcell[57pt]{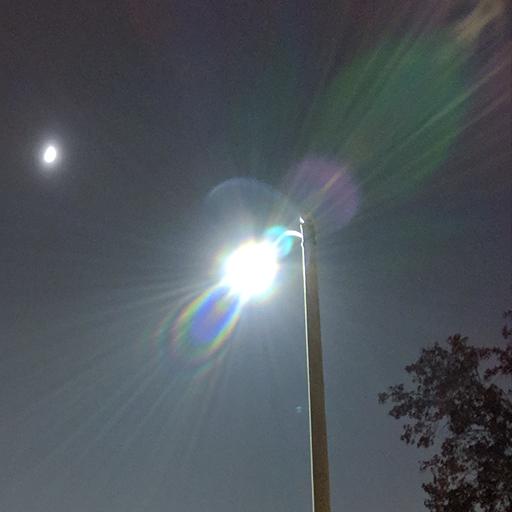} &
     \imgcell[57pt]{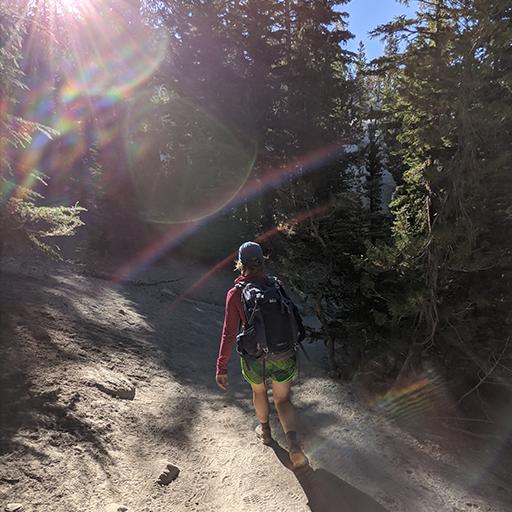} \\[-1pt]
     
     \imgcell[57pt]{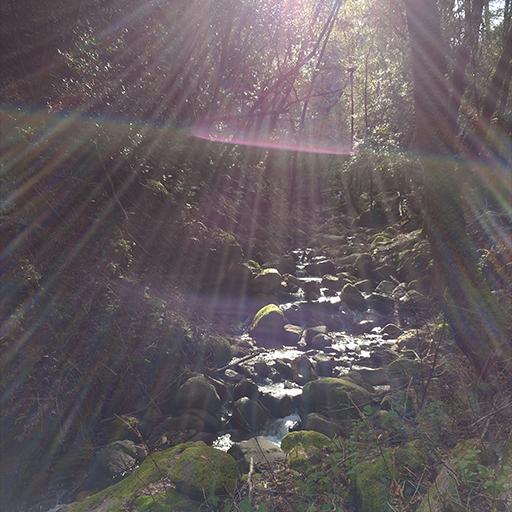} &
     \imgcell[57pt]{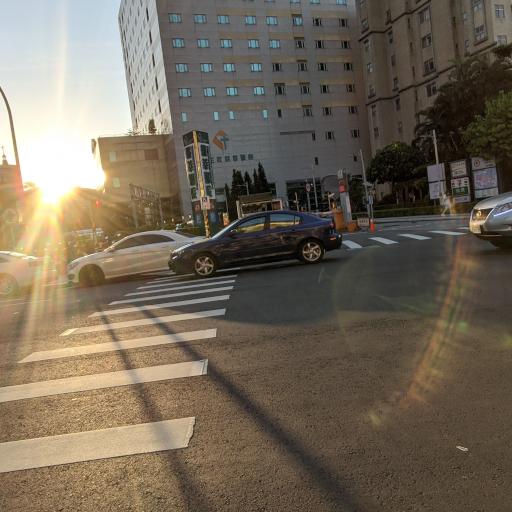} &
     \imgcell[57pt]{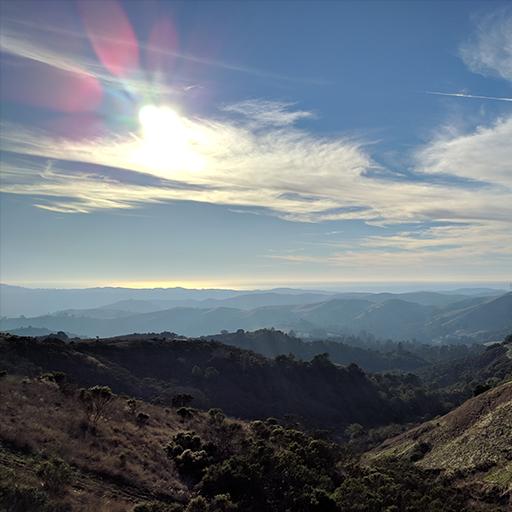} &
     \imgcell[57pt]{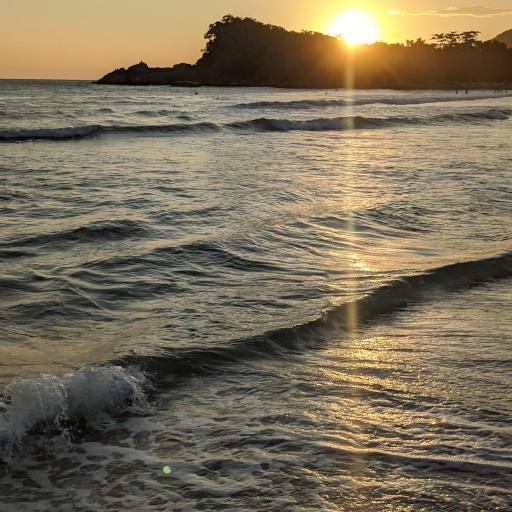}
  \end{tabular}
  \caption{
    Lens flare artifacts frequently occur in photographs with strong light sources. They exhibit a wide range of shapes and colors, which makes them difficult to remove with existing methods.
  }
  \label{fig:flare_diversity}
\end{figure}

Flare patterns depend on the optics of the lens, the location of the light source, manufacturing imperfections, and scratches and dust accumulated through everyday use.
The diversity in the underlying cause of lens flare leads to the diversity in its presentation. As Fig.~\ref{fig:flare_diversity} shows, typical artifacts include halos, streaks, bright lines, saturated blobs, color bleeding, haze, and many others.
This diversity makes the problem of flare removal exceedingly challenging.

Most existing methods for lens flare removal~\cite{asha2019auto,chabert2015automated,vitoria2019automatic} do not account for the physics of flare formation, but rather na\"{i}vely rely on template matching or intensity thresholding to identify and localize the artifact. 
As such, they can only detect and potentially remove limited types of flares, such as saturated blobs, and do not work well in more complex real-world scenarios.

Despite the proliferation of deep neural networks, there seems to be no successful attempt at learning-based flare removal. What is it that makes this problem so hard?

The main challenge is the lack of training data.
Collecting a large number of perfectly-aligned image pairs with and without lens flare would be tedious at best and impossible at worst: the camera and the scene would need to be static (a particularly difficult requirement given most lens flare occurs outdoors and involves the sun), and one would need some mechanism to ``switch'' the artifacts on and off without also changing the illumination of the scene.
With significant effort this can be accomplished by collecting pairs of images taken on a tripod where the photographer manually places an occluder between the illuminant and the camera in one image. But this approach is too labor-intensive to produce the thousands or millions of image pairs usually required to train a neural network.
Furthermore, this approach only works when the flare-causing illuminant lies outside of the camera's field of view (e.g.,~real scenes in Fig.~\ref{fig:compare}), which limits its utility.

To overcome this challenge, we propose to generate \emph{semi-synthetic} data grounded on the principles of physics.
We make the key observation that lens flare is an additive layer on top of the underlying image, and that it is induced by either scattering or internal reflection.
For the scattering case (e.g.,~scratches, dust, other defects), we construct a wave optics model that we demonstrate closely approximates reality.
For the unintended reflections between lens elements, we adopt a rigorous data-driven approach, as an accurate optical model for a commercial camera is often unavailable.
With this formulation, we are able to generate a large and diverse dataset of semi-synthetic flare-corrupted images, paired with ground-truth flare-free images.

Another challenge is removing flare while keeping the visible light source intact. This is hard even with our semi-synthetic data, as we cannot separate the light source from the flare-only layer without affecting the flare it induces. Hence, if trained na\"{i}vely, the network will try to remove the light source along with the flare, leading to unrealistic outputs. To this end, we propose a loss function that ignores the light source region, and a post-processing step to preserve the light source in the output.

To show the effectiveness of our dataset and procedures, we train two distinct convolutional neural networks originally designed for other tasks.
During training, we minimize a loss function on both the predicted flare-free image and the residual (i.e.,~inferred flare).
At test time, the networks require only a single RGB image taken with a standard camera and are able to remove different types of flare across a variety of scenes.
Although trained exclusively on semi-synthetic data, both models generalize well to real-world images.
To the best of our knowledge, this is the \emph{first} general-purpose method for removing lens flare from a single image.

Our code and datasets are publicly available at \href{https://yichengwu.github.io/flare-removal/}{https://yichengwu.github.io/flare-removal/}.

\section{Related work} \label{sec:related}
Existing solutions for flare removal fall into three categories: (a) optical design intended to mitigate the presence of flare, (b) software-only methods that attempt post-capture enhancement, and (c) hardware--software solutions that capture additional information.

\myparagraph{Hardware solutions}
The lenses of high-end cameras often employ sophisticated optical designs and materials to reduce flare.
As each glass element is added to a compound lens to improve image quality, there is also an increased probability that light is reflected from its surface to create flare. One widely used technique is to apply anti-reflective (AR) coating to lens elements, which reduces internal reflection by destructive interference. However, the thickness of this coating can only be optimized for particular wavelengths and angles of incidence and therefore cannot be perfect. Additionally, adding an AR coating to all optical surfaces is expensive, and may preclude or interfere with other coatings (e.g., anti-scratch and anti-fingerprint).

\myparagraph{Computational methods}
Many post-processing techniques have been proposed to remove flare. Deconvolution has been used to remove flare in X-ray imaging~\cite{faulkner1989veiling,seibert1985removal} or HDR photography~\cite{reinhard2010high}. These approaches depend critically on the assumption that the point spread function of the flare does not vary spatially, which is generally not true.
Other solutions~\cite{asha2019auto,chabert2015automated,vitoria2019automatic} adopt a two-stage process: detecting lens flare based on their unique shape, location, or intensity (i.e., by identifying a saturated region), and then recovering the scene behind the flare region using inpainting~\cite{criminisi2004region}. These solutions only work on limited types of flare (e.g., bright spots), and are vulnerable to the misclassification of all bright regions as flare. Additionally, these techniques classify each pixel as either ``flare'' or ``not flare'', ignoring the fact that most lens flares are better modeled as a semitransparent layer on top of the underlying scene.

\myparagraph{Hardware--software co-design}
Researchers have used computational imaging for flare removal, where the camera hardware and post-processing algorithms are designed in conjunction.
Talvala~\etal~\cite{talvala2007veiling} and Raskar~\etal~\cite{raskar2008glare} attempt to selectively block flare-causing light using structured occlusion masks and recover the flare-free scene using either direct--indirect separation or a light field-based algorithm.
Though elegant, they require special hardware and are therefore limited in their practicality.

\myparagraph{Learning-based image decomposition}
While no learning-based flare removal techniques exist, a number of recent works apply learning to similar applications such as reflection removal~\cite{fan2017generic,li2020single,zhang2018single}, rain removal~\cite{qian2018attentive,wei2019semi}, and denoising~\cite{BrooksCVPR2019}.
These methods attempt to decompose an image into ``clean'' and ``corrupt'' components by training a neural network. Their success relies heavily on high-quality domain-specific training datasets, which this work tries to address for the first time.

\section{Physics of lens flare} \label{sec:physics}
\begin{figure}
  \centering
  \subfigure[Optical diagram]{%
    \includegraphics[height=0.33\linewidth]{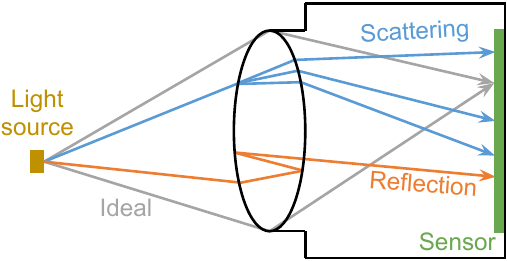}%
    \label{fig:physics:optical_diagram}%
  }\hfil%
  \subfigure[Sample image]{%
    \includegraphics[height=0.33\linewidth]{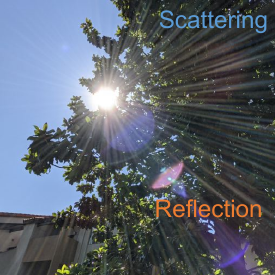}%
    \label{fig:physics:annotated_flare}%
  }%
  \caption{ 
    Cameras are intended to focus light from a point source at exactly one point on the sensor (gray rays).
    However, dust and scratches may cause scattering (blur rays), leading to haze and bright streaks ``emitting'' radially.
    Additionally, internal reflections (orange rays) can occur between the optical surfaces of lens elements (only one element is shown), resulting in disk- or arc-shaped bright spots. This reflective flare may also have a color tint, as the anti-reflective coating only blocks certain wavelengths.
  }
  \label{fig:physics}
\end{figure}

An ideal camera, when in focus, is supposed to refract and converge all rays from a point light source to a single point on the sensor. 
In practice, real lenses scatter and reflect light along unintended paths, resulting in flare artifacts~\cite{kingslake1992optics}, as shown in Fig.~\ref{fig:physics:optical_diagram}.
The scattered and reflected parts only constitute a small fraction of each incident light ray. So although flare is omnipresent, it is imperceptible in most photographs. 
But, when a strong light source is many orders of magnitude brighter than the rest of the scene (e.g., the sun), the small fraction of scattered and reflected rays from this bright light will lead to visible artifacts at other pixels on the image.
The geometry of the scattering from dust and scratches, and that of the multiple reflections, result in characteristic visual patterns.
At a high level, flare can be classified into two principal categories: scattering-induced and reflection-induced.

\myparagraph{Scattering flare}
While an ideal lens is 100\% refractive, real lenses have many imperfections that cause light to scatter. 
The scattering (or diffraction) could be due to manufacturing defects (e.g.,~dents), or normal wear (e.g., dust and scratches).
As a consequence, apart from the primary rays that are refracted, a secondary set of rays are scattered or diffracted instead of following their intended paths. 
While dust adds a rainbow-like effect, scratches introduce streaks that appear to ``emit'' radially from the light source. 
Scattering may also reduce contrast in the region around the light source, leading to a hazy appearance. 

\myparagraph{Reflective flare}
In a practical lens system, each air--glass interface poses an opportunity for a small amount of reflection (typically about 4\%). 
After an even number of reflections, the rays may hit the sensor at an unintended location, forming a reflection pattern.
Even if we assume the light is reflected exactly twice, for a lens module containing $n$ optical elements ($n \approx 5$ for a modern camera), there are $2n$ optical surfaces, and thus $n(2n-1)$ potential flare-causing combinations.
On the image, these reflective flares typically lie on the straight line joining the light source and the principal point. 
They are sensitive to the light source's angle of incidence, as demonstrated by Fig.~\ref{fig:experiment:flares}, but not to rotation about the optical axis. 
The flare's shape depends on the geometry, size, and location of the aperture, which may partially block the reflection more than once, resulting in arc-shaped artifacts.
As mentioned in Sec.~\ref{sec:related}, AR coating may be used to reduce reflection---typically reducing reflections at the air--glass interface to below 1\%.
However, the effectiveness of this coating also depends on wavelength, so lens flare may exhibit a variety of color tints (often blue, purple, or pink).
It is important to note that reflective flare depends on lens design, and therefore cameras with the same design (e.g., all iPhone 12 main camera modules) are expected to produce similar reflective flares when imaging the same scene.

\myparagraph{Challenges in flare removal}
The different types of flare are often difficult to visibly distinguish or separate. 
The appearance of the observed flare may vary significantly based on the properties of the light source (e.g., location, size, intensity, and spectrum), and of the lens (e.g., design and defects).
For these reasons, it is impractical to build a completely physics-based algorithm to analytically identify and remove each type of flare, especially when multiple artifacts are present in the same image. 
Therefore, we propose a data-driven approach.

\section{Physics-based data generation} \label{sec:data}
Unlike many vision problems in a supervised setting, it is hard to obtain a dataset of flare-corrupted and flare-free image pairs. As a result, we build a physically-realistic, semi-synthetic pipeline.

The additive nature of light implies we can model flare as an additive artifact on top of the ``ideal'' image---the reduction in the intensity of the ``ideal'' rays is negligible. We will first explain how the scattering and reflective flares can be modeled and generated, and then use them to synthesize flare-corrupted and flare-free image pairs.

\subsection{Scattering flare} \label{sec:data:scattering}

\myparagraph{Formulation}
Under the thin-lens approximation, an optical imaging system can be characterized by the complex-valued \emph{pupil function} $P(u, v)$: a 2D field describing, for each point $(u, v)$ on the aperture plane, the lens's effect on the amplitude and phase of an incident wave with wavelength $\lambda$:
\begin{equation} \label{eq:general_aperture}
  P_\lambda(u, v) = A(u, v) \cdot \exp \left( i\phi_\lambda(u, v) \right)\,.
\end{equation}

Here, $A$ is an \emph{aperture function}, a property of the optics that represents its attenuation of the incident wave's amplitude.\footnote{Strictly speaking, lens optics can also introduce a phase shift, in which case $A$ becomes a complex-valued function. However, this has shown little difference in our simulation results, so we assume $A$ is real-valued.} In its simplest form, a camera with an aperture of a finite radius $r$ has an aperture function of
\begin{equation} \label{eq:pinhole_aperture}
  A(u, v) = \begin{cases}
     1 & \text{if\,\,} u^2 + v^2 < r^2 \\
     0 & \text{otherwise}
  \end{cases}\,.
\end{equation}

$\phi_\lambda$ in Eq.~\ref{eq:general_aperture} describes the phase shift, which depends on the wavelength as well as the 3D location of the light source $(x, y, z)$. Omitting the aperture coordinates $(u, v)$, $\phi_\lambda$ can be written as:
\begin{equation} \label{eq:phase_shift}
  \phi_\lambda(x, y, z) = \phi_\lambda^\mathrm{S}(x / z, y / z) + \phi_\lambda^\mathrm{DF}(z)
\end{equation}
where the linear term $\phi^\mathrm{S}$ is determined by the angle of incidence, and the defocus term $\phi^\mathrm{DF}$ depends on the depth $z$ of the point light source.
Once fully specified, the pupil function $P$ in Eq.~\ref{eq:general_aperture} can be used to calculate the point spread function (PSF) by a Fourier transform~\cite{goodman2005introduction}:
\begin{equation} \label{eq:psf}
  \PSF_\lambda = \left| \mathcal{F}\{ P_\lambda \}\right|^2
\end{equation}
which, by definition, is the image of a point light source at $(x, y, z)$ formed by a camera with aperture function $A$. This is the flare image that we desire.

\myparagraph{Sampling PSFs}
To mimic dust and scratches on the lens, we add dots and streaks of random size and transparency to the simple aperture function $A$ in Eq.~\ref{eq:pinhole_aperture}. An example synthetic aperture produced by our procedure is shown in Fig.~\ref{fig:simulation:aperture}, and details are included in the appendix. We generate a total of 125 different apertures.

\begin{figure}
  \centering
  \subfigure[Aperture of a dirty lens]{%
    \includegraphics[height=0.36\linewidth]{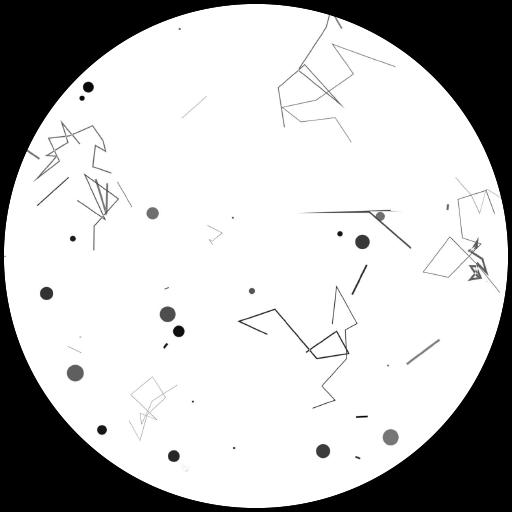}%
    \label{fig:simulation:aperture}%
  }\hfil%
  \subfigure[Simulated flare]{%
    \includegraphics[height=0.36\linewidth]{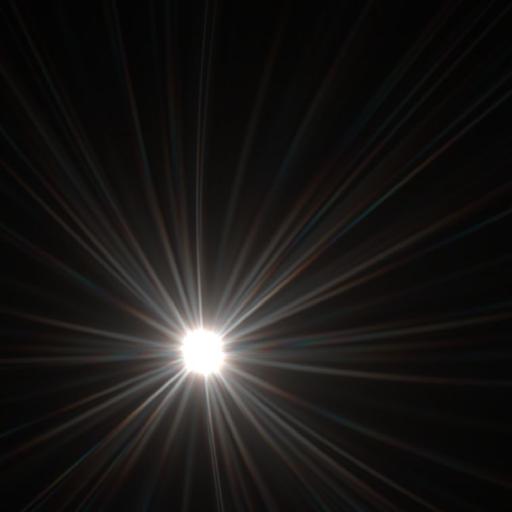}%
    \label{fig:simulation:flare}%
  }%
  \caption{
    To simulate a scattering flare component, we generate a number of apertures \subref{fig:simulation:aperture} with random dots and lines that resemble defects. Wave optics then allows us to compute the image \subref{fig:simulation:flare} of any light source imaged by that synthetic aperture.
  }
  \label{fig:simulation}
\end{figure}

For a given point light source at location $(x, y, z)$ with a single wavelength $\lambda$, the two phase shift terms in Eq.~\ref{eq:phase_shift} can be computed deterministically. The PSF for this light source, $\PSF_\lambda$, can thus be determined by substituting $A$ and $\phi_\lambda$ into Eq.~\ref{eq:psf}.

To simulate a light source across the full visible spectrum, we sample $\PSF_\lambda$ for all wavelengths $\lambda$ from 380nm to 740nm with a spacing of 5nm, resulting in a 73-vector at each pixel of the PSF. The full-spectrum PSF is then left-multiplied by a spectral response function $\SRF$ (a $3 \times 73$ matrix) to derive the PSF measured by the RGB sensor:
\begin{equation}
  \begin{bmatrix}
    \PSF_\mathrm{R}(s, t) \\
    \PSF_\mathrm{G}(s, t) \\
    \PSF_\mathrm{B}(s, t)
  \end{bmatrix} = 
  \SRF
  \begin{bmatrix}
    \PSF_{\lambda=380\text{nm}}(s, t) \\
    \vdots \\
    \PSF_{\lambda=740\text{nm}}(s, t)
  \end{bmatrix}
\end{equation}
where $(s, t)$ are the image coordinates. This produces a flare image for a light source located at $(x, y, z)$.

To construct our dataset of scattering flares, we randomly sample the aperture function $A$, the light source's 3D location $(x, y, z)$, and the spectral response function $\SRF$ (details can be found in the appendix).
We further apply optical distortion (e.g., barrel and pincushion) to augment the $\PSF_\mathrm{RGB}$ images. One example of final output is shown in Fig.~\ref{fig:simulation:flare}. We generate a total of 2,000 such images.

\subsection{Reflective flare} \label{sec:data:reflective}
\label{sec:flare_acq}

\begin{figure}
    \centering
    \subfigure[Capture setup]{%
      \includegraphics[height=0.4\linewidth]{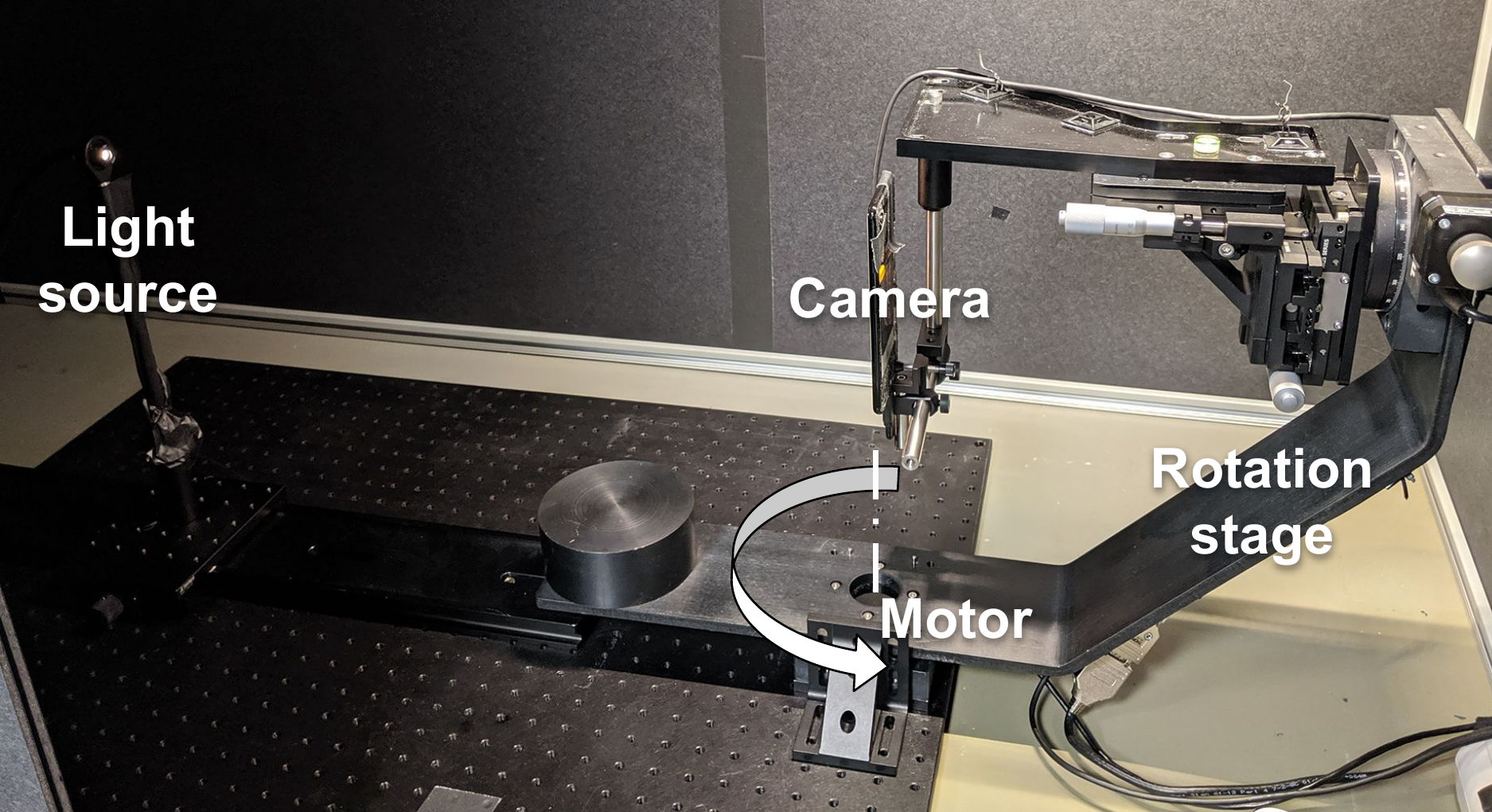}%
      \label{fig:experiment:setup}%
    }\hfil%
    \subfigure[Real flare]{%
      \includegraphics[height=0.4\linewidth]{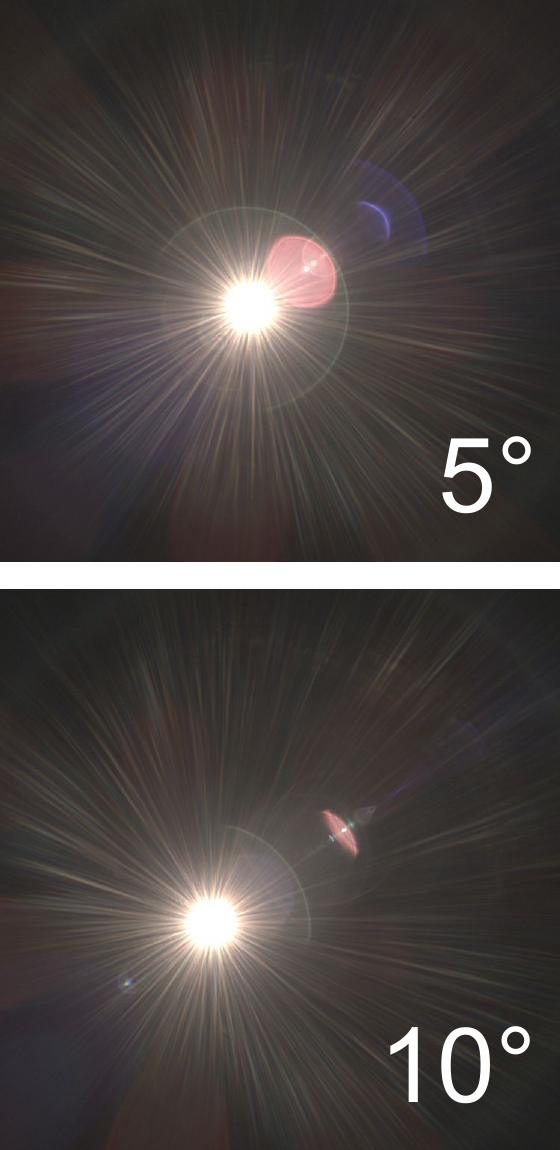}%
      \label{fig:experiment:flares}%
    }%
    \caption{
  We capture images of real lens flare using a strong light source in a dark room. The camera is mounted on a motorized rotation stage that reproduces a wide range of incident angles. Sample images captured at different angles are shown in~\ref{fig:experiment:flares}.
    }
    \label{fig:experiment}
\end{figure}

The reflective flare component is difficult to simulate via rendering techniques~\cite{hullin2011physically, lee2013practical}, as they require an accurate characterization of the optics, which is often unavailable. 
However, lenses of similar design share similar internal reflection paths, so data collected from one instance of the camera often generalizes well to other similar instances.

We capture images of reflective flare in a laboratory setting consisting of a bright light source, a programmable rotation stage, and a fixed-aperture smartphone camera with a $f = $ 13mm lens (35mm equivalent), as shown in Fig.~\ref{fig:experiment:setup}. 
The setup is insulated from ambient light during capture.

The camera is rotated programmatically so that the light source traces (and extends beyond) the diagonal field of view, from $-75^\circ$ to $75^\circ$. 
We capture one HDR image every $0.15^\circ$, resulting in 1,000 samples. Adjacent captures are then interpolated by 2x using the frame interpolation algorithm of \cite{niklaus2018interpolation}, giving us 2,000 images. During training, images are further augmented by random rotation, as reflective flare is rotationally symmetric about the optical axis.

\begin{figure*}
  \centering
  \includegraphics[width=0.9\linewidth]{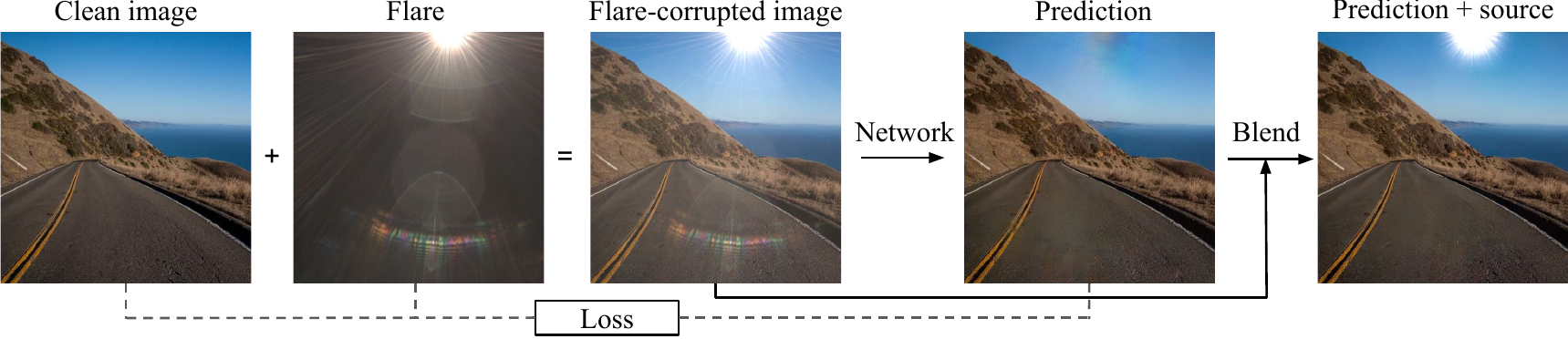}
  \caption{
    Our approach consists of three steps:
    1) We generate training input by randomly compositing a flare-free natural image and a flare image.
    2) A convolutional neural network is trained to recover the flare-free scene (in which the light source may also have been removed, which is undesirable).
    3) After prediction, we blend the input light source back into the output image.
  }
  \label{fig:pipeline}
\end{figure*}

\subsection{Synthesizing flare-corrupted images}

A flare-corrupted image $I_F$ is generated by adding a flare-only image $F$ to a flare-free natural image $I_0$ in linear space (pre-tonemapping raw space where pixel intensities are additive), as illustrated in Fig.~\ref{fig:pipeline}. We also add random Gaussian noise whose variance is sampled once per image from a scaled chi-square distribution $\sigma^2 \sim 0.01\chi^2$, to cover the large range of noise levels we expect to see:
\begin{equation}
  I_F= I_0 + F + N(0,\sigma^2)\,.
\end{equation}

The flare-free image $I_0$ is sampled from the 24k Flickr images in~\cite{zhang2018single}, and is augmented by random flips and brightness adjustments. Since the Flickr images are already gamma-encoded, we approximately linearize them by applying an inverse gamma curve where $\gamma$ is sampled uniformly from $[1.8, 2.2]$ to account for the fact that its exact value is unknown.

The flare-only image $F$ is sampled from both captured and simulated datasets. In Sec.~\ref{sec:ablation}, we show that both are necessary. Random affine transformations (e.g., scaling, translation, rotation, and shear) and white balance are applied as additional augmentations. Randomization details are in the appendix.

\section{Reconstruction algorithm} \label{sec:recon}
Given a flare-corrupted image $I_F \in [0,1]^{512 \times 512 \times 3}$, our goal is to train a neural network $f(I_F,\Theta)$ to predict a flare-free image $I_0$, where $\Theta$ is the trainable network weights.
Many network architectures may be suitable for our task and we evaluate two popular ones. Refer to the appendix for additional details about the network architectures and our training procedure.

\subsection{Losses} \label{sec:loss}

We want to remove \emph{only} the flare caused by a bright light source. But the flare-only images $F$ in our datasets contain \emph{both} the flare and the light source, as it is impractical to physically separate the two during capture or simulation. If we trained the network na\"{i}vely, it would attempt to hallucinate an image with the light source removed, which is not the intended use of this model and a waste of model capacity. To prevent the model from inpainting the scene behind the light source, we ignore the saturated pixels when computing the loss and assume they are due to the light source. In general, saturated pixels are unrecoverable and contain little information about the scene.

Specifically, we modify the raw network output $f(I_F, \Theta)$ with a binary saturation mask $M$ before computing losses. The mask $M$ corresponds to pixels where the luminance of input $I_F$ is greater than a threshold ($0.99$ in our experiments).
We then apply morphological operations so that small saturated regions are excluded from $M$, as they are likely part of the scene or the flare.
For pixels inside $M$, we replace it with the ground truth $I_0$ so that the loss is zero for such regions\footnote{We replace rather than exclude masked pixels because the perceptual loss in Eq.~\ref{eq:image_loss} requires a complete image.}:
\begin{equation} \label{eq:masked_prediction}
  \hat{I}_0 = I_0 \odot M + f(I_F,\Theta) \odot (1-M)\,,
\end{equation}
where $\odot$ denotes element-wise multiplication.

During training, we minimize the sum of two losses: an image loss and a \emph{flare loss}:
\begin{equation} \label{eq:total_loss}
  \lossfun = \lossfun_I + \lossfun_F\,.
\end{equation}

The image loss $\lossfun_I$ encourages the predicted flare-free image $\hat{I}_0$ to be close to the ground truth $I_0$ both photometrically and perceptually. The data term is an L1 loss on the RGB values between $\hat{I}_0$ and $I_0$. The perceptual term is computed by feeding $\hat{I}_0$ and $I_0$ through a pre-trained VGG-19 network~\cite{simonyan2014very}. Like~\cite{zhang2018single}, we penalize the absolute difference between $\Phi_{\level}(\hat{I}_0)$ and $\Phi_{\level}(I_0)$ at selected feature layers \texttt{conv1\_2}, \texttt{conv2\_2}, \texttt{conv3\_2}, \texttt{conv4\_2}, and \texttt{conv5\_2}. In summary, the image loss can be expressed as:
\begin{equation} \label{eq:image_loss}
  \lossfun_I = 
  \left\lVert \hat{I}_0 - I_0 \right\rVert _1 +
  \sum_{\level} \lambda_{\level} \left\lVert \Phi_{\level}( \hat{I}_0 ) - \Phi_{\level}\left(I_0\right) \right\rVert _1\,.
\end{equation}

The flare loss $\lossfun_F$ encourages the predicted flare to be similar to the ground-truth flare $F$, and serves to reduce artifacts in the predicted flare-free image (Sec.~\ref{sec:ablation}).
The expression for $\lossfun_F$ is the same as Eq.~\ref{eq:image_loss}, but with $\hat{I}_0$ and $I_0$ replaced by $\hat{F}$ and $F$, respectively.
Here, the predicted flare $\hat{F}$ is calculated as the difference between the network input and the masked network output:
\begin{equation}
  \hat{F} = I_F - f(I_F,\Theta) \odot (1-M)\,.
\end{equation}

\begin{figure}
  \centering
  \subfigure[Input]{%
    \includegraphics[width=0.24\linewidth]{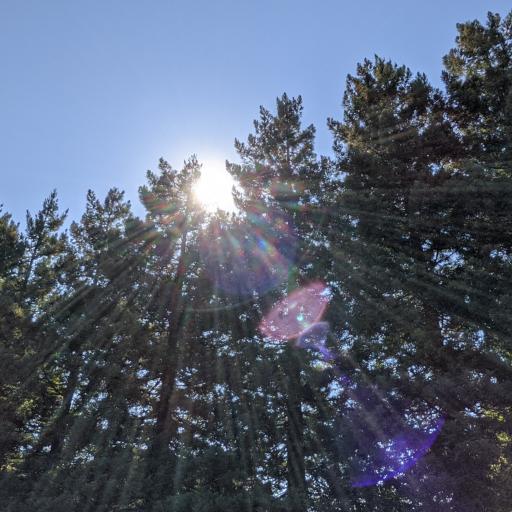}%
    \label{fig:blending:input}%
  }\hfil%
  \subfigure[CNN output]{%
    \includegraphics[width=0.24\linewidth]{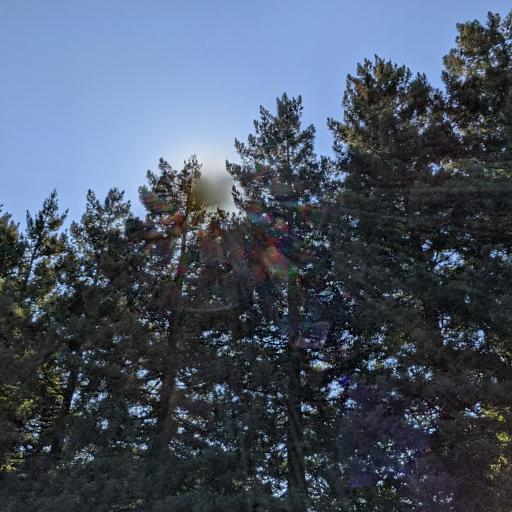}%
    \label{fig:blending:output}%
  }\hfil%
  \subfigure[Mask $M_f$]{%
    \includegraphics[width=0.24\linewidth]{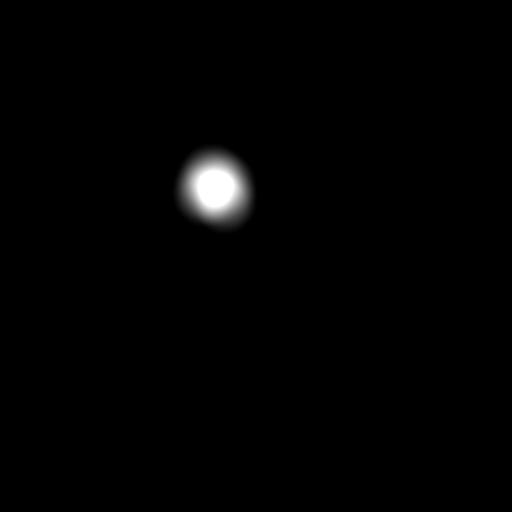}%
    \label{fig:blending:mask}%
  }\hfil%
  \subfigure[Blended]{%
    \includegraphics[width=0.24\linewidth]{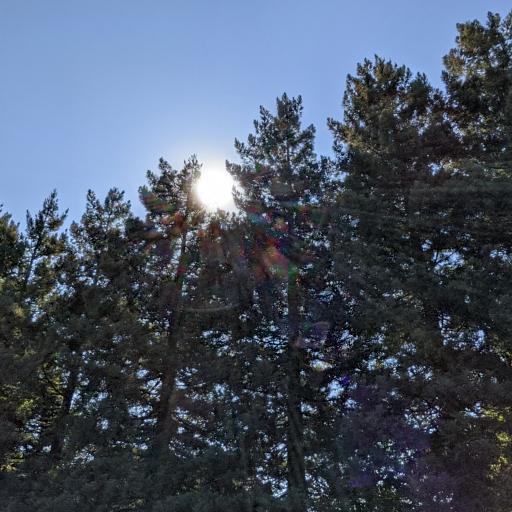}%
    \label{fig:blending:blended}%
  }%
  \caption{
    We deliberately prevent the network from learning to inpaint the saturated regions (illuminants), so its output~\subref{fig:blending:output} is undefined in these regions.
    To preserve highlights, we compute a mask~\subref{fig:blending:mask} for the saturated regions of the input.
    The masked area in the network output is then replaced by the input pixels, producing a more realistic final result~\subref{fig:blending:blended} in which the flare is removed, but not the illuminant that produced it.
  }
  \label{fig:blending}
\end{figure}

\subsection{Post-processing for light source blending} \label{sec:recon:post}

Our losses explicitly prevent the network from ``learning to inpaint'' anything in the saturated regions, so its output there can be arbitrary. In practice, it tends to remove the light source so that it is more similar to the surrounding pixels, as shown in Fig.~\ref{fig:blending:output}. Since the goal of this work is to remove the flare, and not the light source, we post-process the network output to add back the light source.

A key observation is that the flare-causing light source is likely saturated in the input image (otherwise it would not result in a visible flare). Hence, it can be identified easily based on intensity. To create a gradual transition, we feather the mask $M$ defined in Sec.~\ref{sec:loss} at its boundary to construct $M_f$ (details and parameters are in the supplement). We blend the input and output images using the feathered mask in linear space (e.g.,~Fig.~\ref{fig:blending:blended}):
\begin{equation} \label{eq:blended_output}
  I_B = I_F \odot M_f +  f(I_F,\Theta) \odot (1-M_f)\,.
\end{equation}

\section{Results} \label{sec:results}
\begin{figure*}[ht]
  \centering
  \scriptsize
  \setlength{\tabcolsep}{1pt}
  \begin{tabular}{@{}cccccccc@{}}
  
    & Input & Flare spot removal~\cite{asha2019auto} & Dehaze~\cite{he2010single} & Dereflection~\cite{zhang2018single}  & Ours + network~\cite{zhang2018single} & Ours + U-Net~\cite{ronneberger2015u} & Ground truth \\
    
    \rotatebox[origin=c]{90}{Synthetic scene} &
    \imgtextcell[67.5pt]{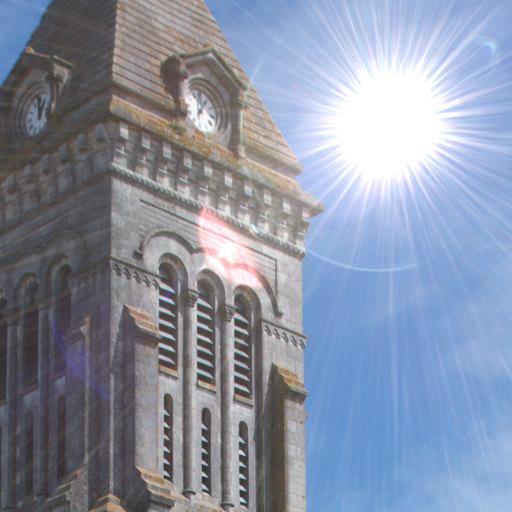}{PSNR=16.64\\SSIM=0.829} &
    \imgtextcell[67.5pt]{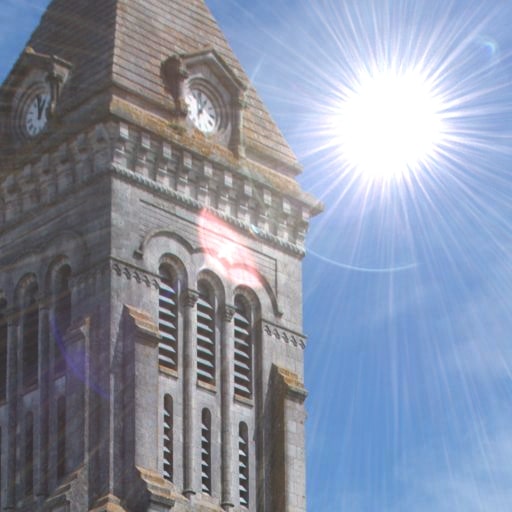}{PSNR=16.64\\SSIM=0.829} &        \imgtextcell[67.5pt]{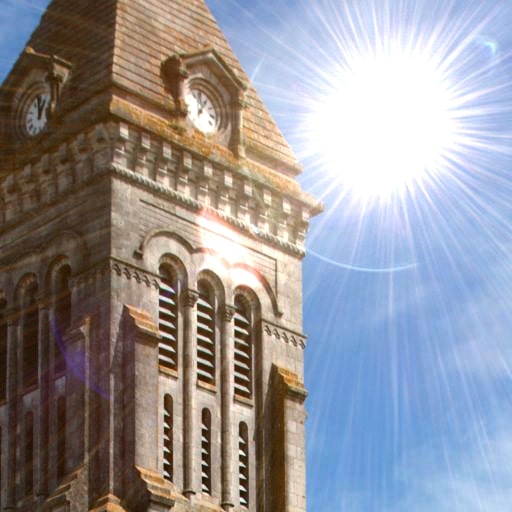}{PSNR=14.68\\SSIM=0.812} &
    \imgtextcell[67.5pt]{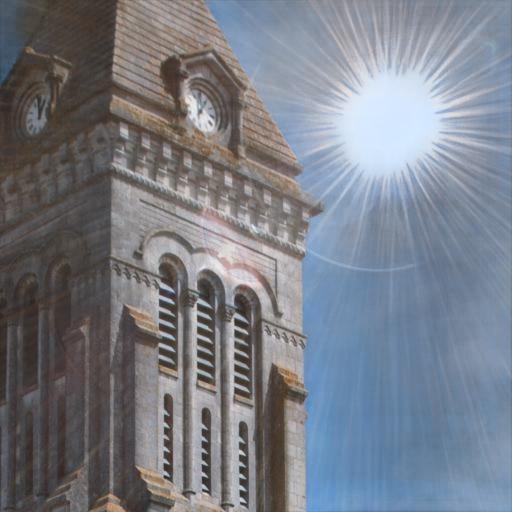}{PSNR=19.95\\SSIM=0.845} &
    \imgtextcell[67.5pt]{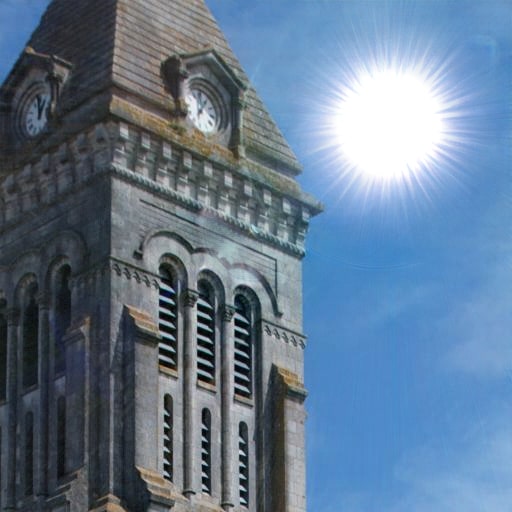}{PSNR=23.89\\SSIM=0.901} &
    \imgtextcell[67.5pt]{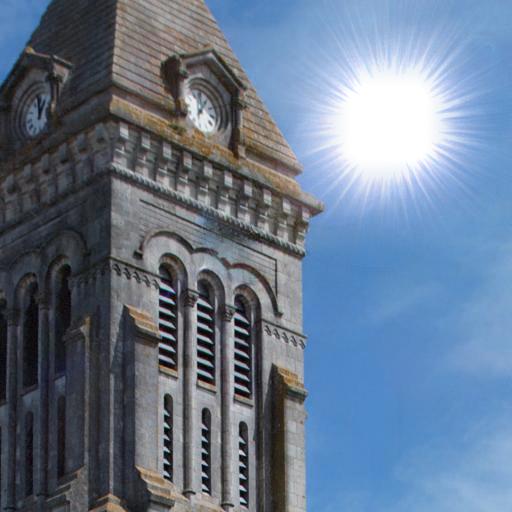}{PSNR=22.99\\SSIM=0.917} &
    \imgcell[67.5pt]{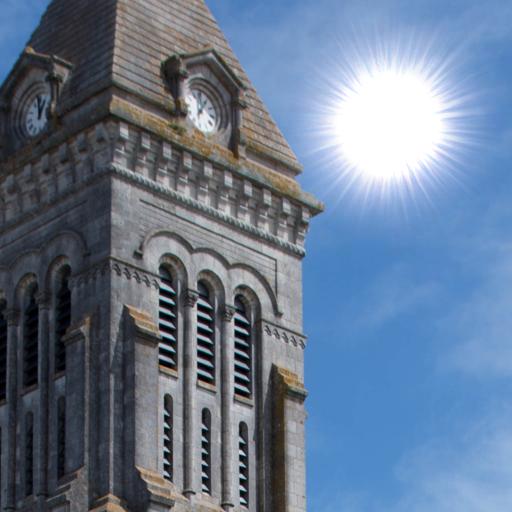}\\[-1pt]
    \rotatebox[origin=c]{90}{Real scene 1} &
    \imgtextcell[67.5pt]{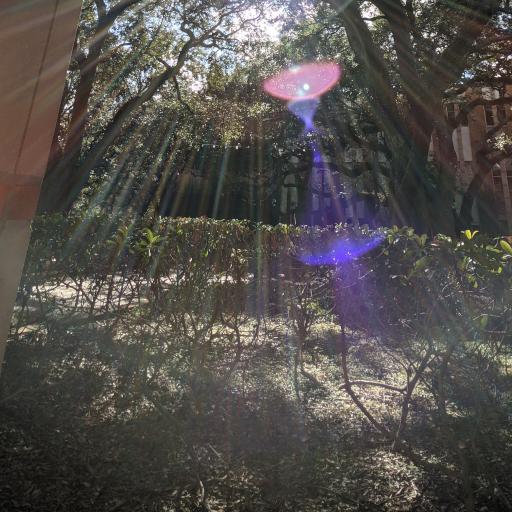}{PSNR=15.04\\SSIM=0.604} &
    \imgtextcell[67.5pt]{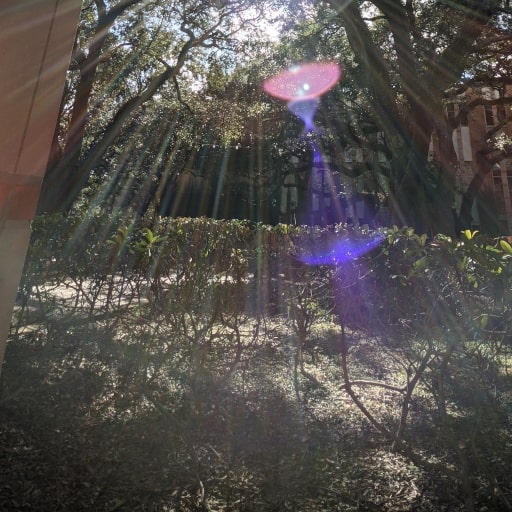}{PSNR=15.04\\SSIM=0.604} &
    \imgtextcell[67.5pt]{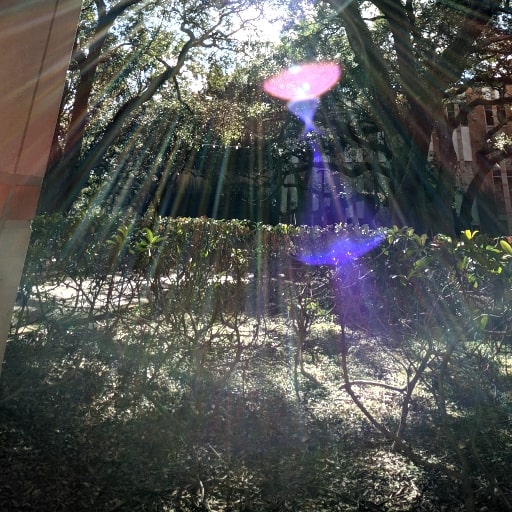}{PSNR=13.67\\SSIM=0.603} &
    \imgtextcell[67.5pt]{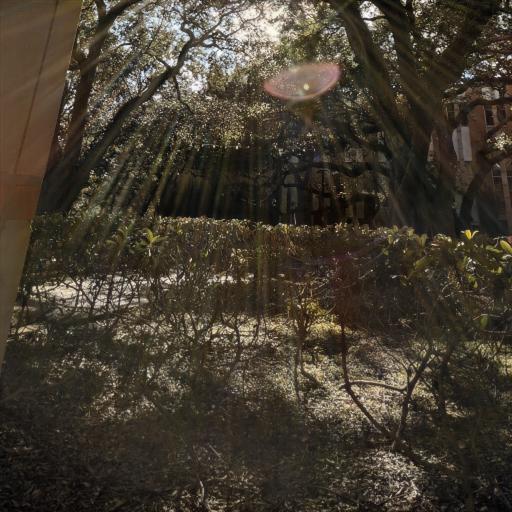}{PSNR=20.63\\SSIM=0.730} &
    \imgtextcell[67.5pt]{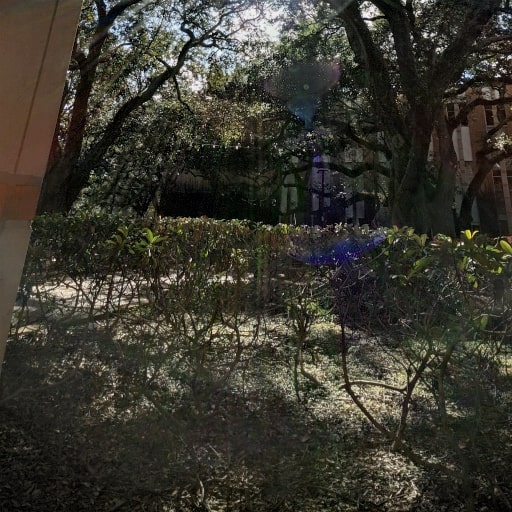}{PSNR=22.37\\SSIM=0.772} &
    \imgtextcell[67.5pt]{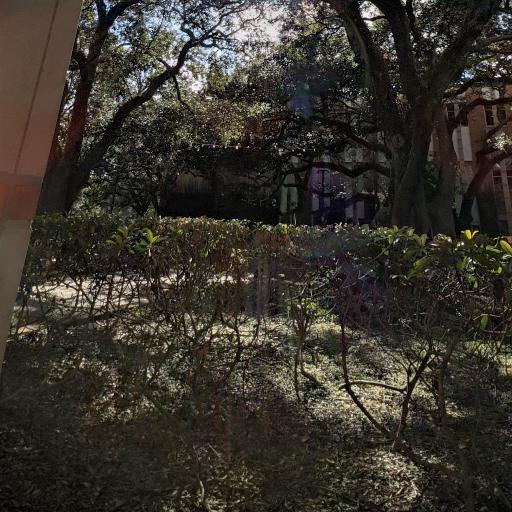}{PSNR=23.41\\SSIM=0.795} &
    \imgcell[67.5pt]{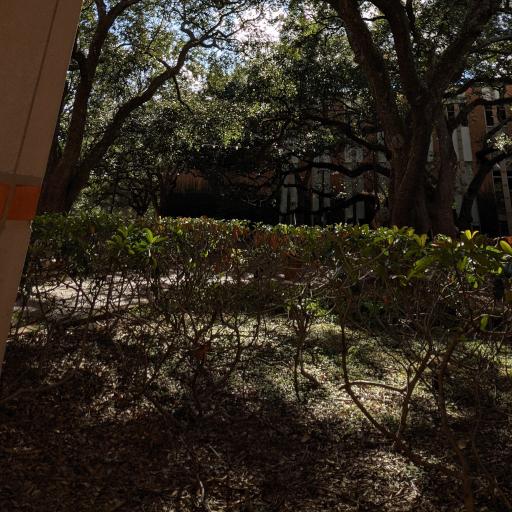}\\[-1pt]

    \rotatebox[origin=c]{90}{Real scene 2} &
    \imgtextcell[67.5pt]{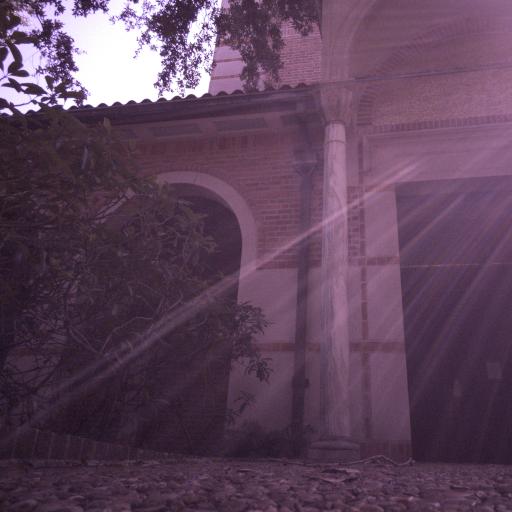}{PSNR=16.96\\SSIM=0.662} &
    \imgtextcell[67.5pt]{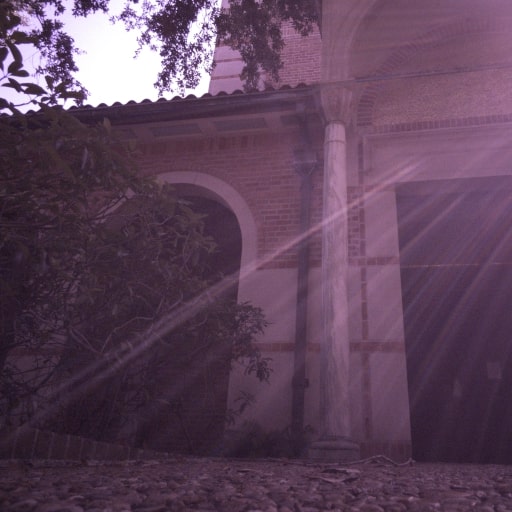}{PSNR=16.96\\SSIM=0.662} &
    \imgtextcell[67.5pt]{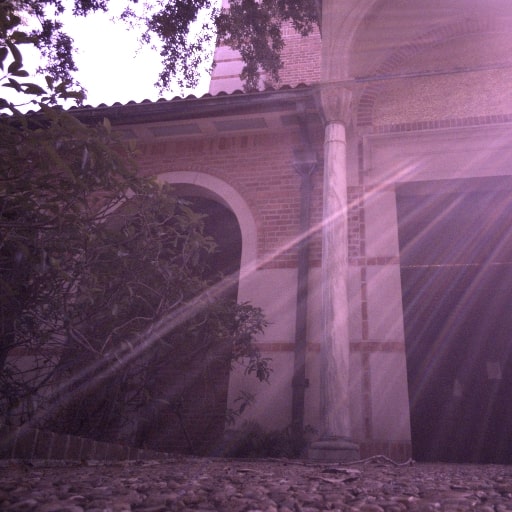}{PSNR=13.35\\SSIM=0.582} &
    \imgtextcell[67.5pt]{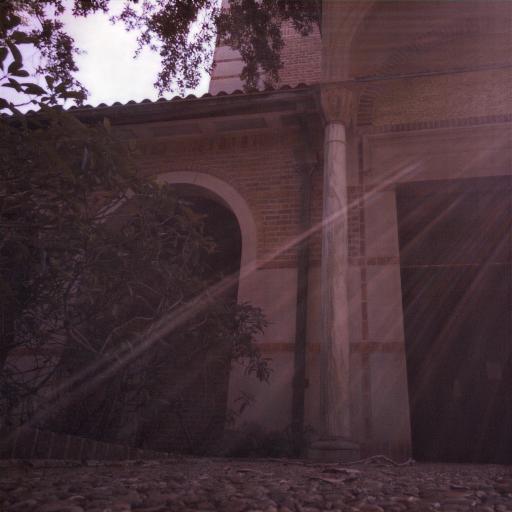}{PSNR=22.85\\SSIM=0.741} &
    \imgtextcell[67.5pt]{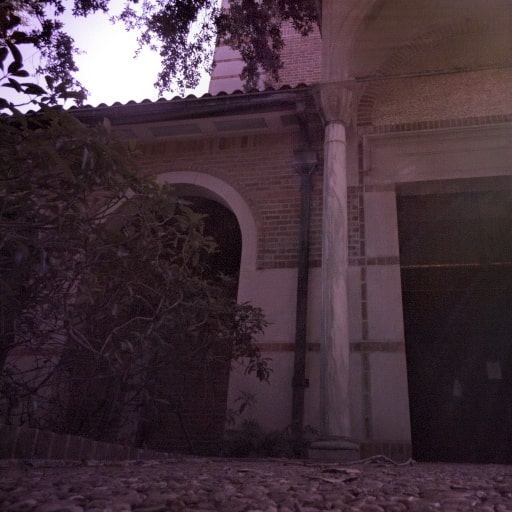}{PSNR=27.02\\SSIM=0.762} &
    \imgtextcell[67.5pt]{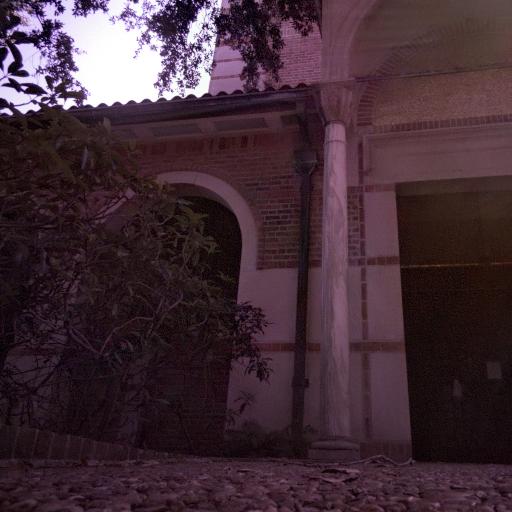}{PSNR=28.39\\SSIM=0.788} &
    \imgcell[67.5pt]{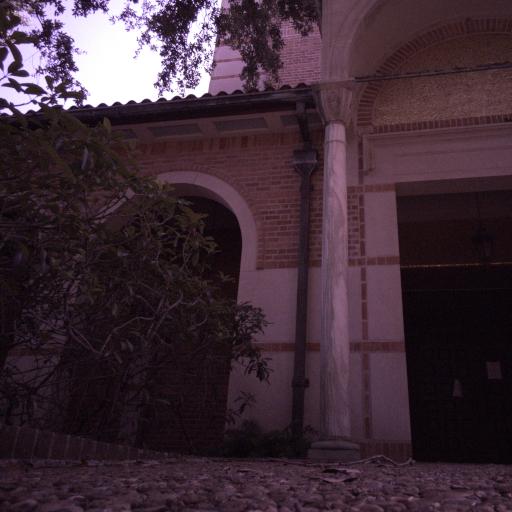}

  \end{tabular}
  
  \caption{
    Visual comparison between three related methods and ours, evaluated on both synthetic and real scenes. Networks trained using our method remove lens flare more accurately and produce cleaner outputs.
  }
  \label{fig:compare}
  \vspace{-0.1in}
\end{figure*}

\begin{figure*}[t]
  \centering
  \footnotesize
  \setlength{\tabcolsep}{1pt}
  \begin{tabular}{@{}ccccccc@{}}
     \rotatebox[origin=c]{90}{Input} &
     \imgcell[70pt]{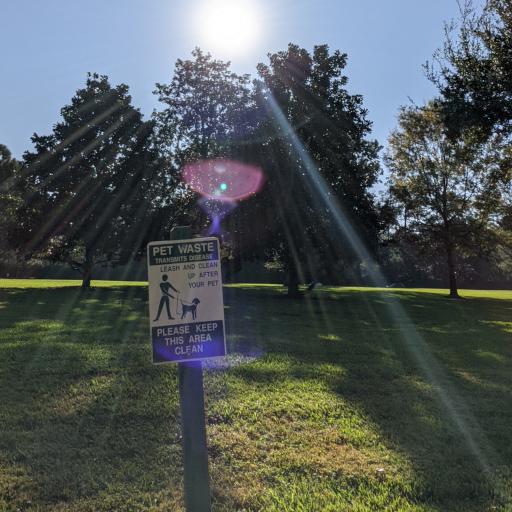} &
     \imgcell[70pt]{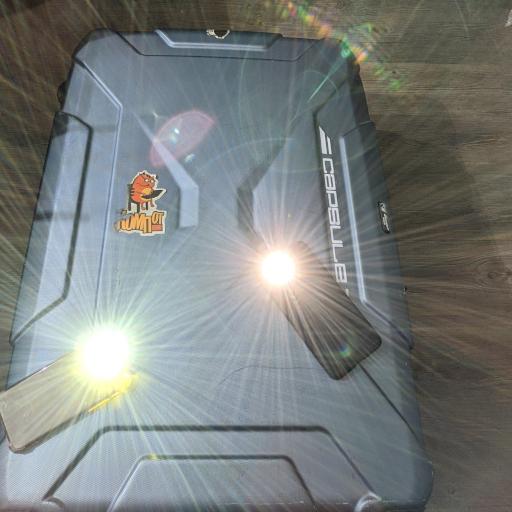} &
     \imgcell[70pt]{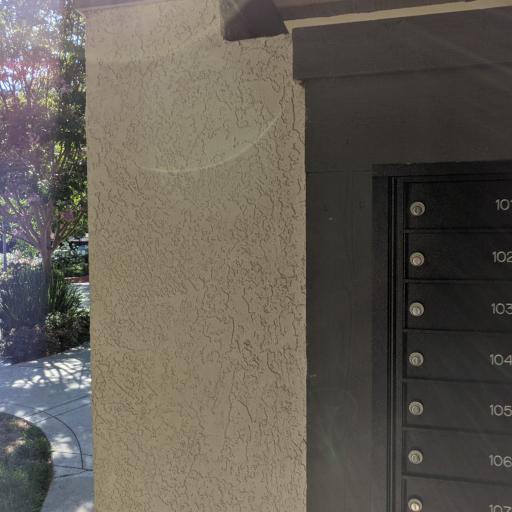} &
     \imgcell[70pt]{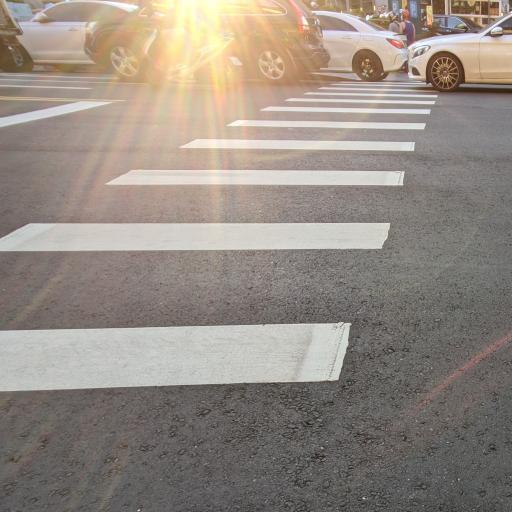} &
     \imgcell[70pt]{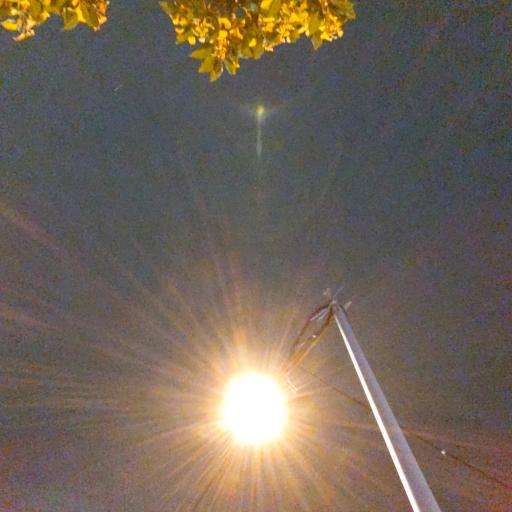} &
     \imgcell[70pt]{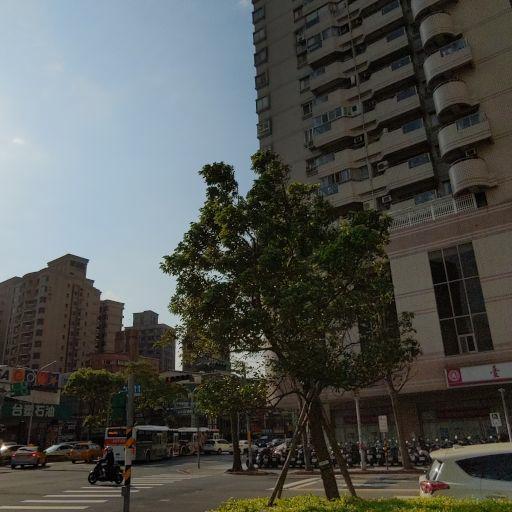} \\[-1pt]
     
     \rotatebox[origin=c]{90}{Output} &
     \imgcell[70pt]{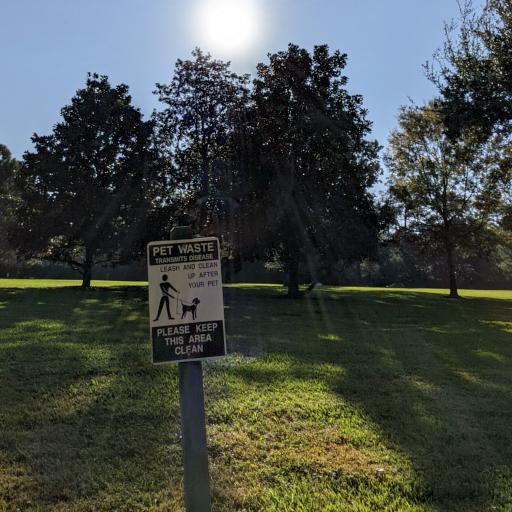} &
     \imgcell[70pt]{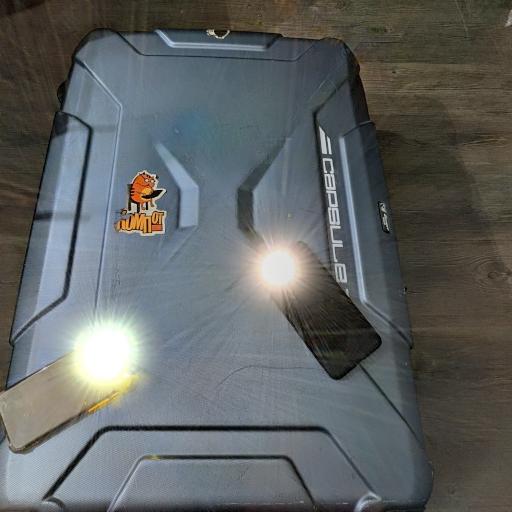} &
     \imgcell[70pt]{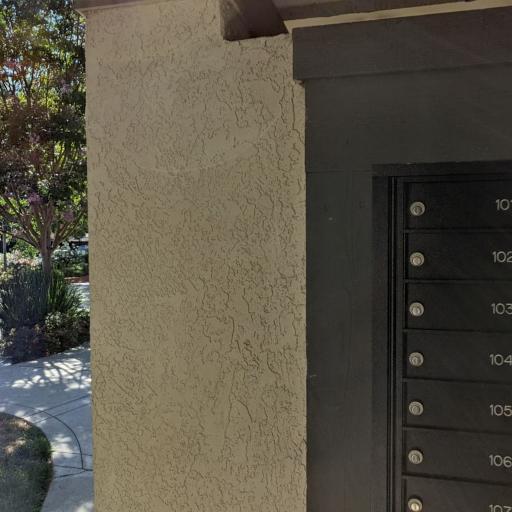} &
     \imgcell[70pt]{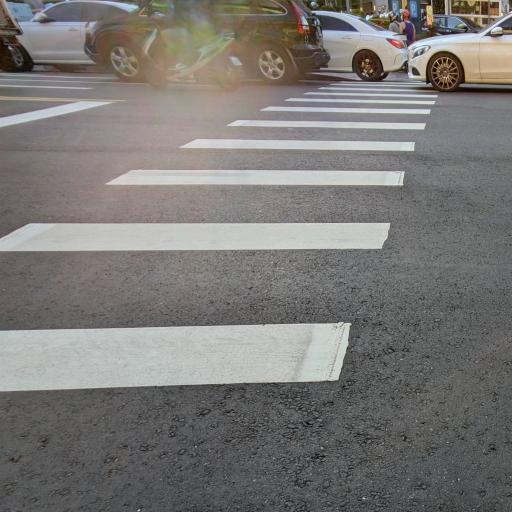} &
     \imgcell[70pt]{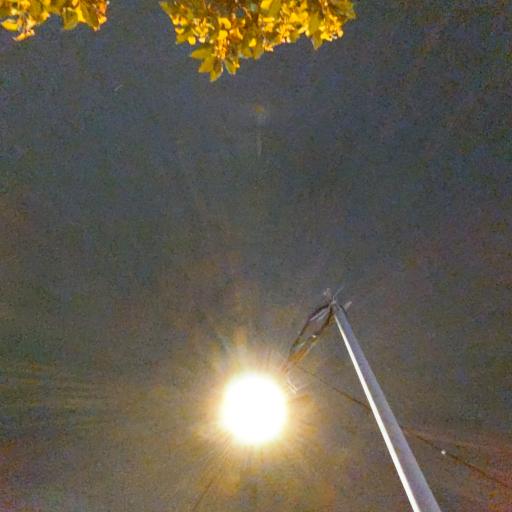} &
     \imgcell[70pt]{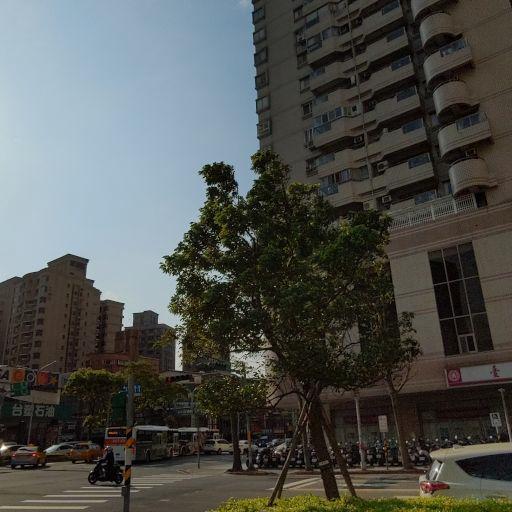}
  \end{tabular}
  \caption{
    Our method robustly removes lens flare of various shapes, colors, and locations on diverse real-world images. It generalizes reasonably to multiple light sources (column 2). When there is no significant flare (last column), the input is kept intact.
  }
  \label{fig:gallery}
  \vspace{-0.1in}
\end{figure*}

To evaluate how the models trained on semi-synthetic data generalizes, 
we use three types of test data: (a)~synthetic images with ground truth (Sec.~\ref{sec:data}), (b)~real images without ground truth, and (c)~real images with ground truth. To obtain (c), we capture a pair of images on a tripod with a bright illuminant just outside the field of view. In one image, bright flare-causing rays can enter the lens, producing artifacts. In the other image, we carefully place an occluder, also out of the field of view (e.g., a lens hood), between the illuminant and the camera, blocking the same rays.

\subsection{Comparison with prior work}

\begin{table}[b]
  \centering
  \small
  \begin{tabular}{@{}lccccc@{}}
  \toprule
                             & \multicolumn{2}{c}{Synthetic} 
                            && \multicolumn{2}{c}{Real} \\
  \cline{2-3} \cline{5-6}
  \multicolumn{1}{c}{Method} & \multicolumn{1}{@{}c@{}}{PSNR} & \multicolumn{1}{@{}c@{}}{SSIM} 
                            && \multicolumn{1}{@{}c@{}}{PSNR} & \multicolumn{1}{@{}c@{}}{SSIM} \\
  \hline
  Input image                                     & 21.13 & 0.843 && 18.57 & 0.787 \\
  \grayline
  Flare spot removal~\cite{chabert2015automated}  & 21.01 & 0.840 && 18.53 & 0.782 \\
  Flare spot removal~\cite{vitoria2019automatic}  & 21.13 & 0.843 && 18.57 & 0.787 \\
  Flare spot removal~\cite{asha2019auto}          & 21.13 & 0.843 && 18.57 & 0.787 \\
  \grayline
  Dehaze~\cite{he2010single}                      & 18.32 & 0.829 && 17.47 & 0.745 \\
  Dereflection~\cite{zhang2018single}             & 20.71 & 0.767 && 22.28 & 0.822 \\
  \grayline
  Ours + network~\cite{zhang2018single}           & 28.49 & 0.920 && 24.21 & 0.834 \\
  Ours + U-Net~\cite{ronneberger2015u}     & \textbf{30.37} & \textbf{0.944} && \textbf{25.55} & \textbf{0.850} \\ \bottomrule
  \end{tabular}
  
  \caption{
    Quantitative comparison with related methods on synthetic and real data.
  } 
  \label{tab:compare}
\end{table}

\begin{table}[b]
  \centering
  \setlength{\tabcolsep}{5pt}
  \footnotesize
  \begin{tabular}{@{}lccc@{}}
    \toprule
    \multicolumn{1}{c}{Comparison}                           & Dataset 1    & Dataset 2     & Dataset 3 \\ \hline
    Ours: Flare spot removal~\cite{chabert2015automated}    & \hphantom{0}98\%:\hphantom{0}2\%
                                                                            & 97\%:\hphantom{0}3\%
                                                                                            & 85\%:15\%  \\
    Ours: Flare spot removal~\cite{vitoria2019automatic}   & \hphantom{0}98\%:\hphantom{0}2\%  & 93\%:\hphantom{0}7\%  & 89\%:11\%  \\
    Ours: Flare spot removal~\cite{asha2019auto}            & 100\%:\hphantom{0}0\%    & 99\%:\hphantom{0}1\%     & 88\%:12\%  \\
    Ours: Dehaze~\cite{he2010single}                        & \hphantom{0}96\%:\hphantom{0}4\% & 91\%:\hphantom{0}9\%  & 92\%:\hphantom{0}8\%  \\
    Ours: Dereflection~\cite{zhang2018single}               & \hphantom{0}83\%:17\%    & 78\%:22\%     & 64\%:36\%  \\
    \grayline
    Average                                                 & \hphantom{0}95\%:\hphantom{0}5\%  & 92\%:\hphantom{0}8\%  & 84\%:16\%\\                                               
    \bottomrule
  \end{tabular}
  
  \caption{
    Percent of images where the users favor our results (ours + U-Net) vs. prior work. Dataset~1 is captured using the same lens design as in Sec.~\ref{sec:data:reflective}. Dataset~2 is captured using five other lens types with different focal lengths. Dataset~3 contains images from \cite{chabert2015automated}. We outperform existing methods in all categories, even on Chabert's own dataset~\cite{chabert2015automated}.
  } 
  \label{tab:user}
\end{table}

We provide quantitative and visual comparisons in Table~\ref{tab:compare} and Fig.~\ref{fig:compare}. To eliminate the influence of the light source when computing metrics, the masked region is replaced by ground truth pixels following Eq.~\ref{eq:masked_prediction}.

We evaluate all recent work in flare removal~\cite{asha2019auto,chabert2015automated,vitoria2019automatic}. Notably, none of them attempt the general flare removal task. Instead, they use hand-crafted heuristics to remove one particular subset of flare (e.g., glare spots). As such, they have little effect on other artifacts such as reflections and streaks and cause the PSNR and SSIM to be close to or even identical to the input.
Since haze and reflections are two common flare artifacts, we also compare with dehazing~\cite{he2010single} and dereflection~\cite{zhang2018single} algorithms on our data.

For our method, we trained two variants, one using the architecture from~\cite{zhang2018single}, and the other using the popular U-Net~\cite{ronneberger2015u}. Our method significantly outperforms existing methods and demonstrates the importance of our pipeline and dataset. We use the U-Net variant for the rest of the paper since it performs better.

Finally we also conducted a user study with 20 participants where each user is presented with a real image with lens flare alongside two predicted flare-free images: one from the U-Net and the other from one of the 5 baselines. We then ask the user to identify which of the two did better at removing lens flare.
We use 52 images from 3 different sets: images captured by the same type of lens as in Sec.~\ref{sec:data:reflective}, images captured using five other lenses with different focal lengths, and images taken from \cite{chabert2015automated}. To avoid bias, we shuffle the images in each instance of the study. As Table~\ref{tab:user} shows, our method outperforms all others by a significant margin on all 3 datasets. Even on the dataset by Chabert~\cite{chabert2015automated}, users strongly preferred our method to theirs (85\% vs. 15\%). Unsurprisingly, it performs slightly worse when tested on lenses not present in our training set.

\subsection{Ablation study} \label{sec:ablation}

In this section, we study two key components in our procedure to demonstrate their impact on the output.

\begin{table}[h]
\centering
\small
\begin{tabular}{ccccc}
\toprule
      & No $\mathcal{L}_F$ & No sim. data & No captured data & \textbf{Full} \\ \hline
PSNR  & 24.84         & 24.44        & 23.77            & \textbf{25.55} \\
SSIM  & 0.841         & 0.843        & 0.828            & \textbf{0.850} \\
\bottomrule
\end{tabular}
\caption{Ablation study on the flare loss and different flare data.}
\vspace{-0.2in}
\end{table}

\begin{figure}
  \centering
  \subfigure[Input]{%
    \begin{tikzpicture}
      \begin{scope}[zoomin]
        \node (n0) {\includegraphics[width=0.32\linewidth]{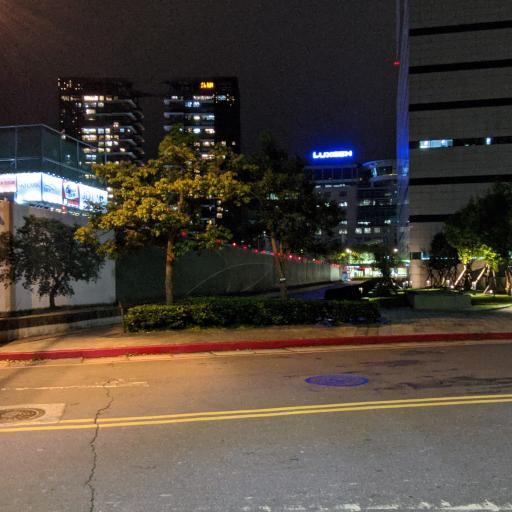}};
        \spy [width=0.22\linewidth, height=0.11\linewidth, magnification=4] on (0.4, 0.36) in node [below left=of n0.north east];
      \end{scope}
    \end{tikzpicture}%
    \label{fig:ablation_loss:input}%
  }\hfil%
  \subfigure[No flare loss]{%
    \begin{tikzpicture}
      \begin{scope}[zoomin]
        \node (n0) {\includegraphics[width=0.32\linewidth]{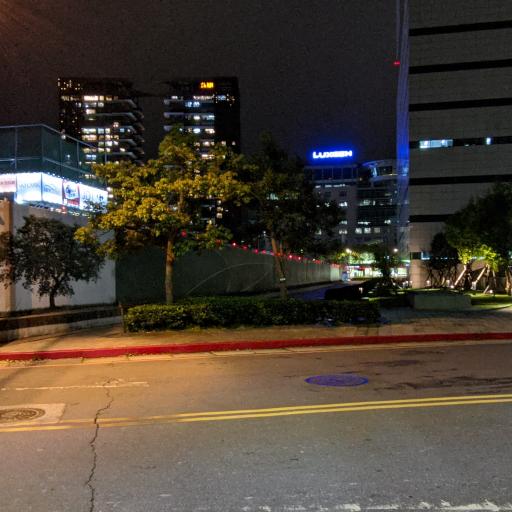}};
        \spy [width=0.22\linewidth, height=0.11\linewidth, magnification=4] on (0.4, 0.36) in node [below left=of n0.north east];
      \end{scope}
    \end{tikzpicture}%
    \label{fig:ablation_loss:no_flare_loss}%
  }\hfil%
  \subfigure[With flare loss]{%
    \begin{tikzpicture}
      \begin{scope}[zoomin]
        \node (n0) {\includegraphics[width=0.32\linewidth]{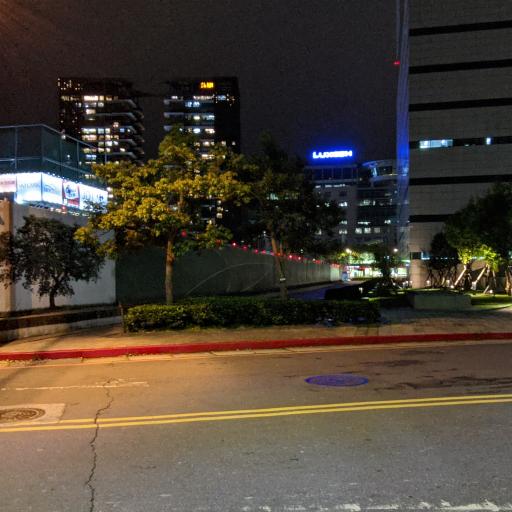}};
        \spy [width=0.22\linewidth, height=0.11\linewidth, magnification=4] on (0.4, 0.36) in node [below left=of n0.north east];
      \end{scope}
    \end{tikzpicture}%
    \label{fig:ablation_loss:with_flare_loss}%
  }%
  \caption{
    Without our flare loss $\lossfun_F$, bright regions in the input \subref{fig:ablation_loss:input} are incorrectly removed, especially on images taken at night \subref{fig:ablation_loss:no_flare_loss}. $\lossfun_F$ makes the model more robust to such errors \subref{fig:ablation_loss:with_flare_loss}.
  }
  \label{fig:ablation_loss}
  \vspace{-0.2in}
\end{figure}

\myparagraph{Flare loss}
Since most flares are brighter than the underlying scene, we need to ensure that the network does not simply learn to darken all bright regions. We explicitly model this in our flare loss $\lossfun_F$. In Fig.~\ref{fig:ablation_loss}, we show test set results from the models trained with and without $\lossfun_F$. Without $\lossfun_F$, the network tends to remove some parts of bright objects even if they are not part of the flare. 

\begin{figure}[t]
  \centering
  \subfigure[Input]{%
    \begin{tikzpicture}
      \begin{scope}[zoomin]
        \node (n0) {\includegraphics[width=0.24\linewidth]{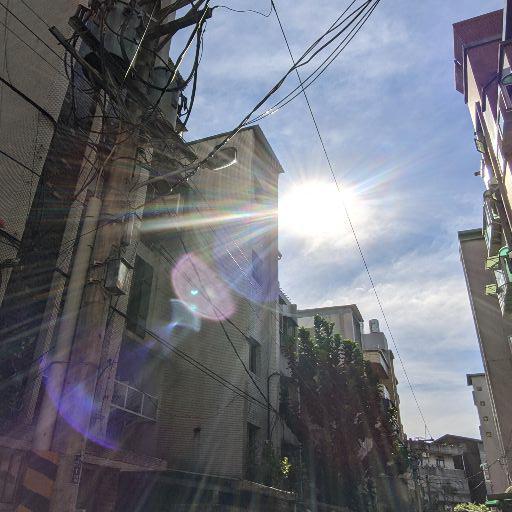}};
        \spy [width=0.12\linewidth, height=0.06\linewidth, magnification=4, yellow] on (0.6, 1.08) in node [below left=of n0.north east];
        \spy [width=0.12\linewidth, height=0.12\linewidth, magnification=4] on (0.67, 0.77) in node [above left=of n0.south east];
      \end{scope}
    \end{tikzpicture}%
    \label{fig:data:input}%
  }\hfil%
  \subfigure[Captured]{%
    \begin{tikzpicture}
      \begin{scope}[zoomin]
        \node (n0) {\includegraphics[width=0.24\linewidth]{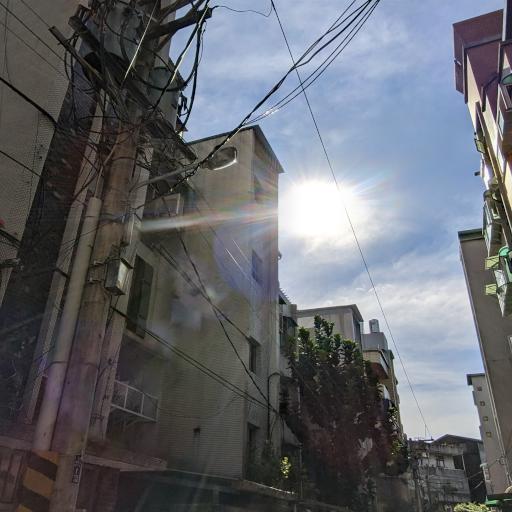}};
        \spy [width=0.12\linewidth, height=0.06\linewidth, magnification=4, yellow] on (0.6, 1.08) in node [below left=of n0.north east];
        \spy [width=0.12\linewidth, height=0.12\linewidth, magnification=4] on (0.67, 0.77) in node [above left=of n0.south east];
      \end{scope}
    \end{tikzpicture}%
    \label{fig:data:captured}%
  }\hfil%
  \subfigure[Simulated]{%
    \begin{tikzpicture}
      \begin{scope}[zoomin]
        \node (n0) {\includegraphics[width=0.24\linewidth]{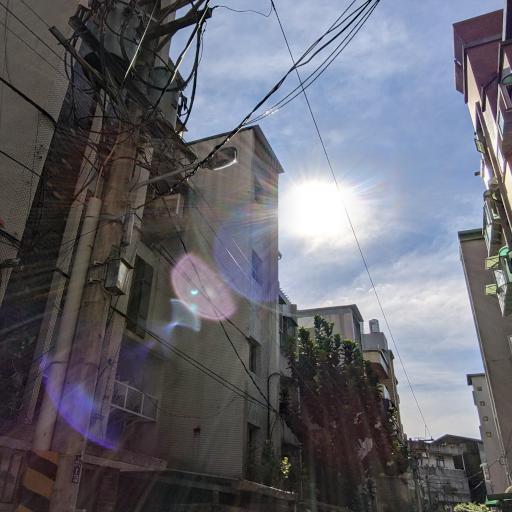}};
        \spy [width=0.12\linewidth, height=0.06\linewidth, magnification=4, yellow] on (0.6, 1.08) in node [below left=of n0.north east];
        \spy [width=0.12\linewidth, height=0.12\linewidth, magnification=4] on (0.67, 0.77) in node [above left=of n0.south east];
      \end{scope}
    \end{tikzpicture}%
    \label{fig:data:simulated}%
  }\hfil%
  \subfigure[Both]{%
    \begin{tikzpicture}
      \begin{scope}[zoomin]
        \node (n0) {\includegraphics[width=0.24\linewidth]{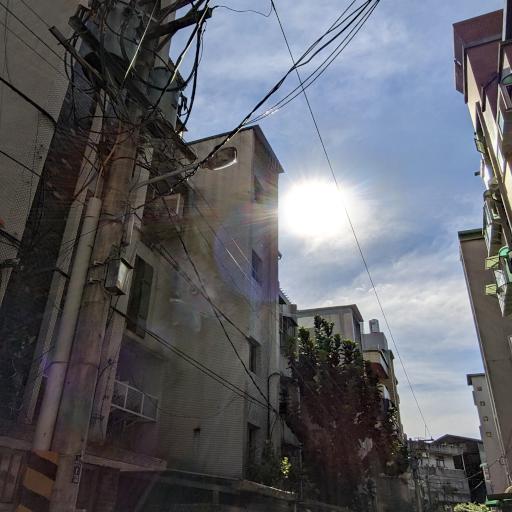}};
        \spy [width=0.12\linewidth, height=0.06\linewidth, magnification=4, yellow] on (0.6, 1.08) in node [below left=of n0.north east];
        \spy [width=0.12\linewidth, height=0.12\linewidth, magnification=4] on (0.67, 0.77) in node [above left=of n0.south east];
      \end{scope}
    \end{tikzpicture}%
    \label{fig:data:both}%
  }%
  \caption{
    The model trained with captured flare only \subref{fig:data:captured} (containing mostly reflective and limited scattering flare) is less effective at removing streak-like scattering artifacts, whereas the simulation-only model \subref{fig:data:simulated} is unable to remove reflective flare patterns. Training with both datasets \subref{fig:data:both} produces superior results.
  }
  \label{fig:ablation_data}
  \vspace{-0.1in}
\end{figure}

\myparagraph{Captured and simulated flare data}
In Sec.~\ref{sec:data}, we mentioned that the captured data mostly accounts for reflective flare, whereas the simulated data covers the scattering case. To show that both components are necessary, we train two ablated models, each with one of the sources excluded. As expected, models trained with flare-only images taken from the captured or simulated dataset alone underperform the model trained with both datasets, as shown in Fig.~\ref{fig:ablation_data}.

\begin{figure}
  \centering
  \footnotesize
  \setlength{\tabcolsep}{1pt}
   \begin{tabular}{@{}ccccc@{}}
    & $f =$ 27mm & $f =$ 44mm & iPhone & Fisheye\footnotemark\\
    
    \rotatebox[origin=c]{90}{Input} &
    \imgcell[55pt]{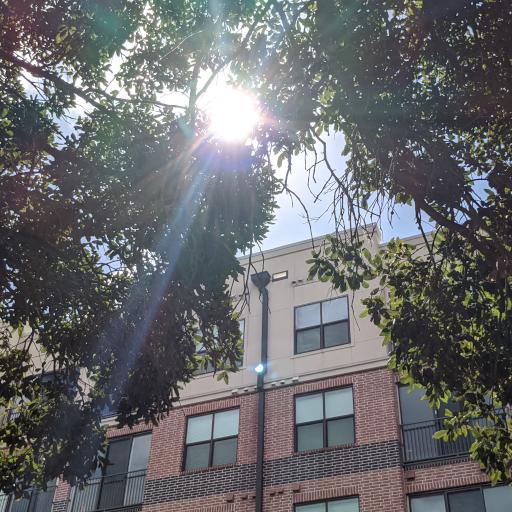} &
    \imgcell[55pt]{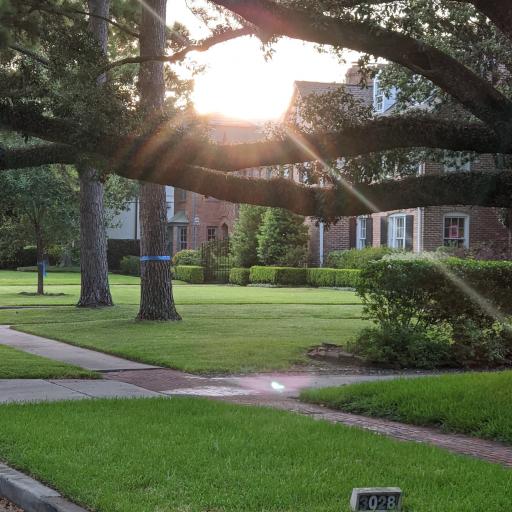} &
    \imgcell[55pt]{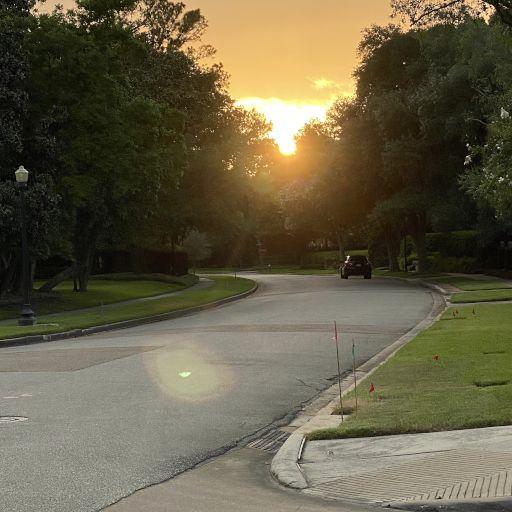} &
    \imgcell[55pt]{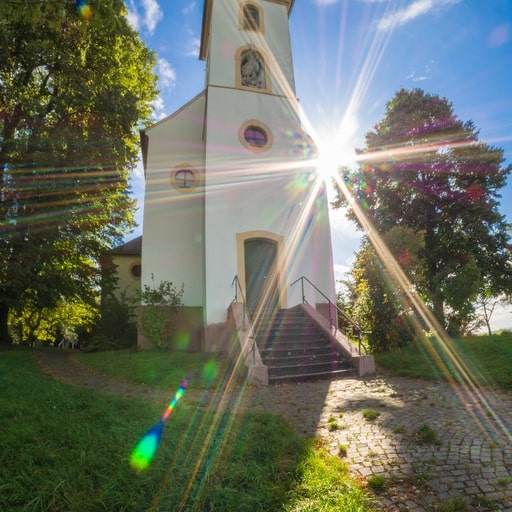}\\[-1pt]
    
    \rotatebox[origin=c]{90}{Output} &
    \imgcell[55pt]{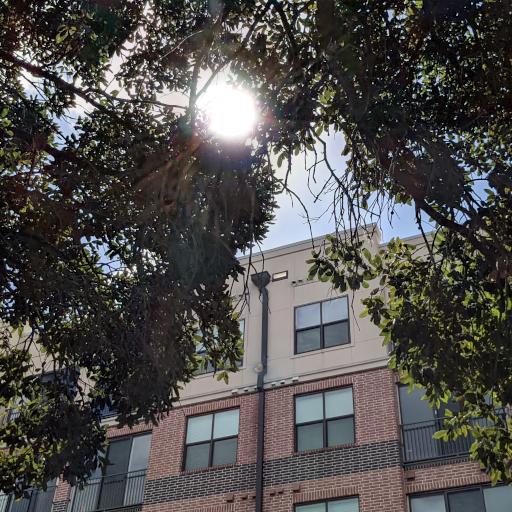} &
    \imgcell[55pt]{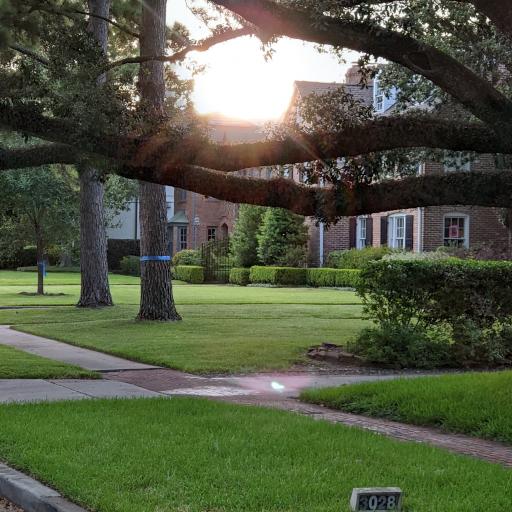} &
    \imgcell[55pt]{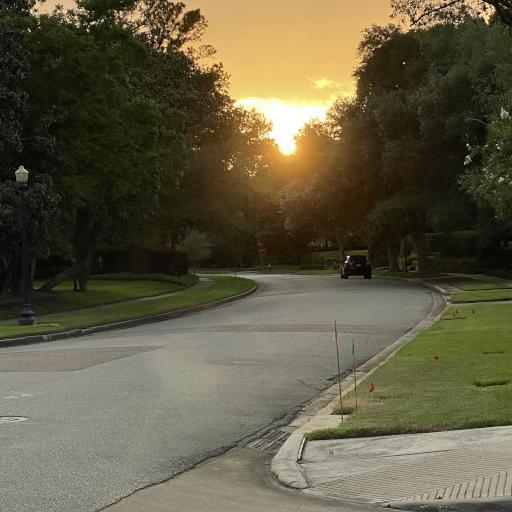} &
    \imgcell[55pt]{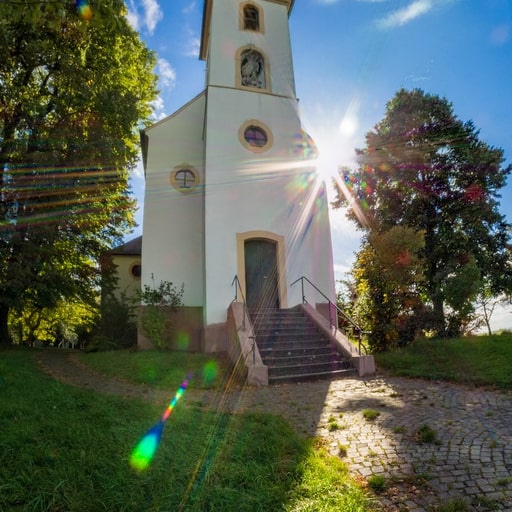} \\
  \end{tabular}
  \caption{
    Although our dataset contains real flare patterns from only one camera of an Android smartphone ($f =$ 13mm), the trained model generalizes effectively to the phone's other lenses ($f =$ 27 and 44mm), and another smartphone camera (iPhone).
    When tested on a drastically different lens design (e.g., a fisheye mirrorless camera), the model performs less well on reflective flare, as expected, and still manages to remove scattering flare.
  }
  \label{fig:general}
    \vspace{-0.13in}
\end{figure}

\subsection{Generalization}

\footnotetext{Photo by Flickr user barit / \href{https://creativecommons.org/licenses/by-sa/2.0/}{CC BY-SA}.}

\myparagraph{Across scenes}
As our semi-synthetic dataset contains diverse flare patterns and scenes, the trained model generalizes well across a wide variety of scene types. As shown in Fig.~\ref{fig:gallery}, the input images contain flares with different shapes, colors, and locations, as well as scenes with varying subjects and lighting conditions. The model produces a high-quality output in most scenarios. When there is no flare in the input image, the network correctly performs a no-op.

\myparagraph{Across cameras}
As mentioned in Sec.~\ref{sec:data:reflective}, all of our reflective flare training images come from one smartphone camera with focal length $f =$ 13mm. We also test our model on other camera designs excluded from training. As shown in Fig.~\ref{fig:general}, the model is still able to reduce lens flare effectively, echoing the findings of the user study in Table~\ref{tab:user}.

That said, there is a limit on how far the model can generalize. For example, the model performs less well on images taken with an extremely different lens, such as a fisheye (Fig.~\ref{fig:general}, last column). This is especially true for the lens-dependent reflective component, as discussed in Sec.~\ref{sec:data:reflective}.
Domain adaptation for drastically different camera designs is an interesting avenue for future work.

\subsection{High-resolution images}

Our network is trained on $512 \times 512$ images. The na\"{i}ve way to apply our method to a higher-resolution input is to train at the desired resolution (e.g., $2048 \times 2048$), which requires 16x more bandwidth at both training and test time.

\begin{figure}[t]
  \centering
  \subfigure[Pipeline: blue and red blocks represent high- and low-res respectively.]{%
    \includegraphics[width=\linewidth]{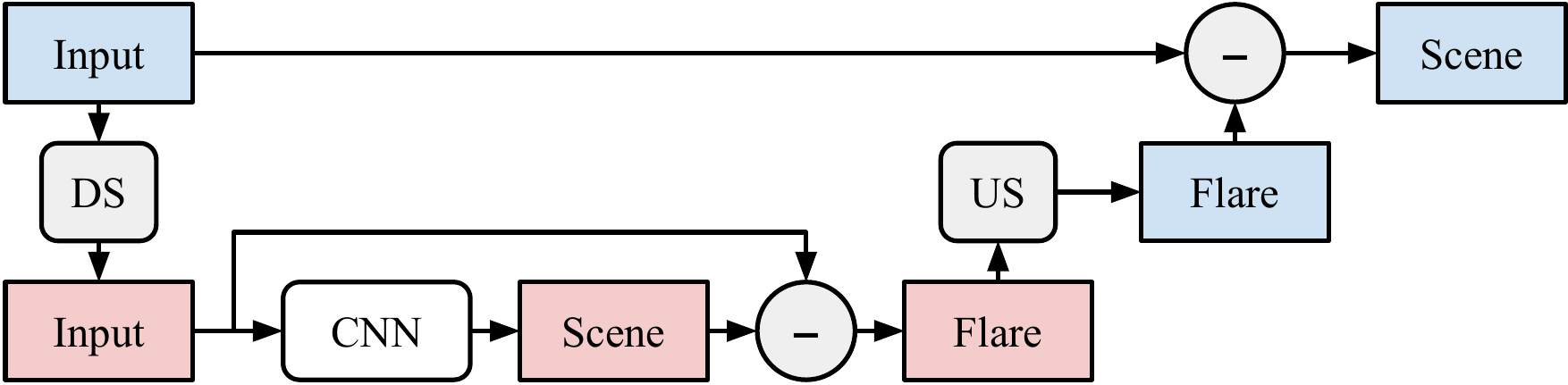}%
    \label{fig:high_res:pipeline}%
  }\\%
  \subfigure[Input]{%
    \begin{tikzpicture}
      \begin{scope}[zoomin]
        \node (n0) {\includegraphics[width=0.3\linewidth]{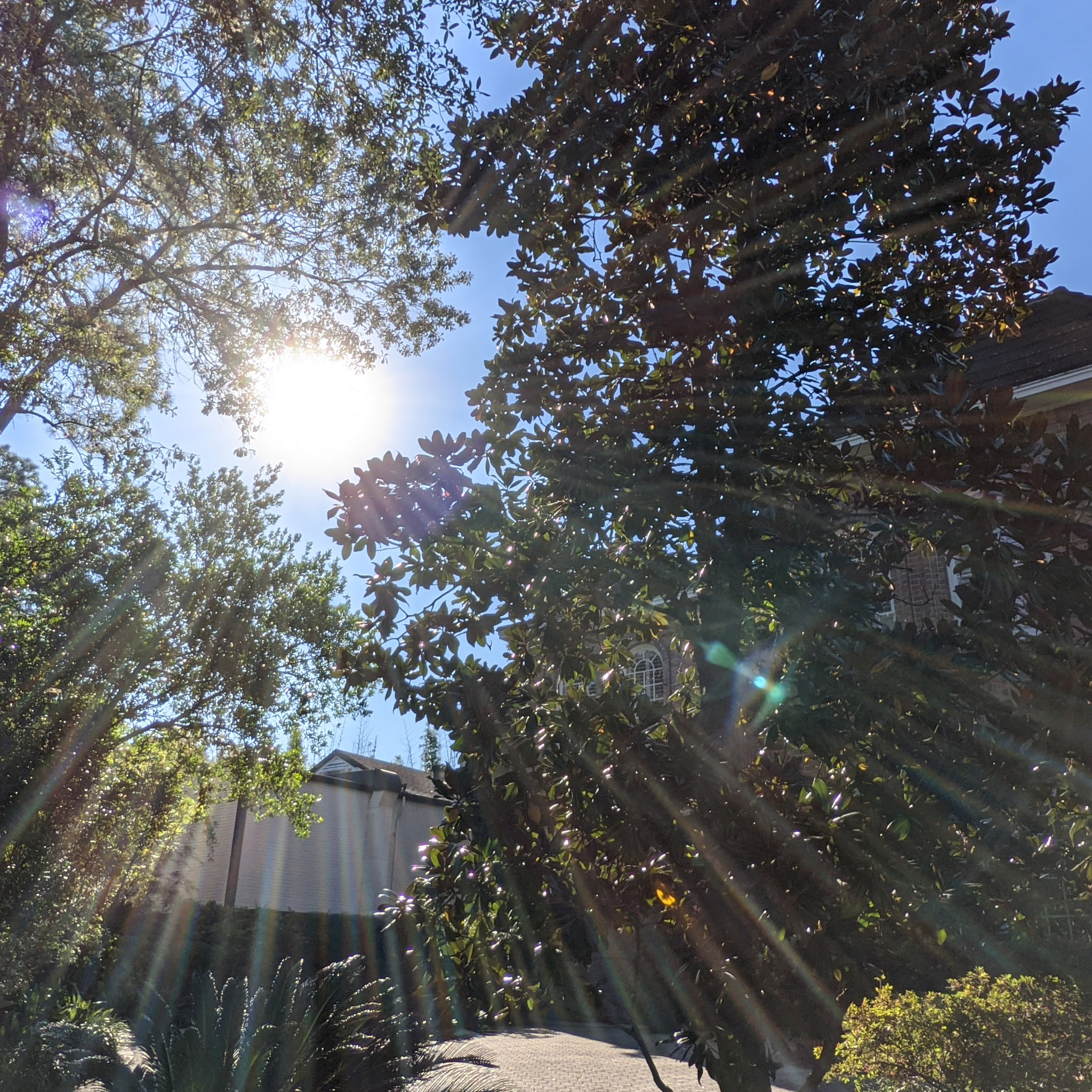}};
        \spy [width=0.2\linewidth, height=0.13\linewidth, magnification=6] on (1.45, 0.88) in node [below left=of n0.north east];
      \end{scope}
    \end{tikzpicture}%
    \label{fig:high_res:input}%
  }\hfil%
  \subfigure[Low-res output]{%
    \begin{tikzpicture}
      \begin{scope}[zoomin]
        \node (n0) {\includegraphics[width=0.3\linewidth]{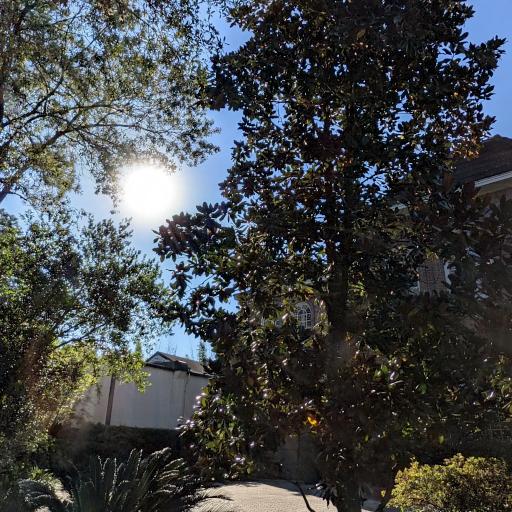}};
        \spy [width=0.2\linewidth, height=0.13\linewidth, magnification=6] on (1.45, 0.88) in node [below left=of n0.north east];
      \end{scope}
    \end{tikzpicture}%
    \label{fig:high_res:output_low}%
  }\hfil%
  \subfigure[High-res output]{%
    \begin{tikzpicture}
      \begin{scope}[zoomin]
        \node (n0) {\includegraphics[width=0.3\linewidth]{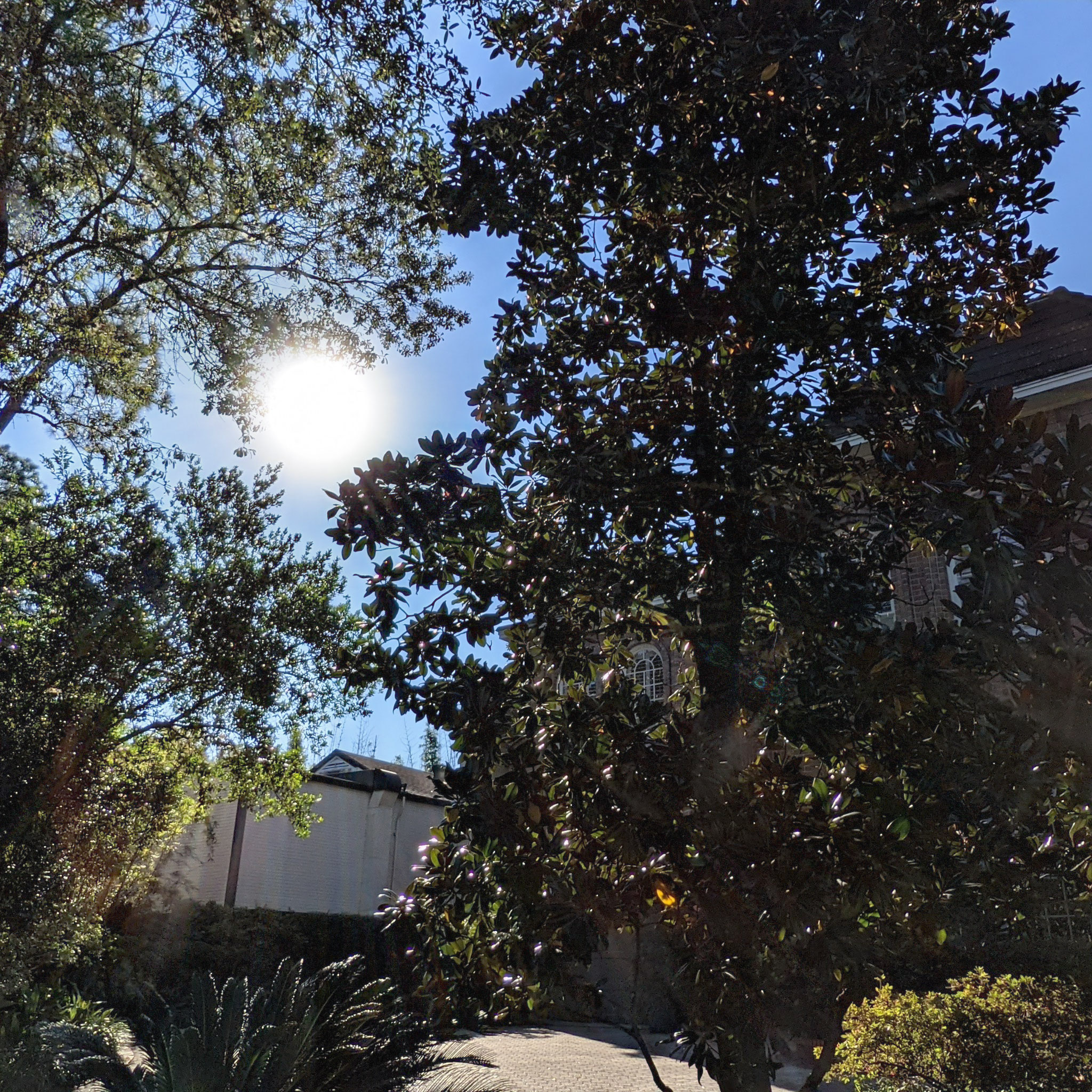}};
        \spy [width=0.2\linewidth, height=0.13\linewidth, magnification=6] on (1.45, 0.88) in node [below left=of n0.north east];
      \end{scope}
    \end{tikzpicture}%
    \label{fig:high_res:output_high}%
  }%
  \caption{
    Because lens flares are mostly low-frequency, our algorithm can be trivially extended to high-res inputs. As \subref{fig:high_res:pipeline} shows, we downsample (DS) a high-res input \subref{fig:high_res:input}, predict a flare-free image \subref{fig:high_res:output_low} with our network (CNN), and compute the flare as the difference. This predicted flare is then upsampled (US) and subtracted from the input image to produce a high-res flare-free output \subref{fig:high_res:output_high}.
  }
  \label{fig:high_res}
  \vspace{-0.2in}
\end{figure}

Fortunately, we can leverage the fact that lens flare is predominantly a low-frequency artifact. For a high-resolution image, we bilinearly downsample the input, predict a low-resolution flare-only image, bilinearly upsample it back to full resolution, and subtract it from the original input (see Fig.~\ref{fig:high_res}). This allows a network trained at a fixed low resolution to process high-resolution images without significant quality loss. On a Xeon E5 CPU, processing time for a $2048 \times 2048$ input is reduced from 8s to 0.55s when running inference at $512 \times 512$.

\section{Conclusion} \label{sec:con}
We introduced a novel, tractable, and physically-realistic model for lens flare. By building a semi-synthetic data generation pipeline using a principled image formation model, we are able to train convolutional neural networks to recover a clean flare-free image from a single flare-corrupted image without the need for real training data. Our method is shown to achieve accurate results across a range of scenes and cameras. To our knowledge, this is the \emph{first} general-purpose lens flare removal technique.

\myparagraph{Acknowledgments}
We thank Sam Huynh, Chung-Po Huang, Xi Chen, and Lu Gao for their help in setting up the  lab, and acknowledge the support from NSF grants IIS-1652633 and IIS-1730574.

\newpage

\appendix 

\section{Simulating random scattering flare}

    To simulate scattering flare with our wave optics model, we randomly sample the aperture function $A$, the light source's 3D location $(x, y, z)$, and the spectral response function $\SRF$.
    
    \myparagraph{Notations}
    \begin{itemize}
        \item $N(\mu, \sigma^2)$: normal distribution with mean $\mu$ and standard deviation $\sigma$.
        \item $U(a, b)$: uniform distribution on the interval $[a, b]$.
        \item $k_\lambda$: the wavenumber corresponding to wavelength $\lambda$. $k_\lambda = 2\pi / \lambda$.
    \end{itemize}
    
    \subsection{Aperture function}
    
    As Fig.~\ref{fig:simulation:aperture} illustrates, we simulate dust and scratches with random dots and polylines on a clean disk-shaped aperture of radius $R = 3000$ pixels.
    
    For each aperture function, we randomly generate $n_d \sim N(30, 5^2)$ dots with maximum radius $r_d^\text{max} \sim N(100, 50^2)$. Each individual dot's radius $r_d$ is drawn independently from $U(0, r_d^\text{max})$.
    
    Additionally, we also generate $n_p \sim N(30, 5^2)$ polylines, each of which consists $n_l \sim U(1, 16)$ line segments connected from end to end. The maximum line width for each aperture function is $w_p^\text{max} \sim N(20, 5^2)$, and the width of each individual line is drawn independently from $U(0, w_p^\text{max})$.
    
    The opacity of each of the dots and polylines is sampled independently from $U(0, 1)$, and its location $(u, v)$ on the aperture plane is also drawn uniformly.
    
    \subsection{Phase shift}
    
    The light source's 3D location $(x, y, z)$ determines the phase shift $\phi_\lambda$ of the pupil function $P_\lambda$. As shown in Eq.~\ref{eq:phase_shift}, it consists of a linear term $\phi_\lambda^\mathrm{S}$ and a defocus term $\phi_\lambda^\mathrm{DF}$.
    
    The linear phase shift $\phi_\lambda^\mathrm{S}$ on the aperture plane becomes a spatial shift on the sensor plane due to the Fourier transform $\mathcal{F}\{\cdot\}$ in Eq.~\ref{eq:psf}. For an $800\times800$ image sensor with the center as the origin, the center of the light source $(x, y)$ is sampled from $x, y \sim U(-500, 500)$.
    
    The term $\phi_\lambda^\mathrm{DF}(z)$ is the defocus aberration due to the mismatch between the in-focus depth $z_0$ of the lens and the actual depth $z$ of the light source. The analytical expression for $\phi_\lambda^{DF}(z)$ is
    \begin{align}
        \phi_\lambda^\mathrm{DF}(z) 
        &= {k_\lambda } \frac{{u^2 + v^2}}{2}\left( {\frac{1}{z} - \frac{1}{z_0}} \right) \nonumber \\
        &= {k_\lambda } \cdot r(u,v)^2 \cdot W_m(z)
    \end{align}
    where $k_\lambda$ is the wavenumber,
    $(u,v)$ are aperture coordinates, and
    $r(u,v) = {\sqrt {u^2 + v^2} }/{R}$ is the relative displacement on the aperture plane. $W_m(z)$ is defined as
    \begin{equation}
    \label{eq:Wm}
        {W_m}(z) = \frac{{R^2}}{2}\left( {\frac{1}{z} - \frac{1}{{z_0}}} \right)
    \end{equation}
    and quantifies the amount of defocus in terms of the depth $z$.
    Since we do not target any specific camera devices (i.e., specific $R$ and $z_0$ values), it suffices to sample the value of $W_m$ directly from $N(0, \sigma^2)$, with $\sigma = 5 / {k_{\lambda=550\text{nm}}}$.
    
    \subsection{Spectral response function}
    
    The spectral response $\SRF_c(\lambda)$ describes the sensitivity of color channel $c$ to wavelength $\lambda$, where $c \in \{\mathrm{R}, \mathrm{G}, \mathrm{B}\}$.
    
    We model $\SRF_c(\lambda)$ as a Gaussian probability density function $N(\mu_c, \sigma_c^2)$. The mean $\mu_c$ is the central wavelength of each channel in nanometers, and is drawn from $\mu_\mathrm{R} \sim U(620, 640)$, $\mu_\mathrm{G} \sim U(540, 560)$, and $\mu_\mathrm{B} \sim U(460, 480)$. The standard deviation $\sigma_c$ represents the width of the passband of each color channel, and is drawn independently for each channel from $U(50, 60)$.
    
    We then discretize the wavelength $\lambda$ in $\SRF_c(\lambda)$ for each of the RGB channels at 5nm intervals from $\lambda = 380\text{nm}$ to $740\text{nm}$, resulting in a 73-vector for each channel. The 3 vectors are stacked to produce an instance of $\SRF$ in the form of a $3 \times 73$ matrix.

    \begin{figure}
        \includegraphics[width=\linewidth]{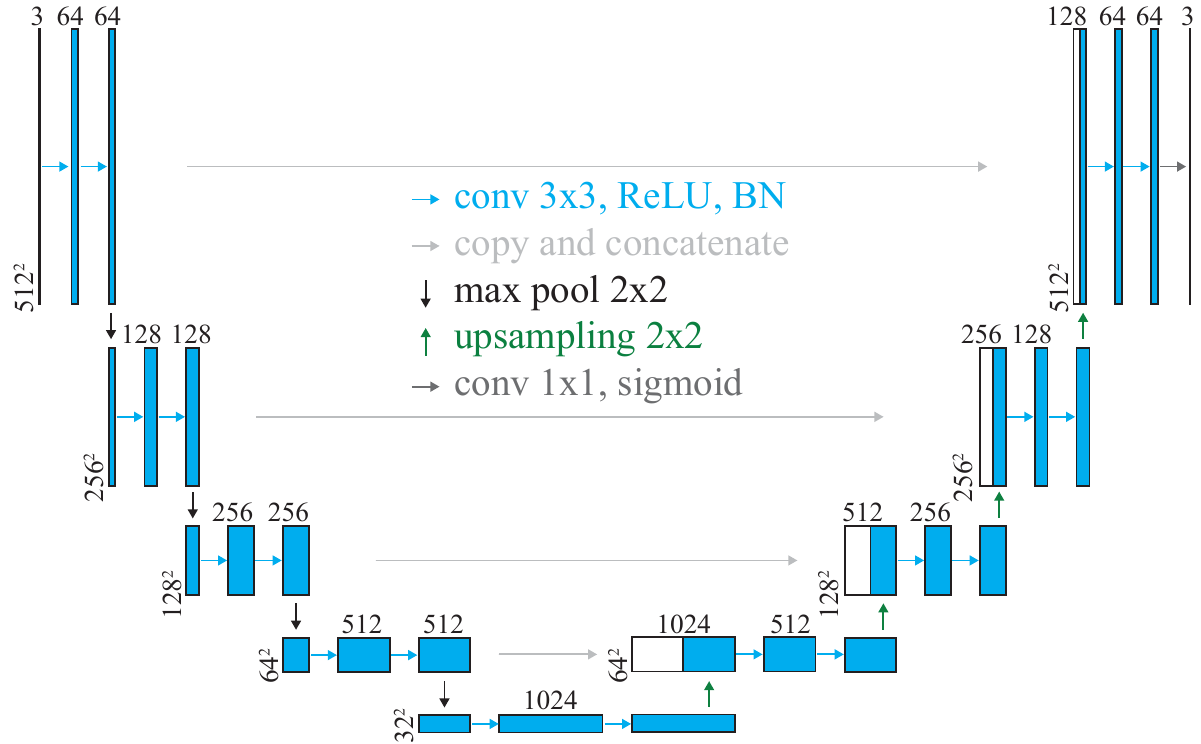}
        \caption{
          The U-Net architecture~\cite{ronneberger2015u} we use in the flare removal task. 
        }
        \label{fig:unet}%
    \end{figure}
    
    \begin{figure*}
        \centering
        \subfigure[Input and our output]{\includegraphics[height=1.9in]{%
          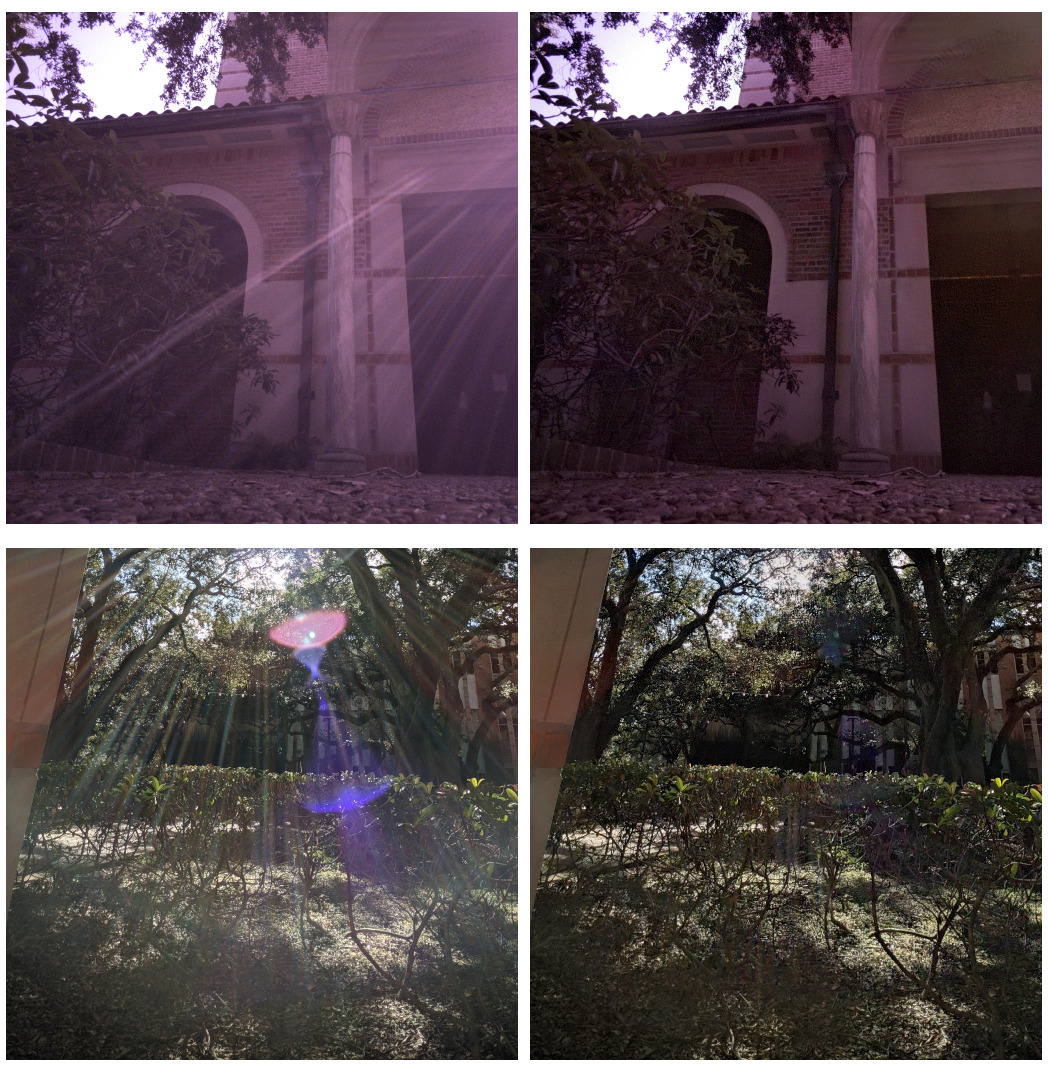}%
          \label{subfig:rgb}%
        }\hfil%
        \subfigure[Semantic segmentation \cite{deeplab2018}]{%
          \includegraphics[height=1.9in]{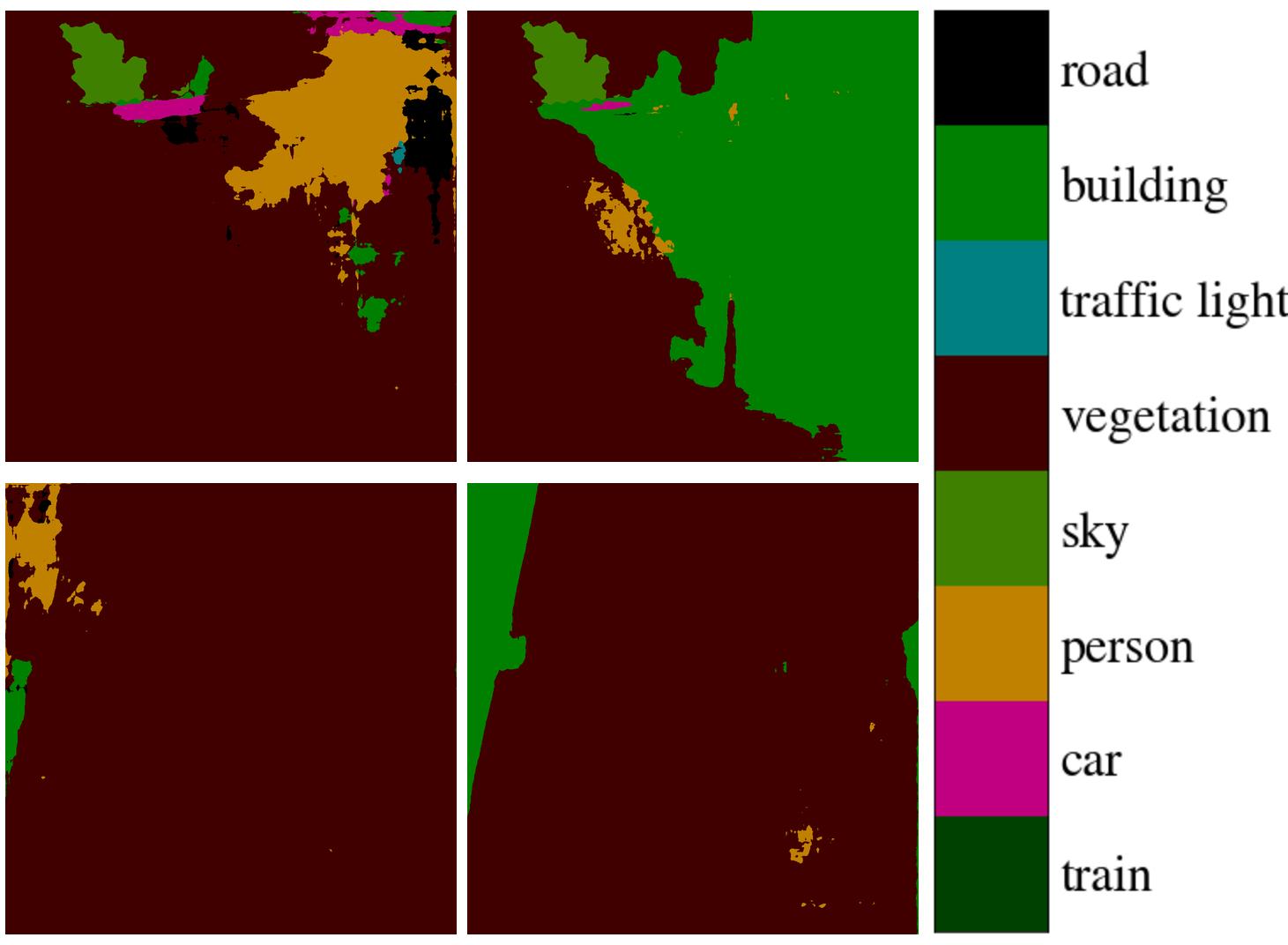}%
          \label{subfig:segmentation}%
        }\hfil%
        \subfigure[Monocular depth \cite{midas2020}]{%
          \includegraphics[height=1.9in]{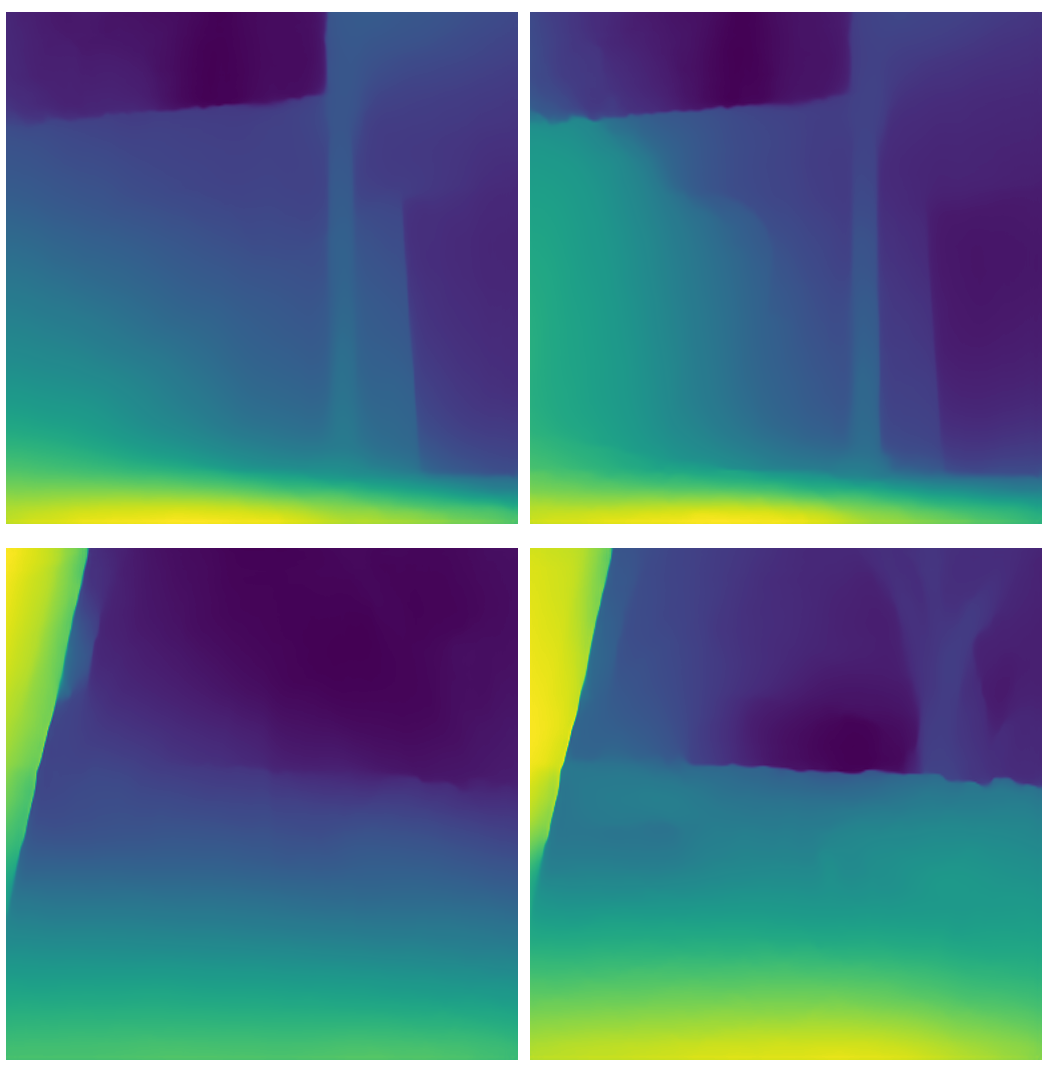}%
          \label{subfig:depth}%
        }%
        \caption{
          A pair of images with flare and our output \subref{subfig:rgb}. Semantic segmentation \subref{subfig:segmentation} has much less misclassification once flare is removed using our method. Similarly, monocular depth estimation \subref{subfig:depth} is more accurate (e.g.,~the bush on the left in the top image and the tree in the distance in the lower image).
        } 
        \label{fig:downstream_tasks}
    \end{figure*}

\section{Flare-only image augmentation}

    We augment the captured flare-only images $F$ by applying random geometric and color transformations. For geometric augmentation, we generate a $3 \times 3$ affine transform matrix with random rotation ($\sim U(0, 2\pi)$), random translation ($\sim U(-10, 10)$ pixels), random shear angle ($\sim U(-\pi/9,\pi/9)$), and random scale ($\sim U(0.9, 1.2)$ for $x$ and $y$ independently). For color augmentation, we multiply each color channel with a random weight in $U(0, 10)$. Pixel values are clipped to $[0, 1]$ after the transformed flare image is composited with the clean image.

\section{Networks and training details}

    To show the effectiveness of our method, we train two popular neural networks with distinct architectures. These networks have not previously been used for the task of removing lens flare. Using our method, both networks produce satisfactory results. We detail the network architectures and training configurations below.

    \subsection{Context aggregation network (CAN)}
    
    We repurpose a network originally designed for reflection removal~\cite{zhang2018single}, which is a variant of the original CAN~\cite{yu2015multi}. Starting from the $512 \times 512 \times 3$ input image, the network first extracts features from a pre-trained VGG-19 network~\cite{simonyan2014very} at layers \texttt{conv1\_2}, \texttt{conv2\_2}, \texttt{conv3\_2}, \texttt{conv4\_2}, and \texttt{conv5\_2}. Next, these features are combined with the input image to form a 1475-channel tensor, which is subsequently reduced to 64 channels with a $1 \times 1$ convolution.
    It is then passed through eight $3 \times 3$ convolution layers with 64 output channels at dilation rates of 1 -- 64.
    The last layer is a $1\times 1$ convolution with 3 output channels.
    
    \subsection{U-Net}
    
    The U-Net~\cite{ronneberger2015u} is shown in Fig.~\ref{fig:unet}. The input image $I_F $ is $512 \times 512 \times 3$. Each convolution operator consists of a $3 \times 3$ convolution and ReLU activation. We use $2 \times 2$ max pooling for downsampling and resize--convolution for upsampling. Concatenation is applied between the encoder and decoder to avoid the vanishing gradient problem. At the final layer, we use a sigmoid function to squeeze the activations to the $[0, 1]$ range.

    \subsection{Training details}
    
    We train both CAN and U-Net using our semi-synthetic dataset.
    The clean images come from the Flickr dataset of~\cite{zhang2018single}.
    The flare-only images are generated as described in the previous Sections, and augmented on the fly during training.
    Training lasted 1.2M iterations (approximately 60 epochs over the 20k images in the Flickr dataset) with batch size 2 on an Nvidia V100 GPU. We use the Adam optimizer~\cite{kingma2017adam} with default parameters and a fixed learning rate of $10^{-4}$.

\section{Mask feathering}

    As mentioned in Sec.~\ref{sec:recon:post} of the main text, we detect the light source and create a feathered mask $M_f$ to smoothly blend the bright illuminants into the network output.
    
    Starting from a binary mask $M$ (the pixels where the input luminance is greater than 0.99), we apply morphological opening with a disk kernel of size equal to 0.5\% of the image size. To find the primary illuminant, we partition the mask into connected components and find the equivalent diameter $D$ of the largest region (the diameter of a circle with the same area as the region). We then blur the binary mask $M$ using a disk kernel of diameter $D$. Finally, we scale the blurred mask intensity by 3 in order to guarantee that all the pixels inside the illuminant are saturated after blurring, and clip the mask to $[0,1]$. This is the $M_f$ used in Eq.~\ref{eq:blended_output} which has a feathered edge.

\section{Flare removal for downstream tasks}

    In Fig.~\ref{fig:downstream_tasks}, we show that removing lens flare may benefit downstream tasks such as semantic segmentation and depth estimation. We look forward to thoroughly investigating the effect of reduced flare on a range of computer vision algorithms.

\section{More results}

\begin{figure*}[]
      \centering
      \scriptsize
      \setlength{\tabcolsep}{1pt}
      \begin{tabular}{@{}cccccccc@{}}
      
        & Input & Flare spot removal~\cite{asha2019auto} & Dehaze~\cite{he2010single} & Dereflection~\cite{zhang2018single}  & Ours + network~\cite{zhang2018single} & Ours + U-Net~\cite{ronneberger2015u} & Ground truth \\

        \rotatebox[origin=c]{90}{Real scene 3} &
        \imgtextcell[67.5pt]{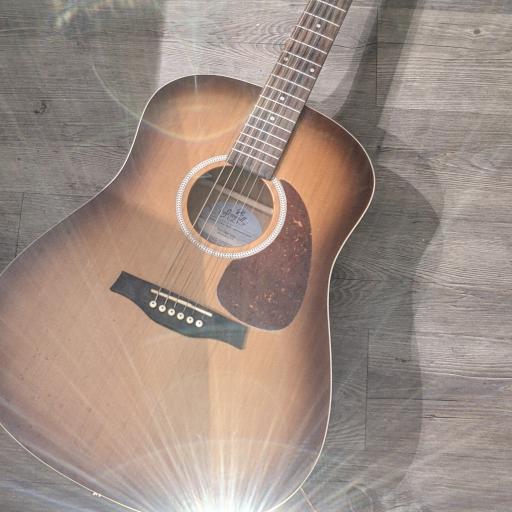}
        {PSNR=17.67\\SSIM=0.866} &
        \imgtextcell[67.5pt]{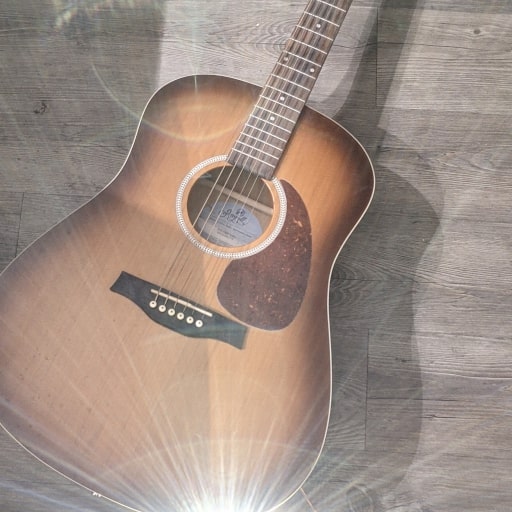}
        {PSNR=17.67\\SSIM=0.866} &
        \imgtextcell[67.5pt]{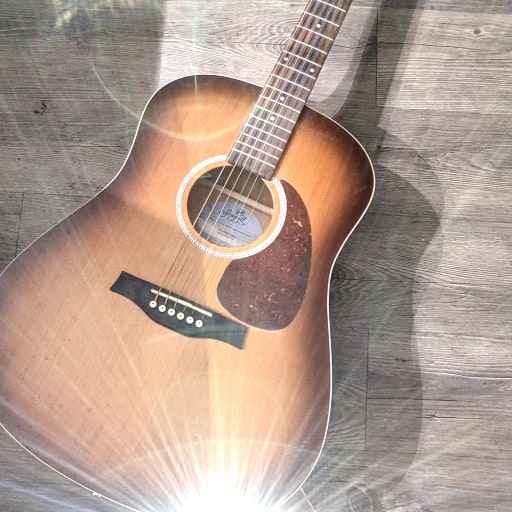}
        {PSNR=14.58\\SSIM=0.823} &
        \imgtextcell[67.5pt]{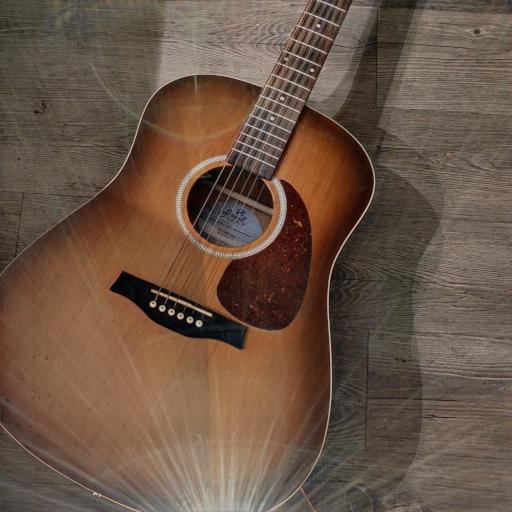}
        {PSNR=16.69\\SSIM=0.813} &
        \imgtextcell[67.5pt]{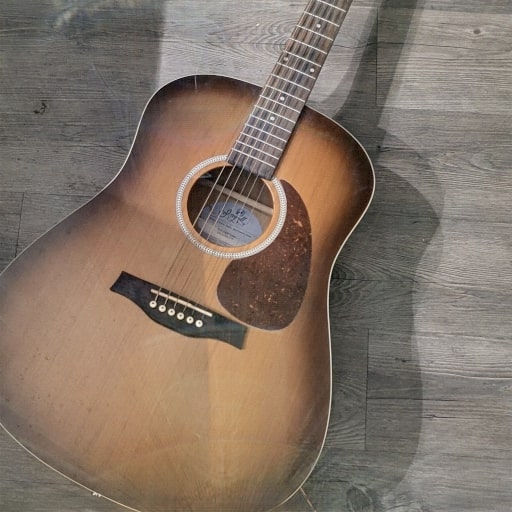}
        {PSNR=25.94\\SSIM=0.899} &
        \imgtextcell[67.5pt]{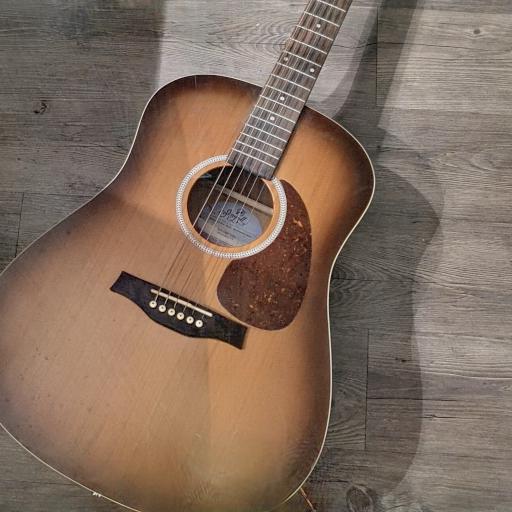}
        {PSNR=25.58\\SSIM=0.905} &
        \imgcell[67.5pt]{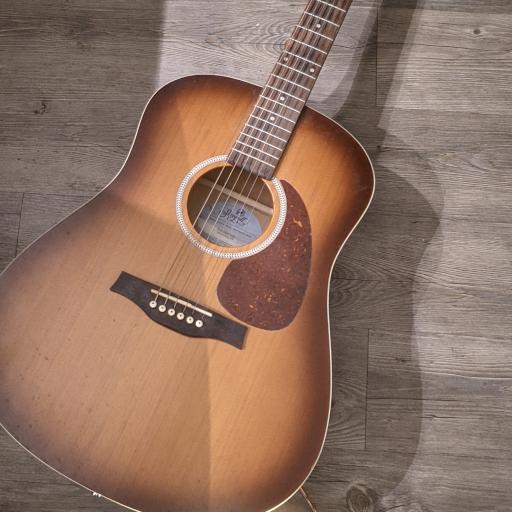}\\[12pt]

        \rotatebox[origin=c]{90}{Real scene 4} &
        \imgtextcell[67.5pt]{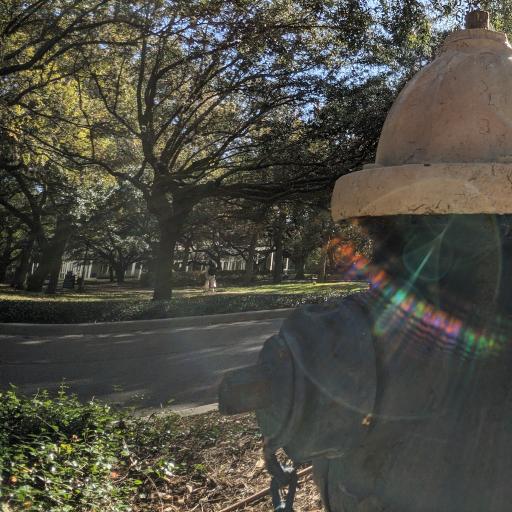}
        {PSNR=20.02\\SSIM=0.780} &
        \imgtextcell[67.5pt]{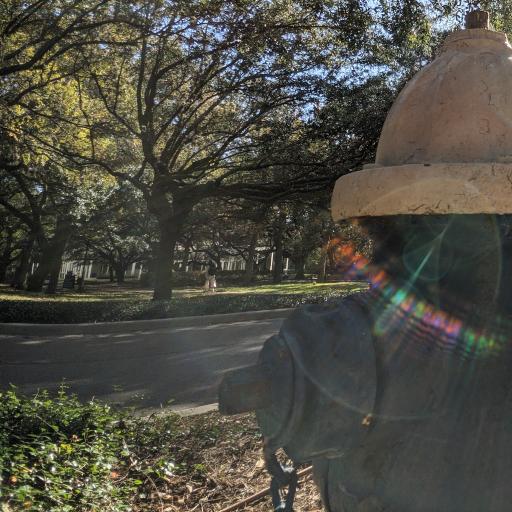}
        {PSNR=20.02\\SSIM=0.780} &
        \imgtextcell[67.5pt]{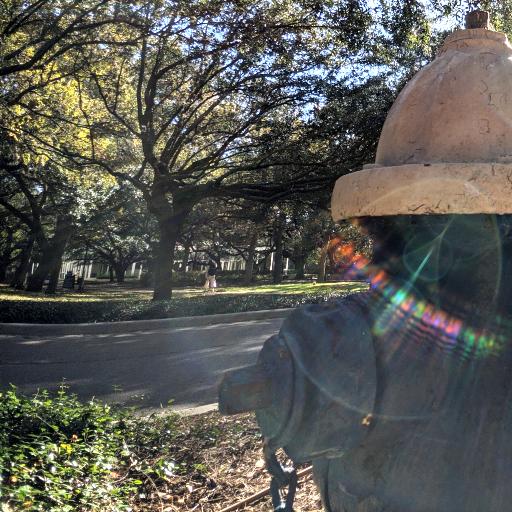}
        {PSNR=16.09\\SSIM=0.703} &
        \imgtextcell[67.5pt]{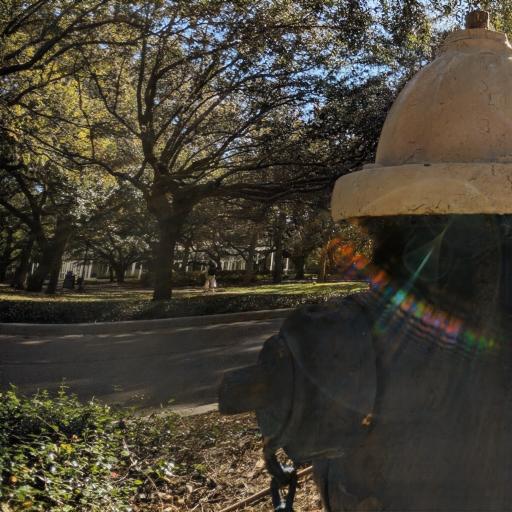}
        {PSNR=23.71\\SSIM=0.836} &
        \imgtextcell[67.5pt]{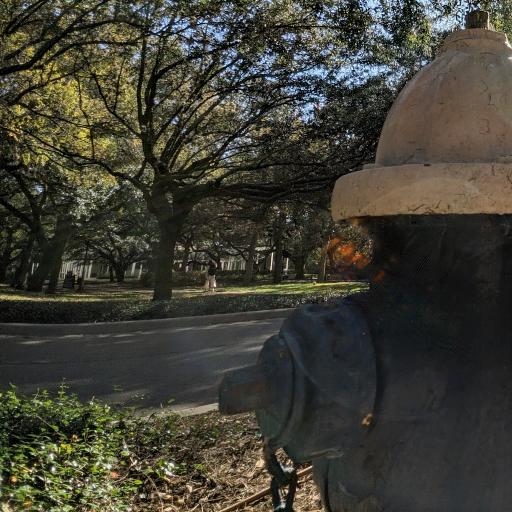}
        {PSNR=24.27\\SSIM=0.838} &
        \imgtextcell[67.5pt]{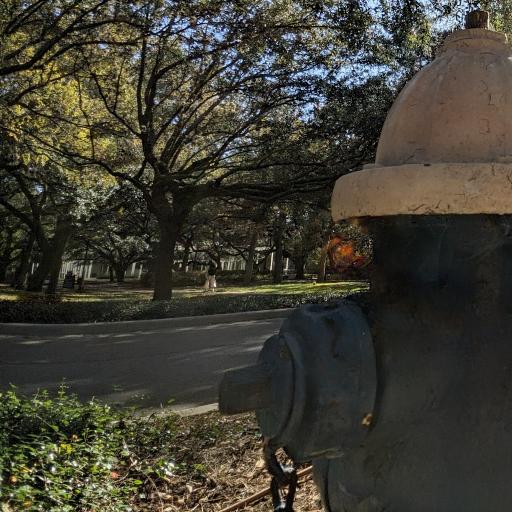}
        {PSNR=24.57\\SSIM=0.847} &
        \imgcell[67.5pt]{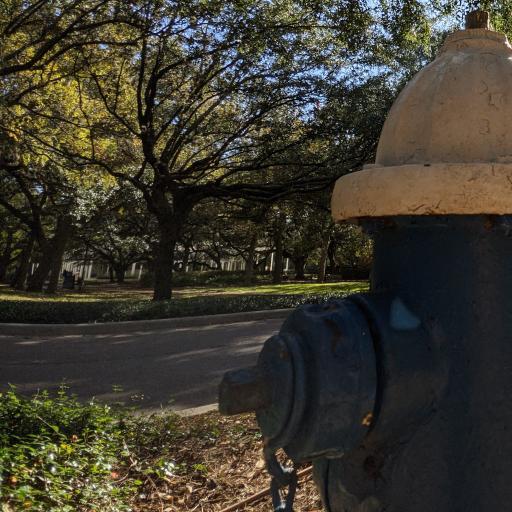}\\[12pt]
        
        \rotatebox[origin=c]{90}{Real scene 5} &
        \imgtextcell[67.5pt]{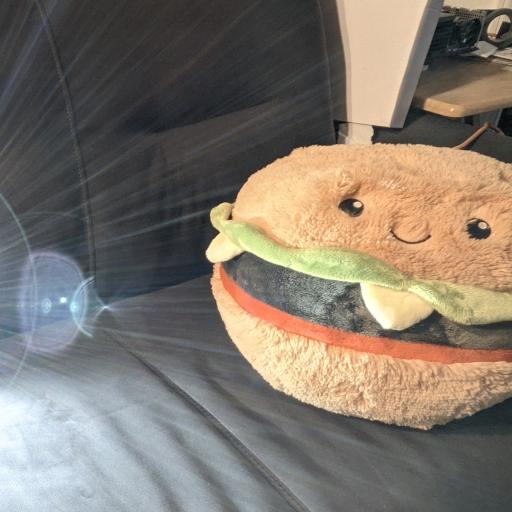}
        {PSNR=15.70\\SSIM=0.750} &
        \imgtextcell[67.5pt]{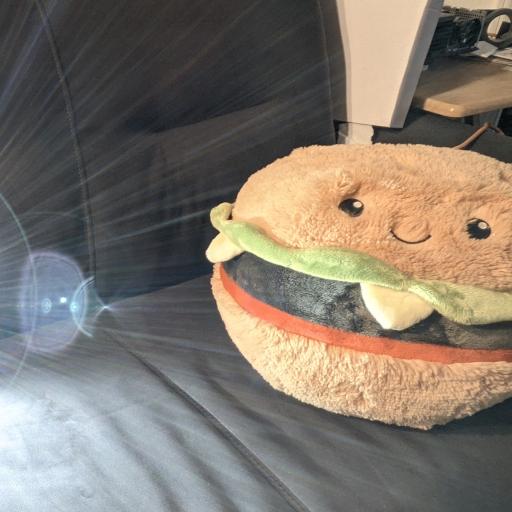}
        {PSNR=15.70\\SSIM=0.750} &
        \imgtextcell[67.5pt]{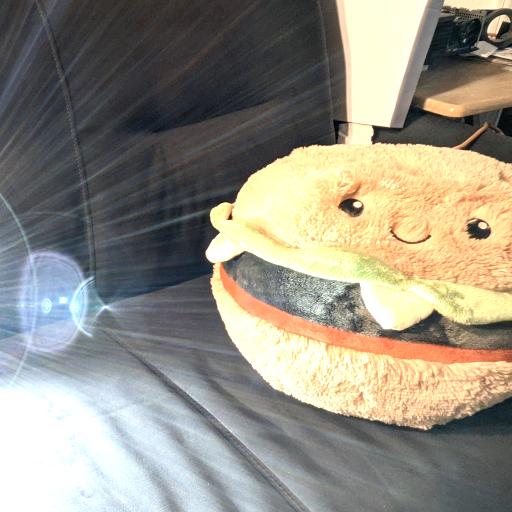}
        {PSNR=11.68\\SSIM=0.624} &
        \imgtextcell[67.5pt]{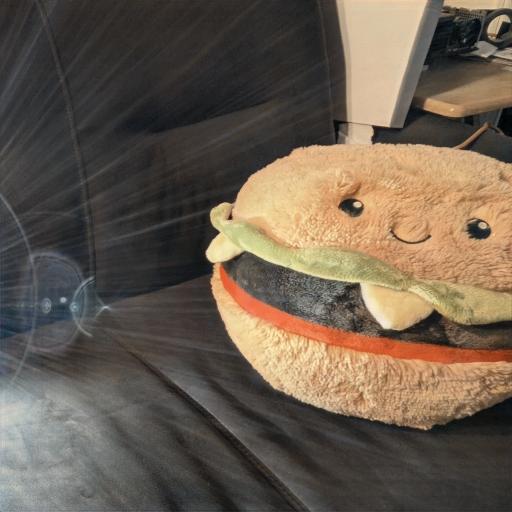}
        {PSNR=19.67\\SSIM=0.796} &
        \imgtextcell[67.5pt]{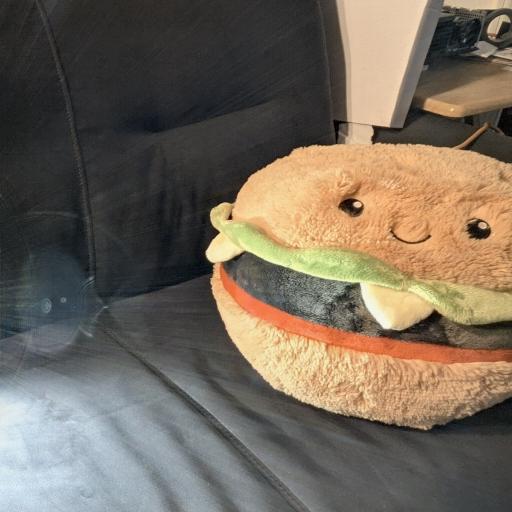}
        {PSNR=19.94\\SSIM=0.800} &
        \imgtextcell[67.5pt]{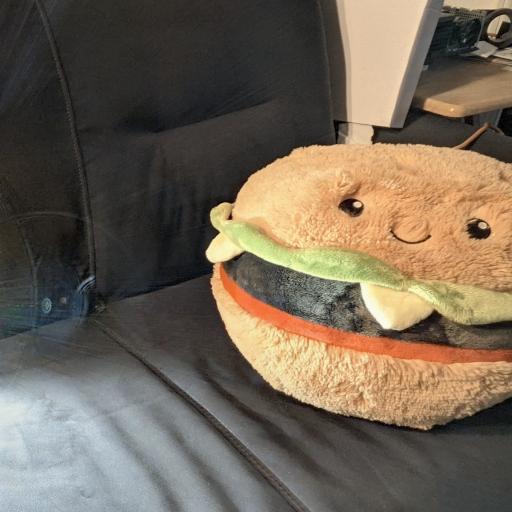}
        {PSNR=21.29\\SSIM=0.825} &
        \imgcell[67.5pt]{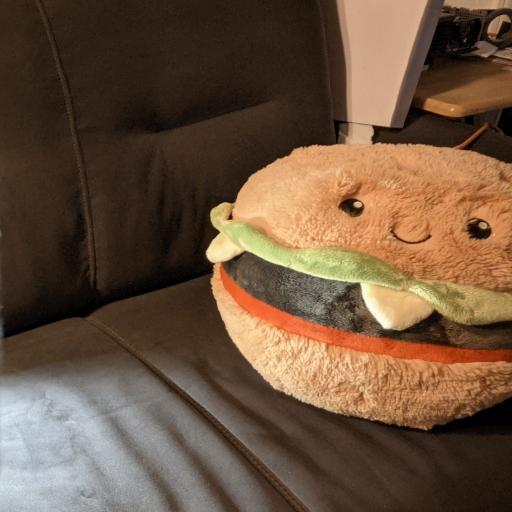}\\[12pt]
        
        \rotatebox[origin=c]{90}{Real scene 6} &
        \imgtextcell[67.5pt]{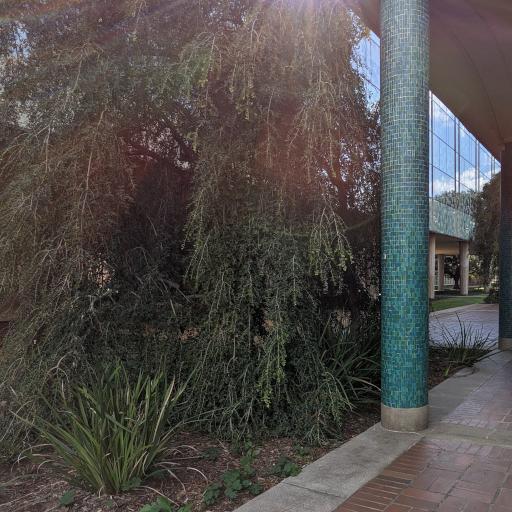}
        {PSNR=18.61\\SSIM=0.848} &
        \imgtextcell[67.5pt]{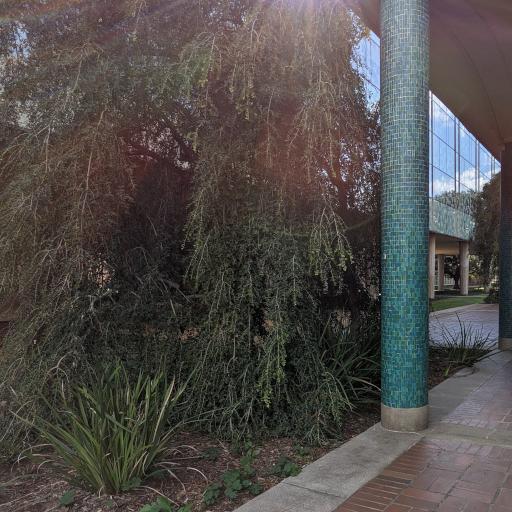}
        {PSNR=18.61\\SSIM=0.848} &
        \imgtextcell[67.5pt]{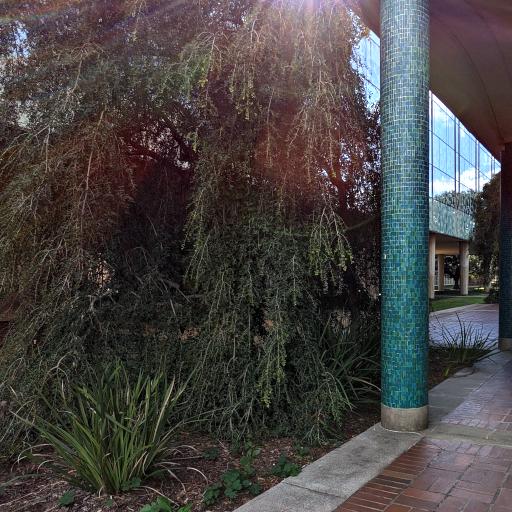}
        {PSNR=19.68\\SSIM=0.888} &
        \imgtextcell[67.5pt]{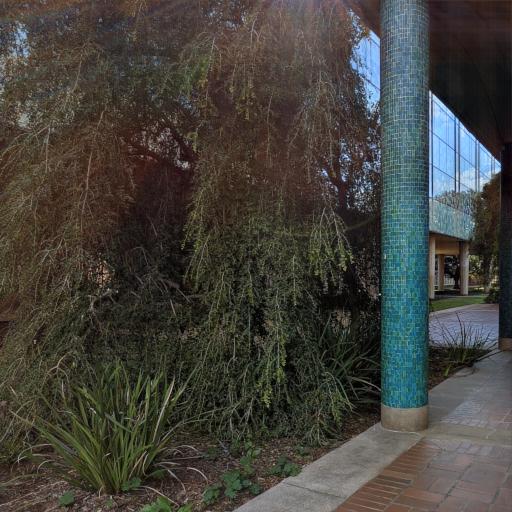}
        {PSNR=23.56\\SSIM=0.873} &
        \imgtextcell[67.5pt]{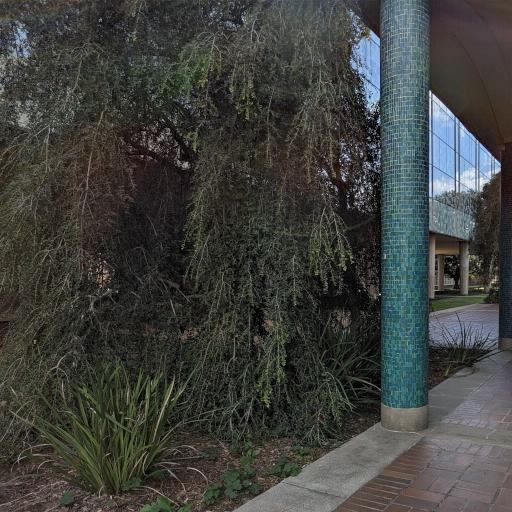}
        {PSNR=25.68\\SSIM=0.915} &
        \imgtextcell[67.5pt]{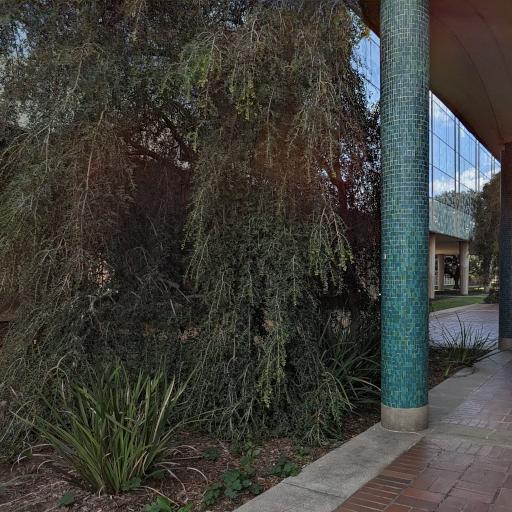}
        {PSNR=26.90\\SSIM=0.924} &
        \imgcell[67.5pt]{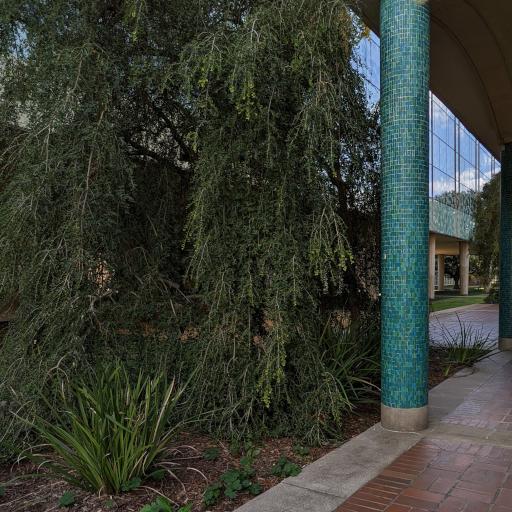}\\[12pt]
        
        \rotatebox[origin=c]{90}{Real scene 7} &
        \imgtextcell[67.5pt]{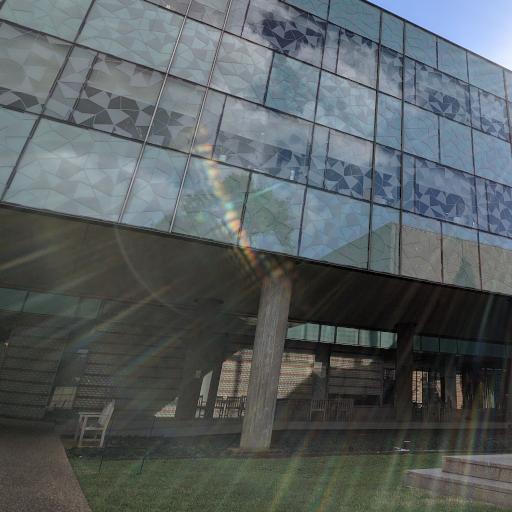}
        {PSNR=20.51\\SSIM=0.814} &
        \imgtextcell[67.5pt]{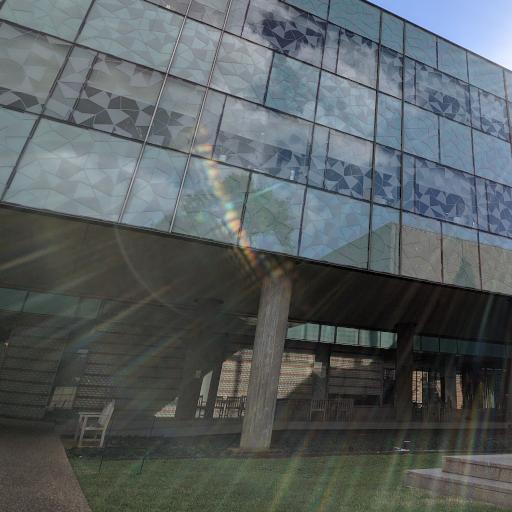}
        {PSNR=20.51\\SSIM=0.814} &
        \imgtextcell[67.5pt]{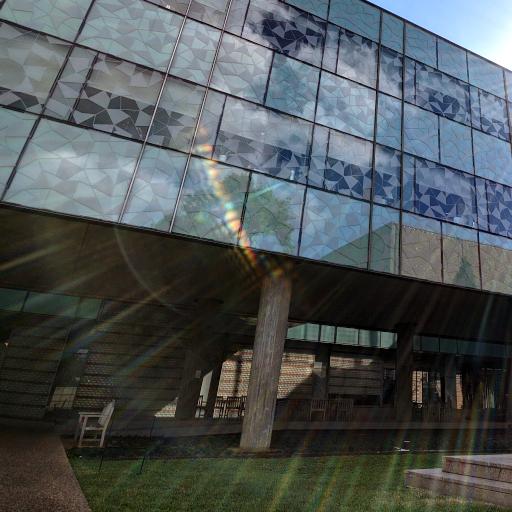}
        {PSNR=24.05\\SSIM=0.859} &
        \imgtextcell[67.5pt]{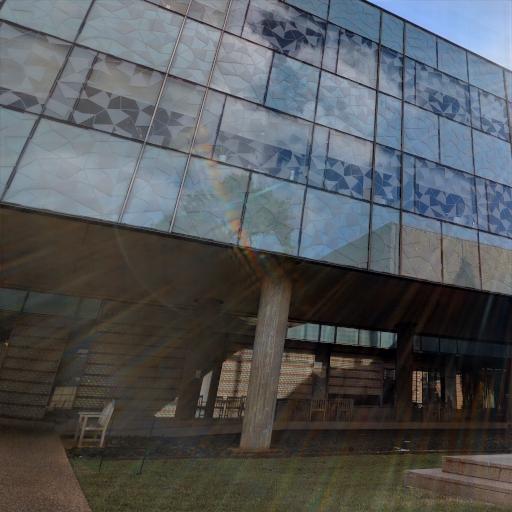}
        {PSNR=23.45\\SSIM=0.849} &
        \imgtextcell[67.5pt]{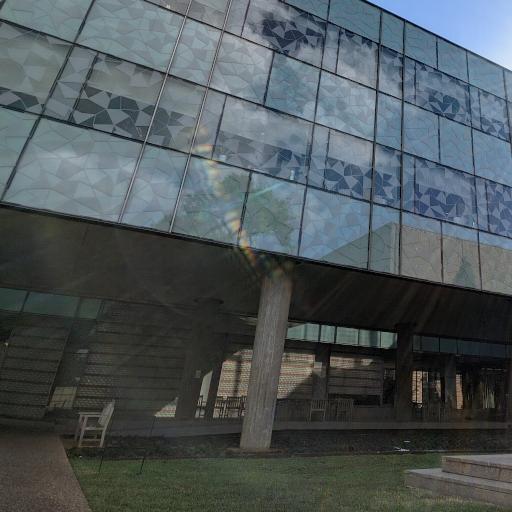}
        {PSNR=24.47\\SSIM=0.855} &
        \imgtextcell[67.5pt]{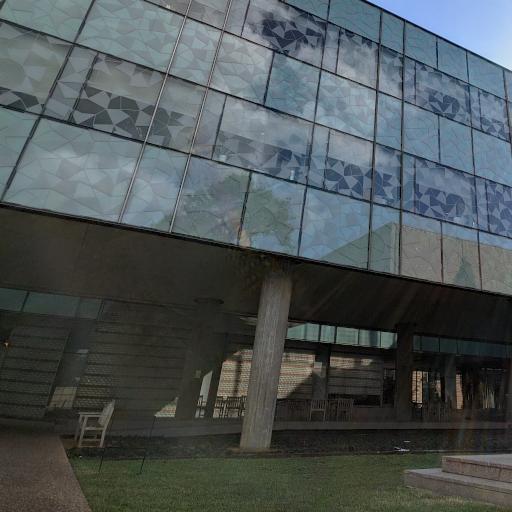}
        {PSNR=25.51\\SSIM=0.875} &
        \imgcell[67.5pt]{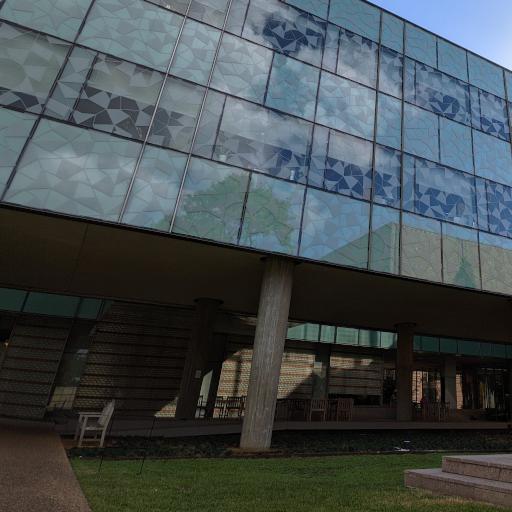}  \\[12pt]

        \rotatebox[origin=c]{90}{Real scene 8} &
        \imgtextcell[67.5pt]{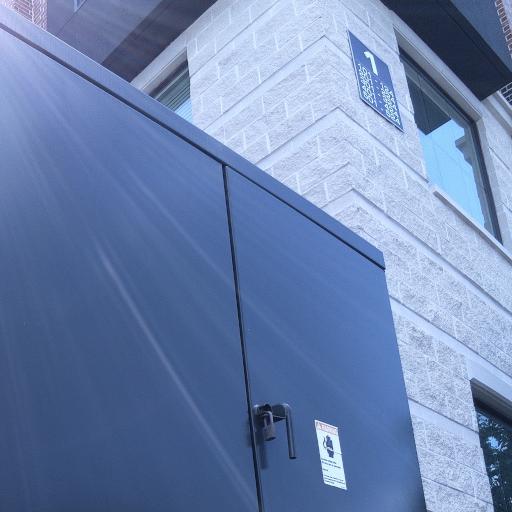}
        {PSNR=24.35\\SSIM=0.902} &
        \imgtextcell[67.5pt]{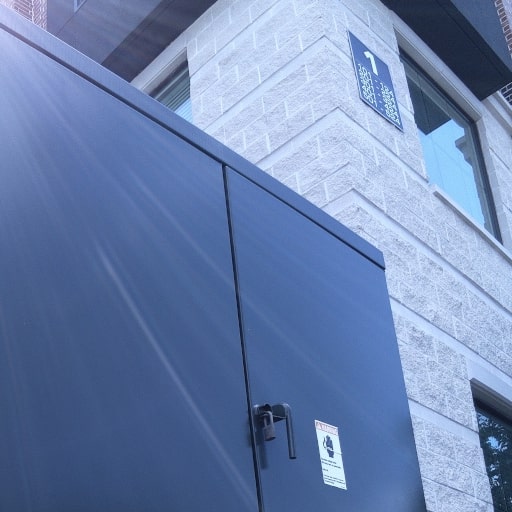}
        {PSNR=24.35\\SSIM=0.902} &
        \imgtextcell[67.5pt]{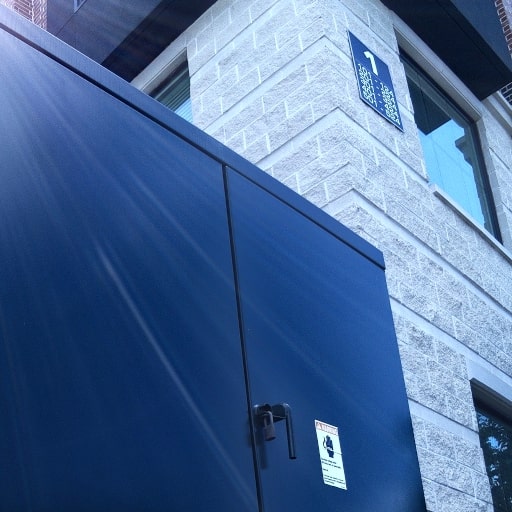}
        {PSNR=20.68\\SSIM=0.807} &
        \imgtextcell[67.5pt]{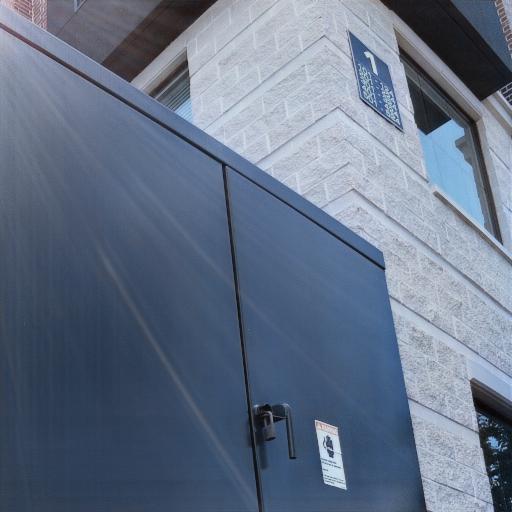}
        {PSNR=22.00\\SSIM=0.898} &
        \imgtextcell[67.5pt]{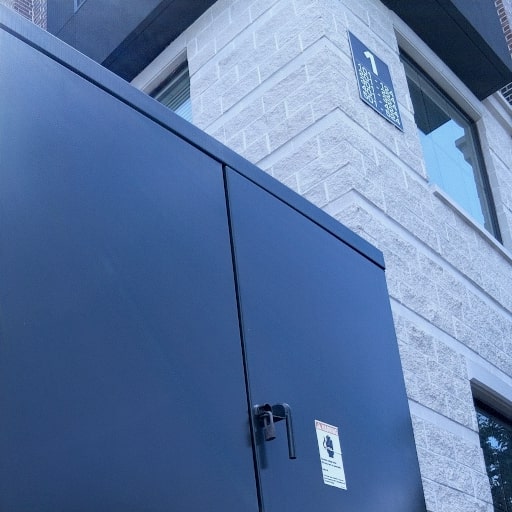}
        {PSNR=29.65\\SSIM=0.905} &
        \imgtextcell[67.5pt]{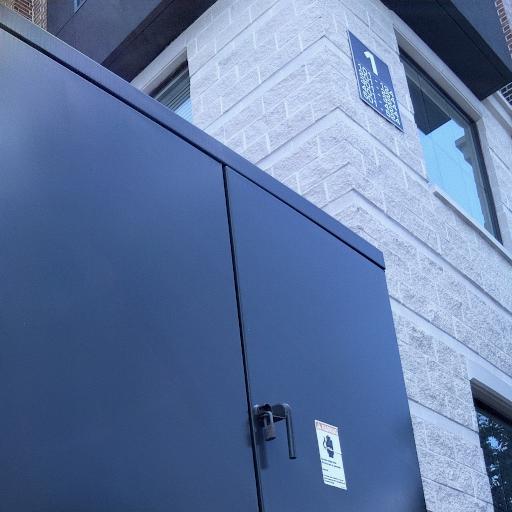}
        {PSNR=32.64\\SSIM=0.919} &
        \imgcell[67.5pt]{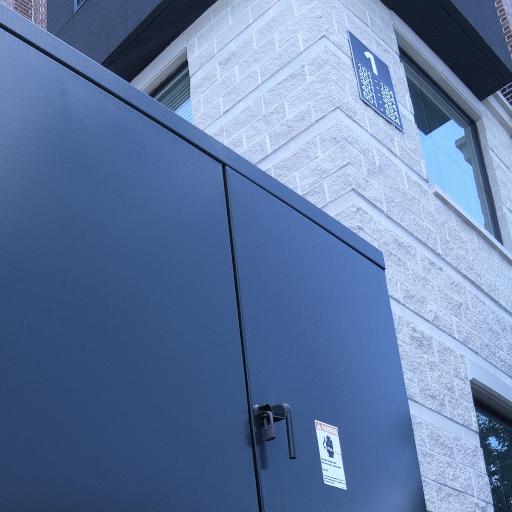}\\[12pt]

        \rotatebox[origin=c]{90}{Real scene 9} &
        \imgtextcell[67.5pt]{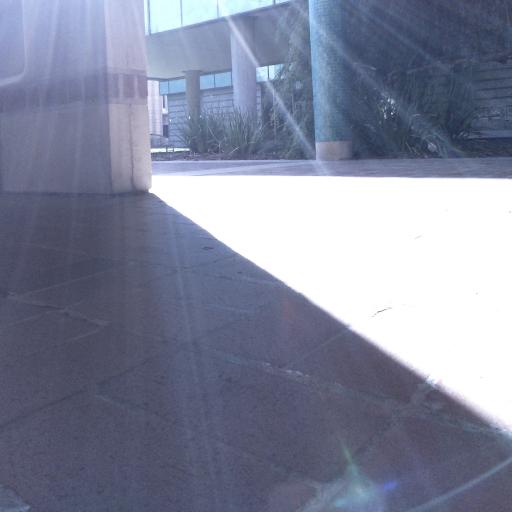}
        {PSNR=15.90\\SSIM=0.802} &
        \imgtextcell[67.5pt]{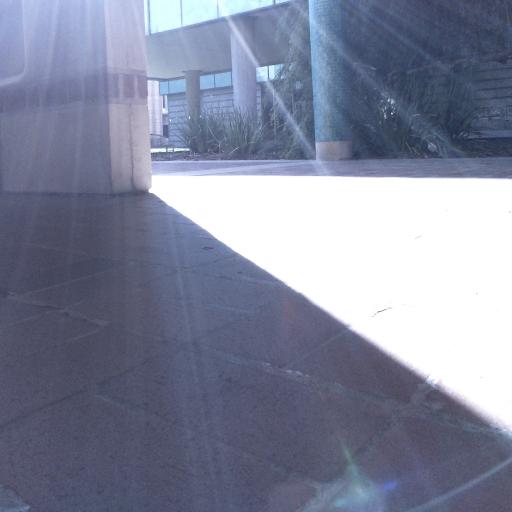}
        {PSNR=15.90\\SSIM=0.802} &
        \imgtextcell[67.5pt]{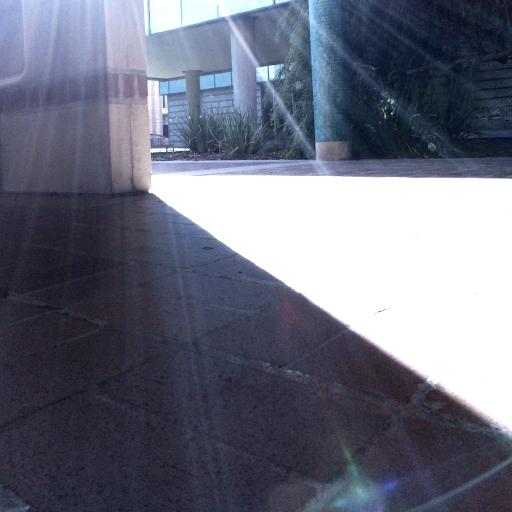}
        {PSNR=18.83\\SSIM=0.810} &
        \imgtextcell[67.5pt]{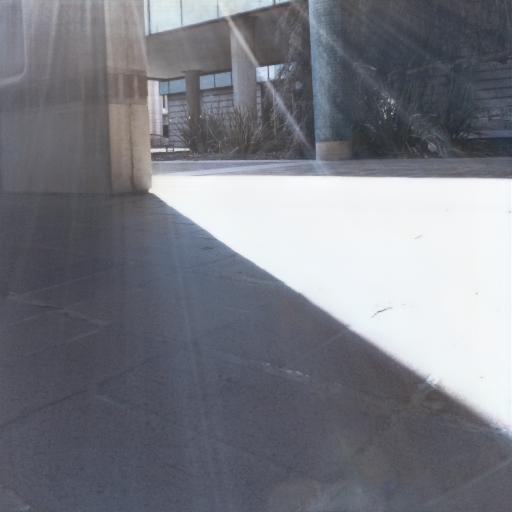}
        {PSNR=20.10\\SSIM=0.838} &
        \imgtextcell[67.5pt]{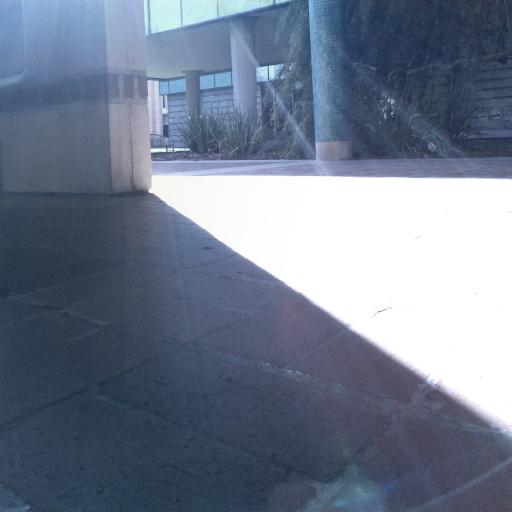}
        {PSNR=19.07\\SSIM=0.839} &
        \imgtextcell[67.5pt]{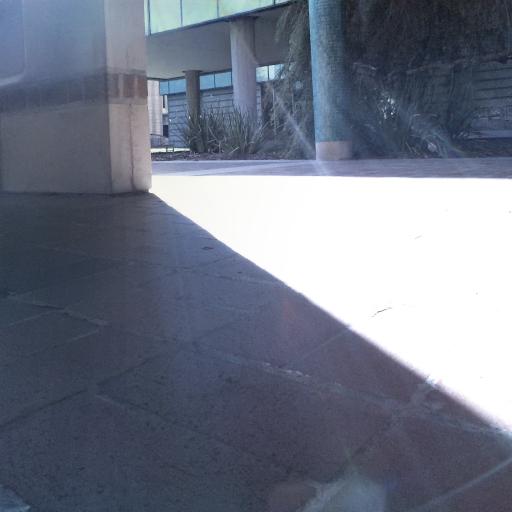}
        {PSNR=21.74\\SSIM=0.863} &
        \imgcell[67.5pt]{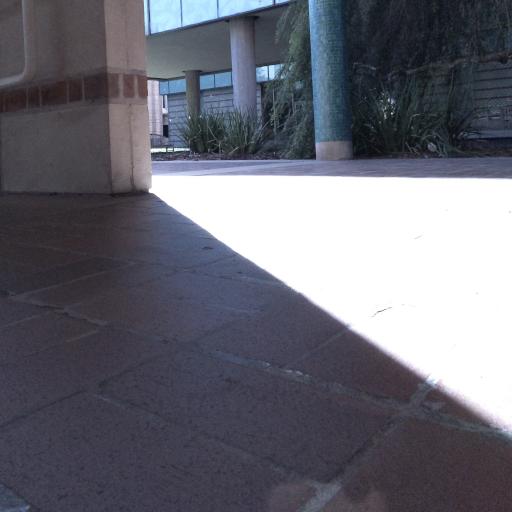} \\[12pt]

        \rotatebox[origin=c]{90}{Real scene 10} &
        \imgtextcell[67.5pt]{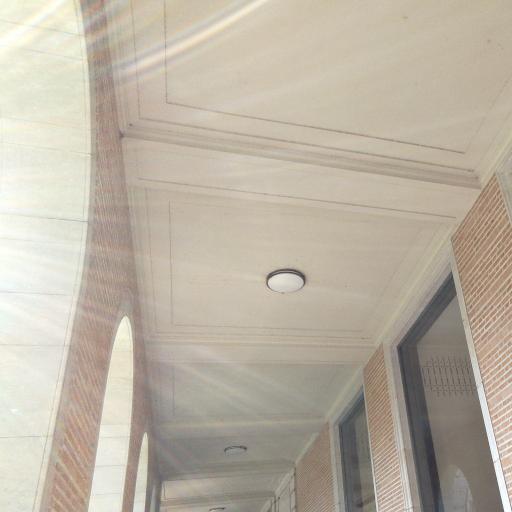}
        {PSNR=20.04\\SSIM=0.867} &
        \imgtextcell[67.5pt]{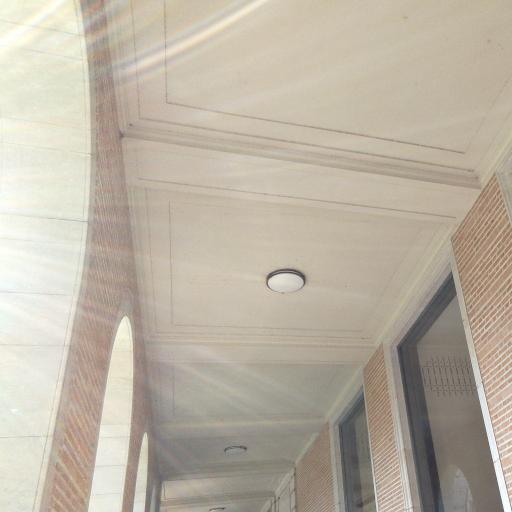}
        {PSNR=20.04\\SSIM=0.867} &
        \imgtextcell[67.5pt]{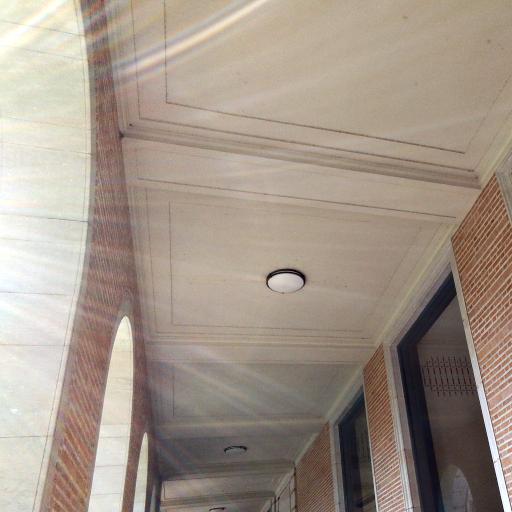}
        {PSNR=22.26\\SSIM=0.825} &
        \imgtextcell[67.5pt]{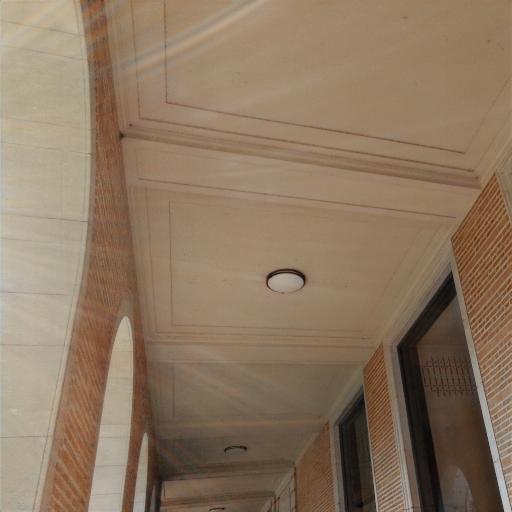}
        {PSNR=16.93\\SSIM=0.853} &
        \imgtextcell[67.5pt]{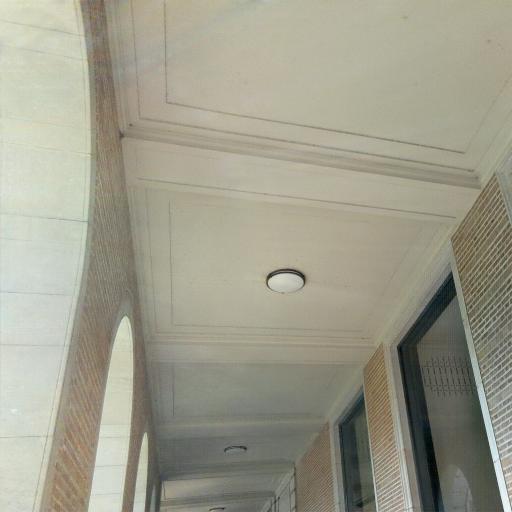}
        {PSNR=25.95\\SSIM=0.878} &
        \imgtextcell[67.5pt]{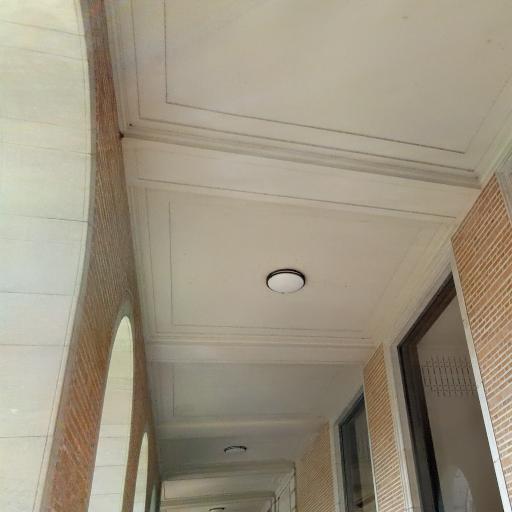}
        {PSNR=26.12\\SSIM=0.895} &
        \imgcell[67.5pt]{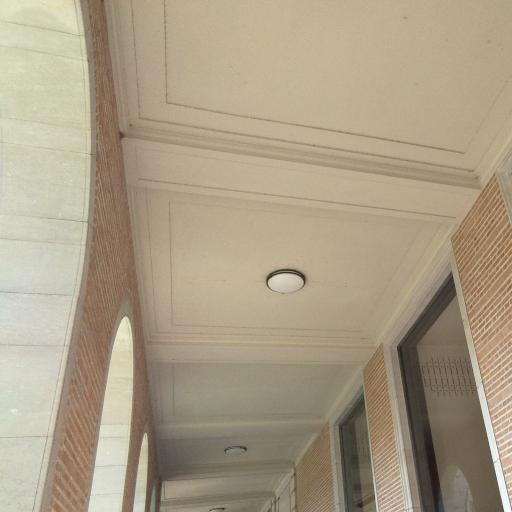}

      \end{tabular}
      \vspace{2pt}
      \caption{ More visual comparison between three related methods and ours on real scenes, with PSNR and SSIM values. }
      \label{fig:supp_compare}
    \end{figure*}

    Fig.~\ref{fig:supp_compare} provides more visual comparisons between our method and prior work. Overall, we have 20 real test images with ground truth. The complete results, including the input images and output images from different methods, can be found at \href{https://yichengwu.github.io/flare-removal/}{https://yichengwu.github.io/flare-removal/}.
    
\begin{figure*}
  \centering
  \footnotesize
  \setlength{\tabcolsep}{2pt}
  \begin{tabular}{@{}ccccccc@{}}
    \rotatebox[origin=c]{90}{Input} &
    \imgcell[75pt]{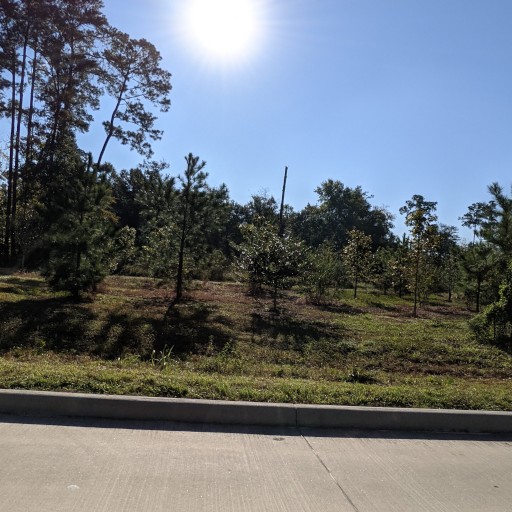} &
    \imgcell[75pt]{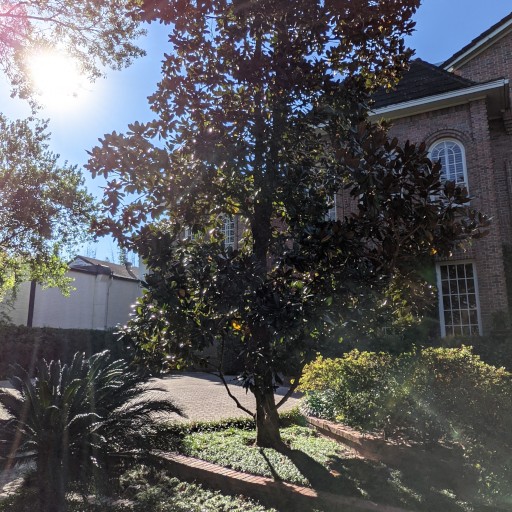} &
    \imgcell[75pt]{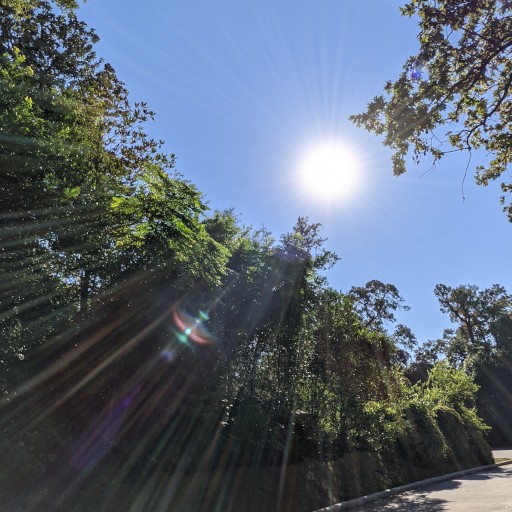} &
    \imgcell[75pt]{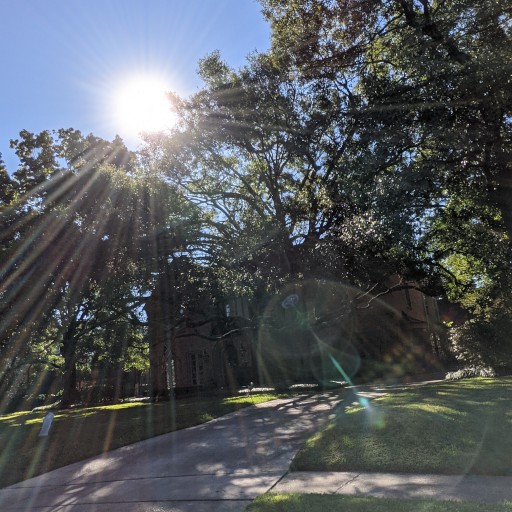} &
    \imgcell[75pt]{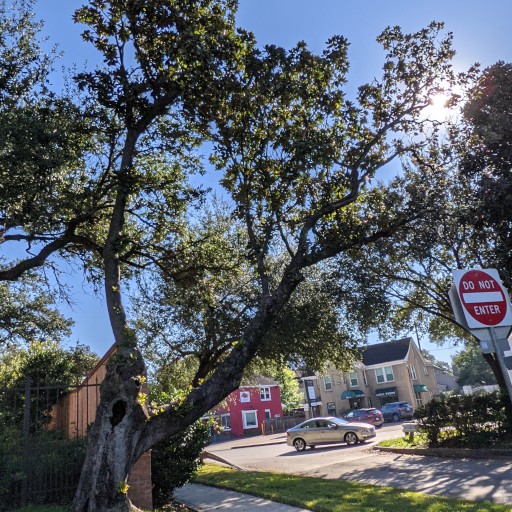} &
    \imgcell[75pt]{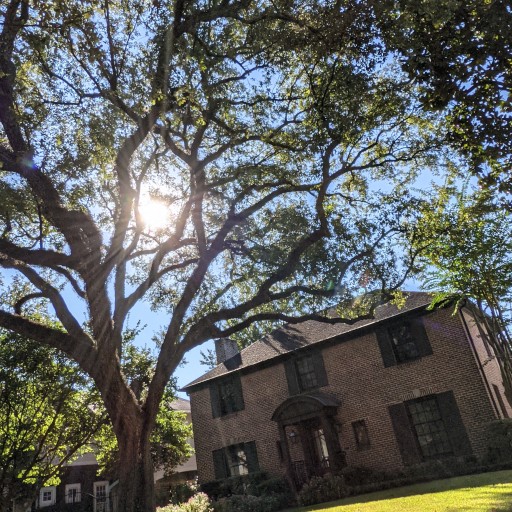} \\[-1pt]
    
    \rotatebox[origin=c]{90}{Output} &
    \imgcell[75pt]{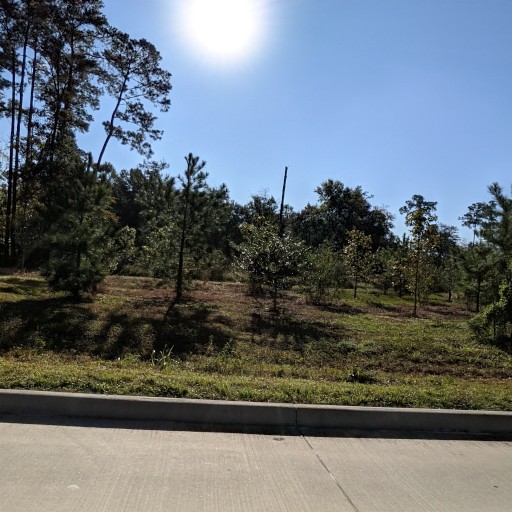} &
    \imgcell[75pt]{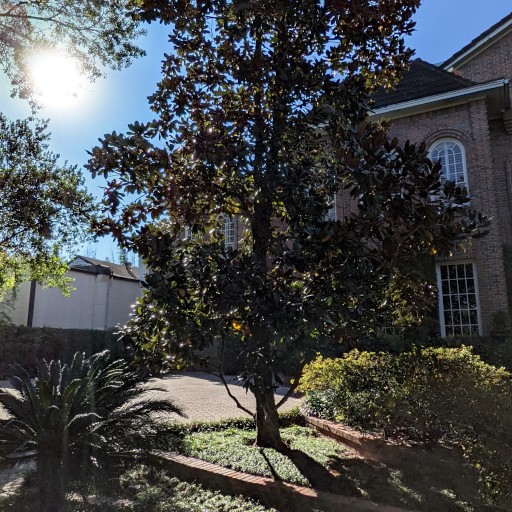} &
    \imgcell[75pt]{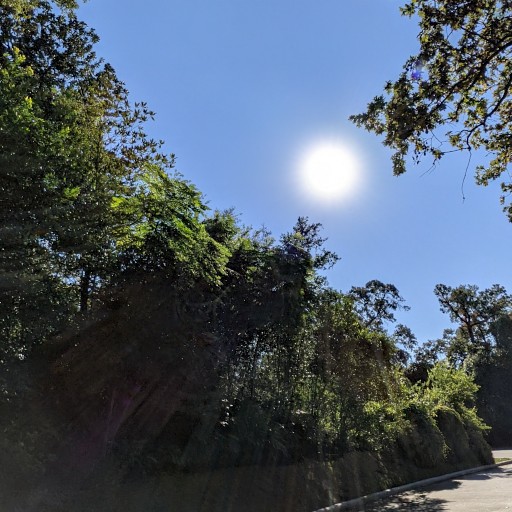} &
    \imgcell[75pt]{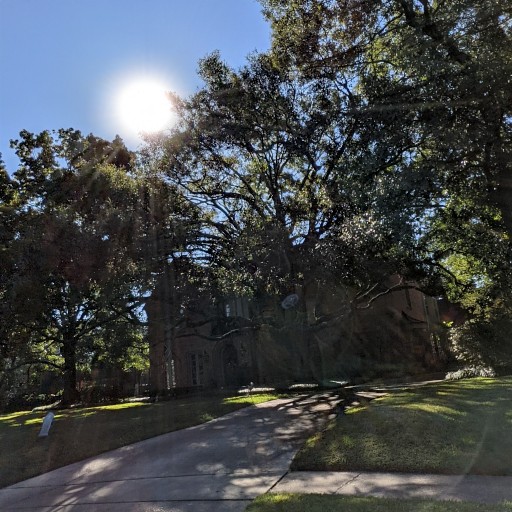} &
    \imgcell[75pt]{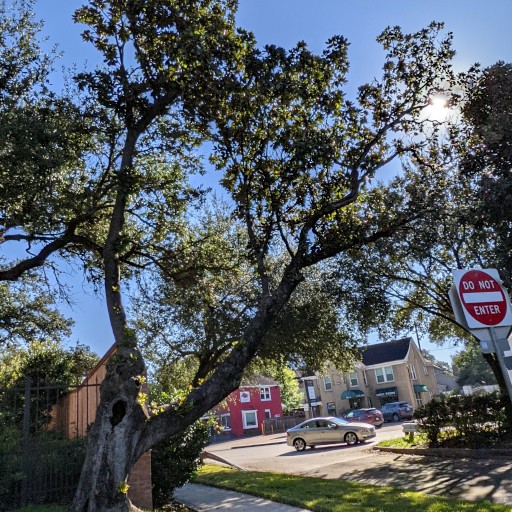} &
    \imgcell[75pt]{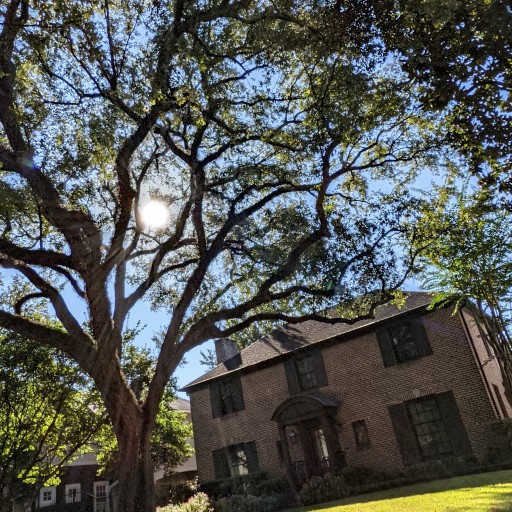} \\[35pt]
    
    \rotatebox[origin=c]{90}{Input} &
    \imgcell[75pt]{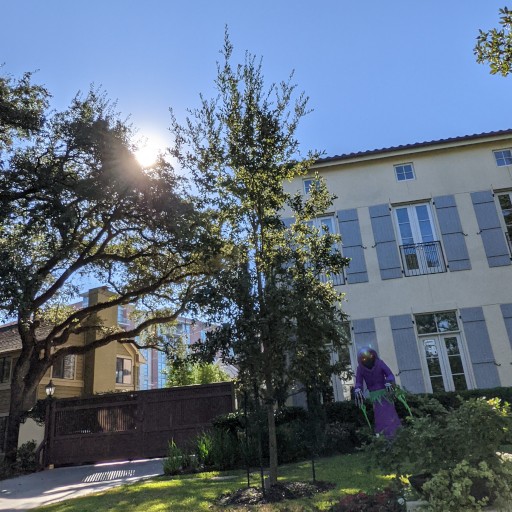} &
    \imgcell[75pt]{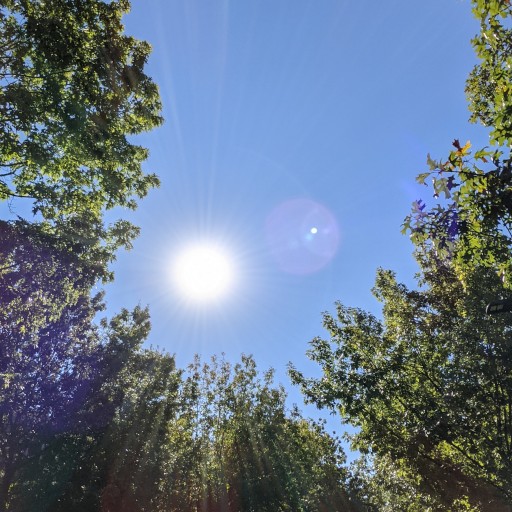} &
    \imgcell[75pt]{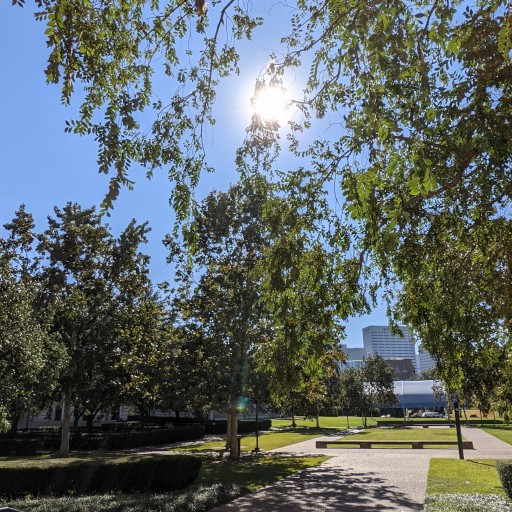} &
    \imgcell[75pt]{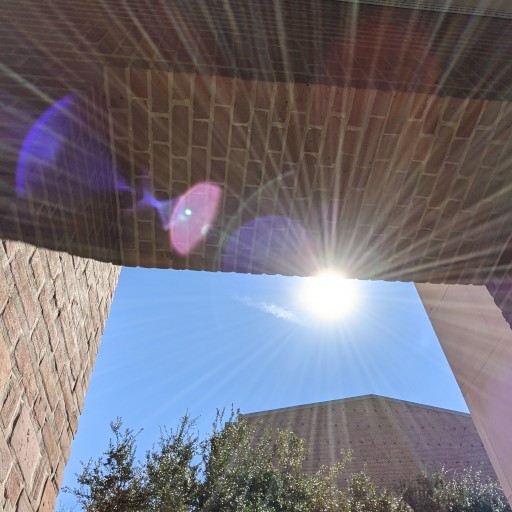} &
    \imgcell[75pt]{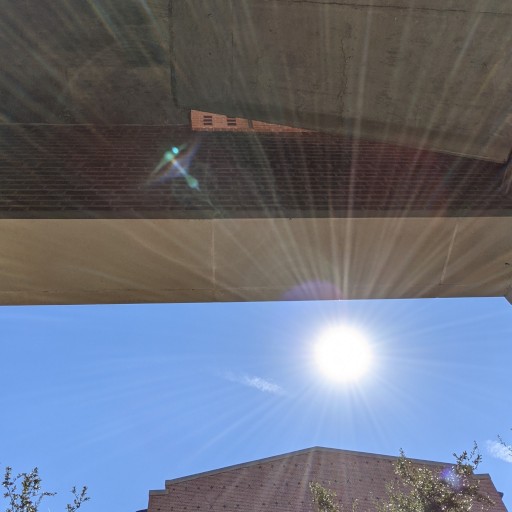} &
    \imgcell[75pt]{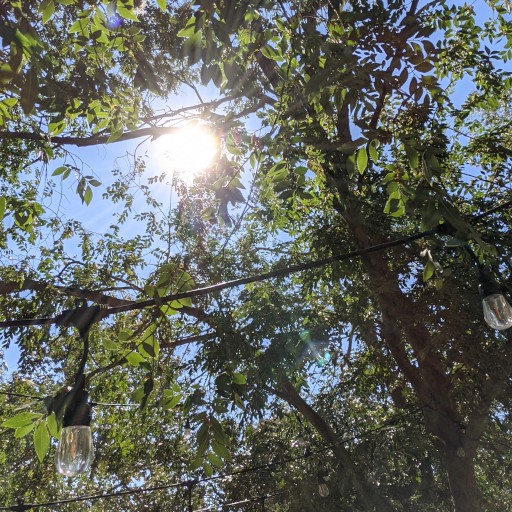} \\[-1pt]
    
    \rotatebox[origin=c]{90}{Output} &
    \imgcell[75pt]{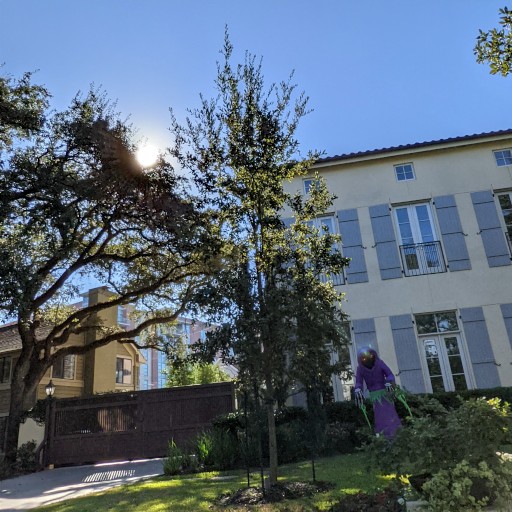} &
    \imgcell[75pt]{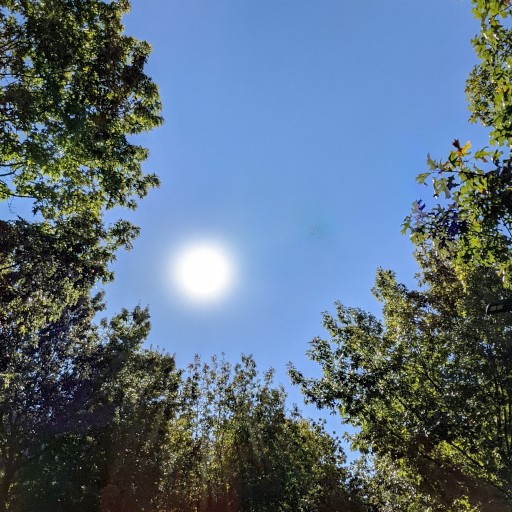} &
    \imgcell[75pt]{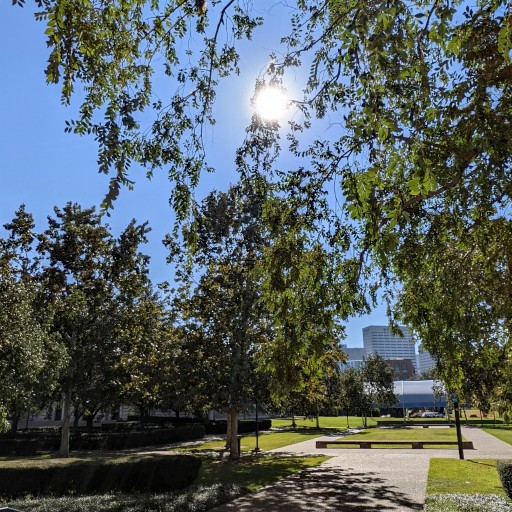} &
    \imgcell[75pt]{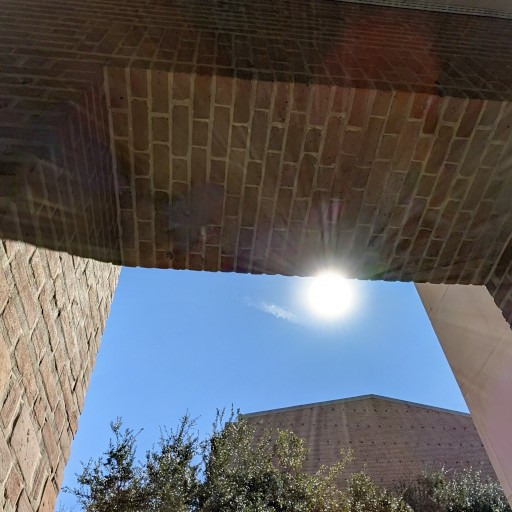} &
    \imgcell[75pt]{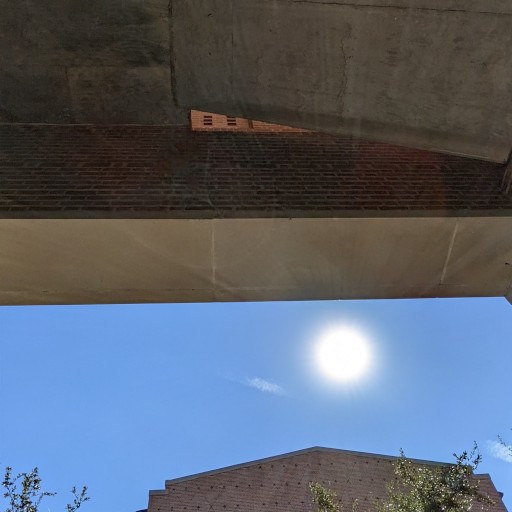} &
    \imgcell[75pt]{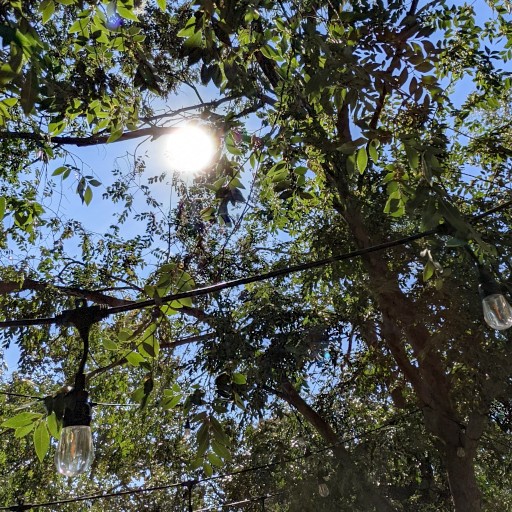} \\[35pt]
    
    \rotatebox[origin=c]{90}{Input} &
    \imgcell[75pt]{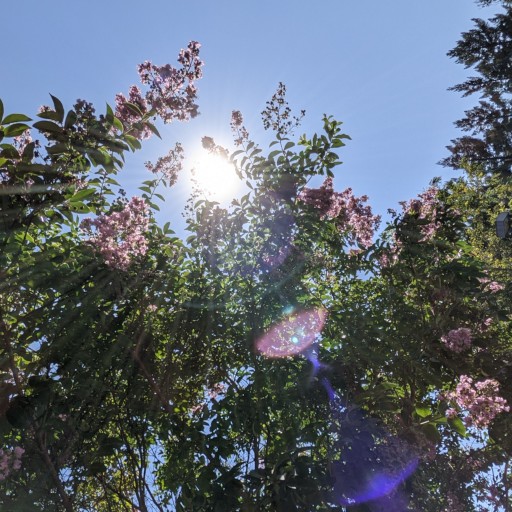} &
    \imgcell[75pt]{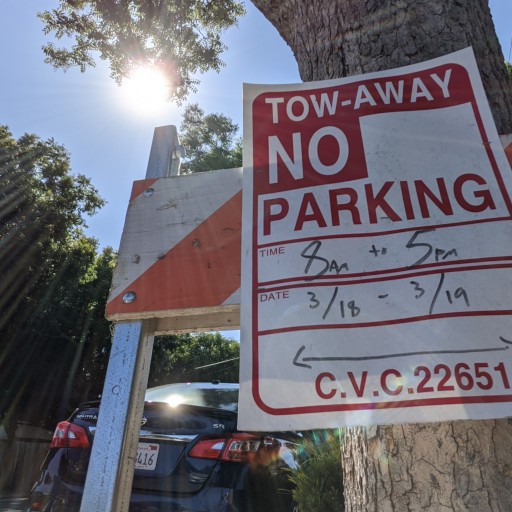} &
    \imgcell[75pt]{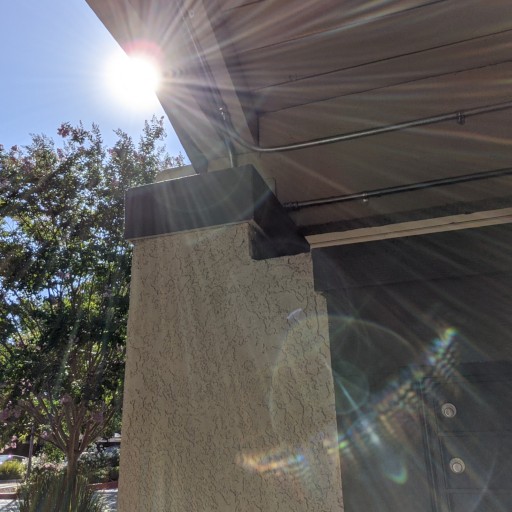} &
    \imgcell[75pt]{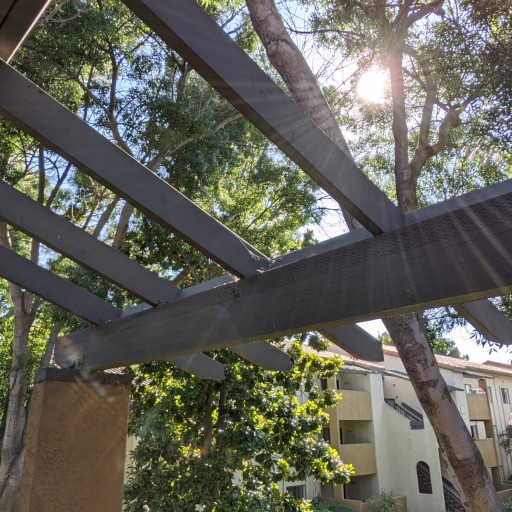} &
    \imgcell[75pt]{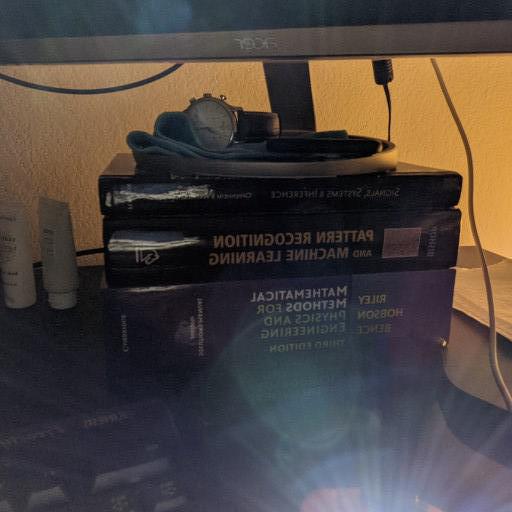} &
    \imgcell[75pt]{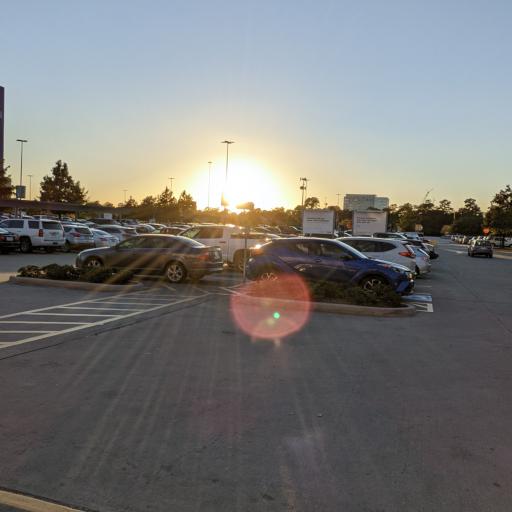} \\[-1pt]
    
    \rotatebox[origin=c]{90}{Output} &
    \imgcell[75pt]{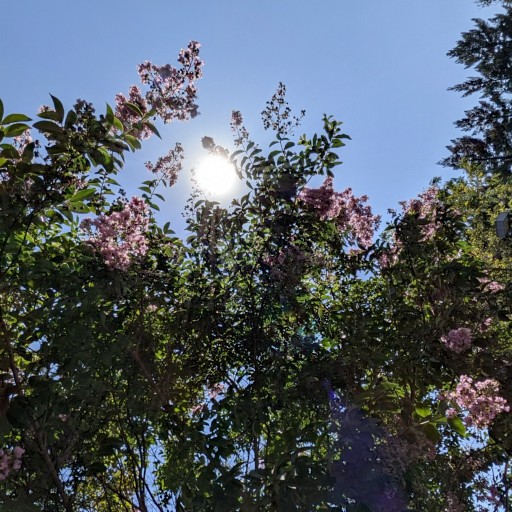} &
    \imgcell[75pt]{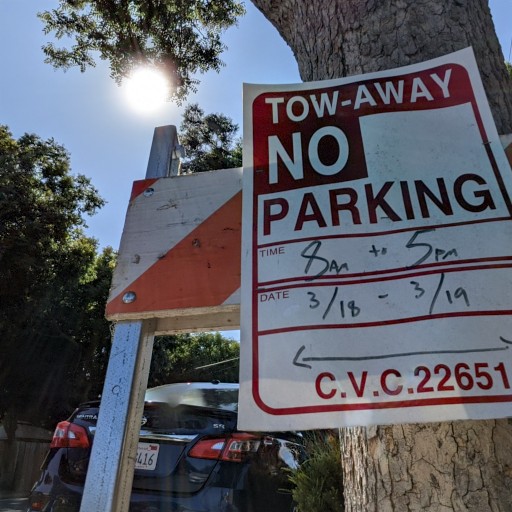} &
    \imgcell[75pt]{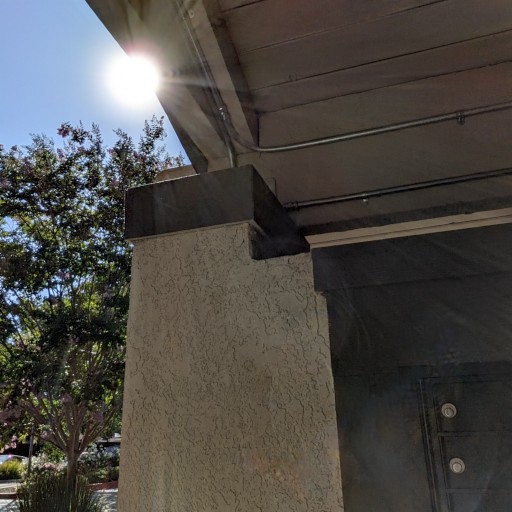} &
    \imgcell[75pt]{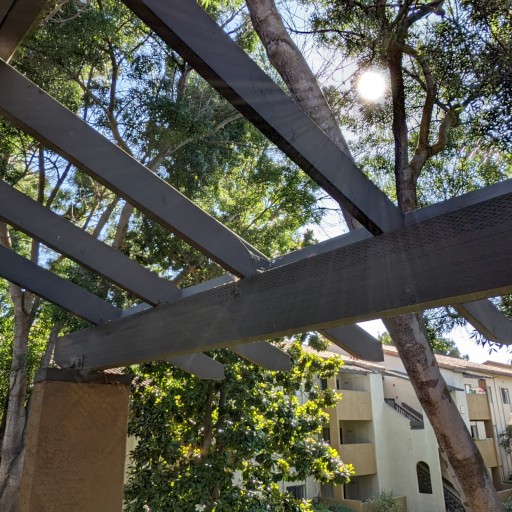} &
    \imgcell[75pt]{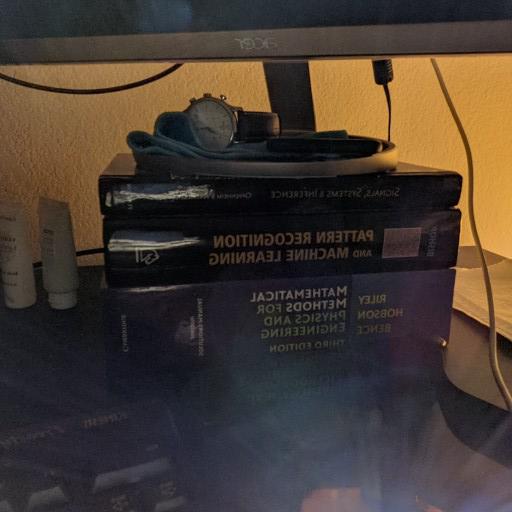} &
    \imgcell[75pt]{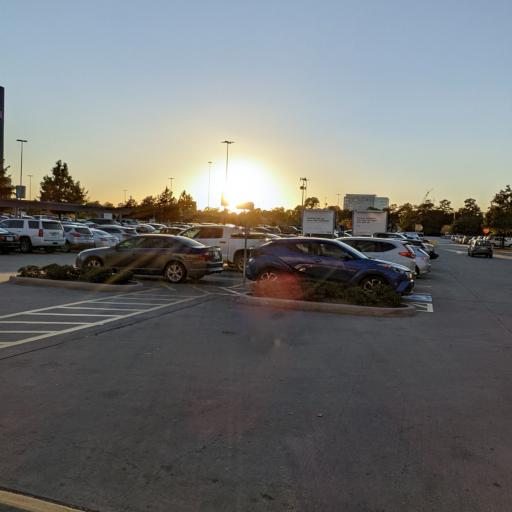} \\[35pt]
    
    \Cline{5-5}
    
    \rotatebox[origin=c]{90}{Input} &
    \imgcell[75pt]{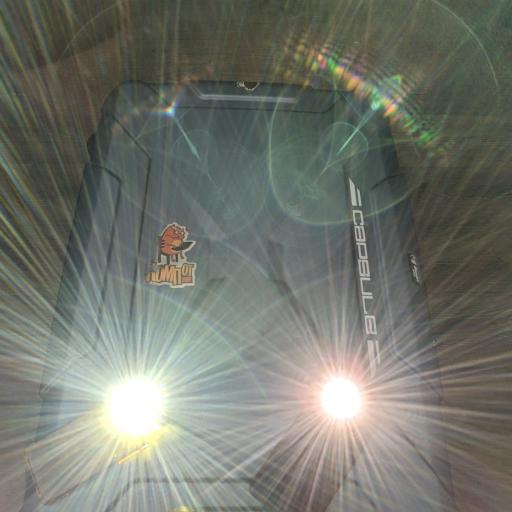} &
    \imgcell[75pt]{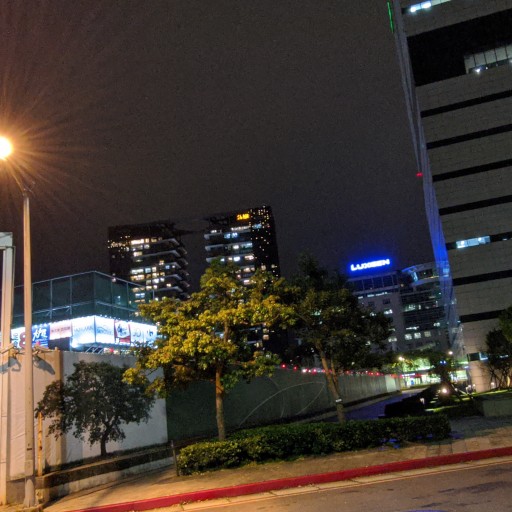} &
    \imgcell[75pt]{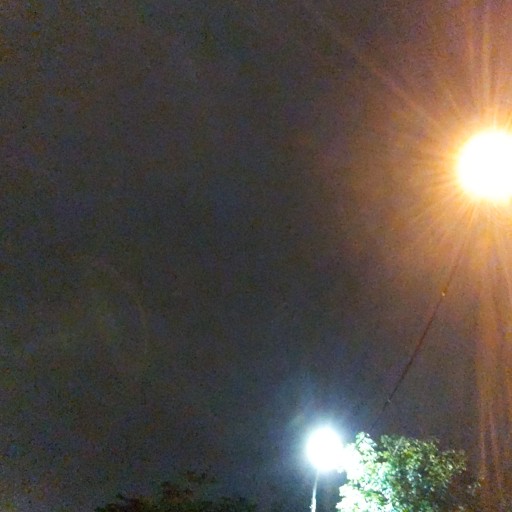} &
    \Thickvrule{%
    \imgcell[75pt]{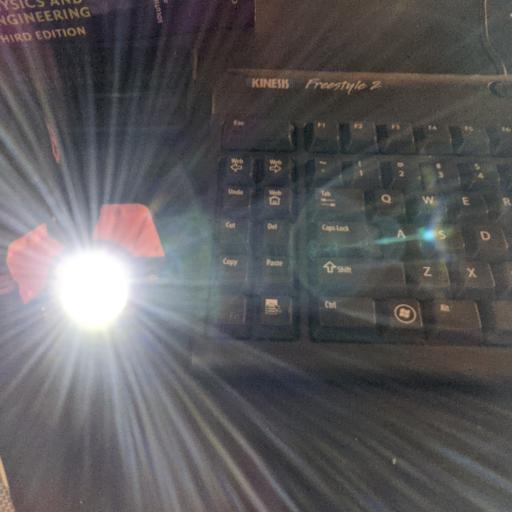}%
    } &
    \imgcell[75pt]{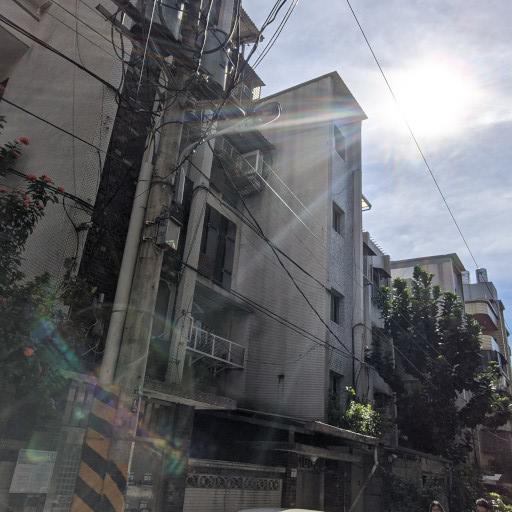} &
    \imgcell[75pt]{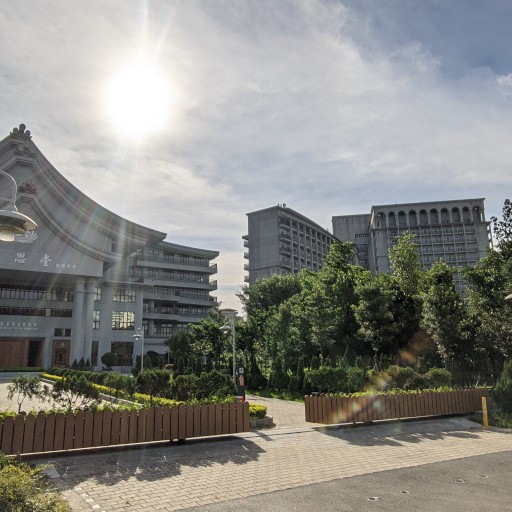} \\[-1pt]
    
    \rotatebox[origin=c]{90}{Output} &
    \imgcell[75pt]{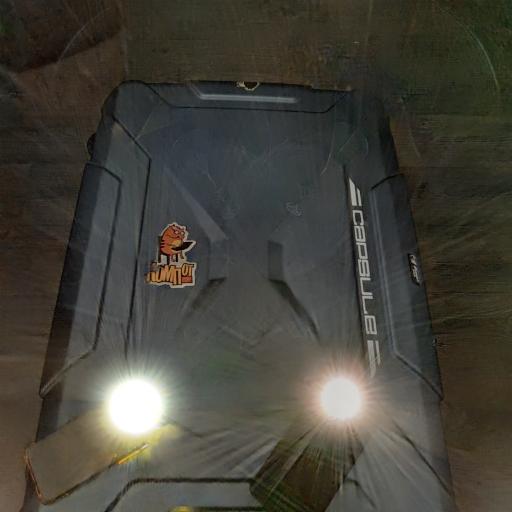} &
    \imgcell[75pt]{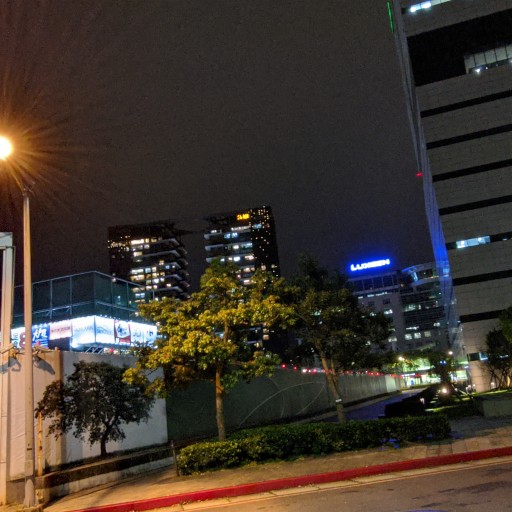} &
    \imgcell[75pt]{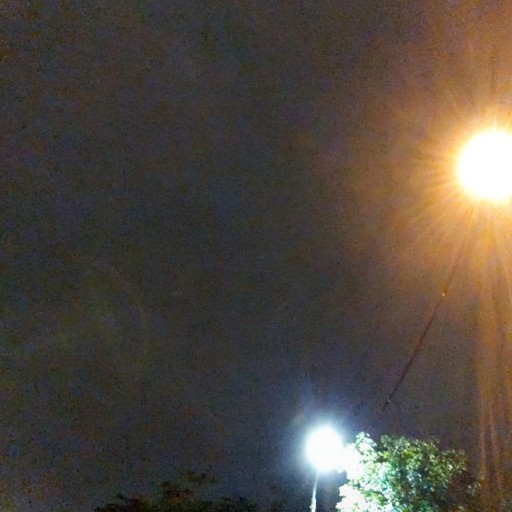} &
    \Thickvrule{%
    \imgcell[75pt]{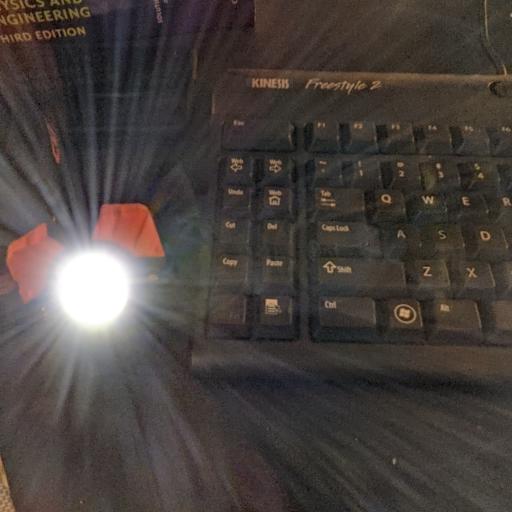}%
    } &
    \imgcell[75pt]{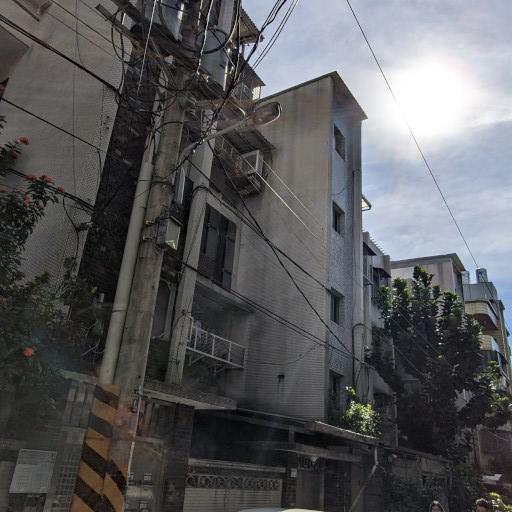} &
    \imgcell[75pt]{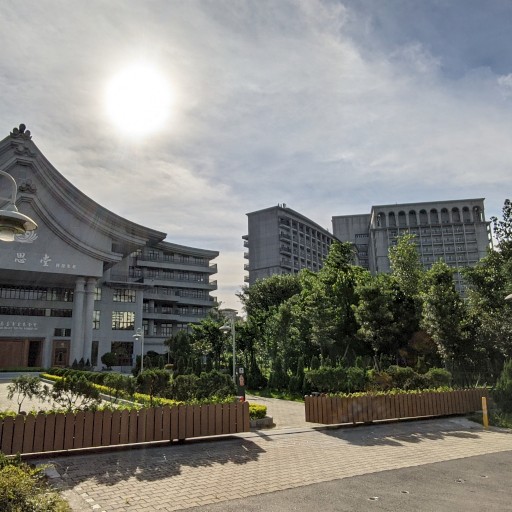} \\
    
    \Cline{5-5}

  \end{tabular}
  \vspace{2pt}
  \caption{
    24 testing images captured by the same type of lens as in Sec.~\ref{sec:data:reflective}. Our method is effective in removing most of the lens flare, with occasional failures where it only removes a part of the flare (red box).
  }
  \label{fig:supp_gallery_same_camera}
\end{figure*}

\begin{figure*}
  \centering
  \footnotesize
  \setlength{\tabcolsep}{2pt}
  \begin{tabular}{@{}ccccccc@{}}
  
    \Cline{3-3}\Cline{5-5}
    
    \rotatebox[origin=c]{90}{Input} &
    \imgcell[75pt]{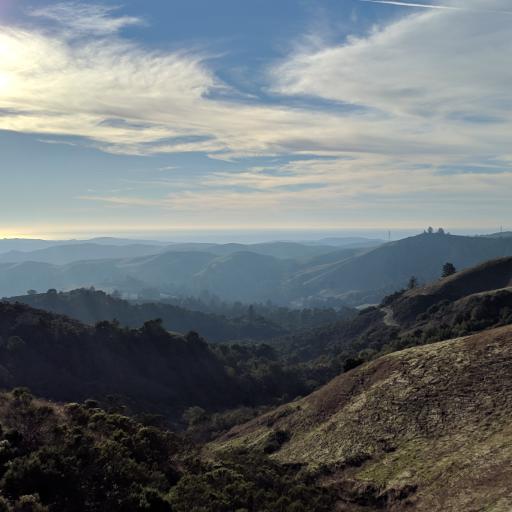} &
    \Thickvrule{%
    \imgcell[75pt]{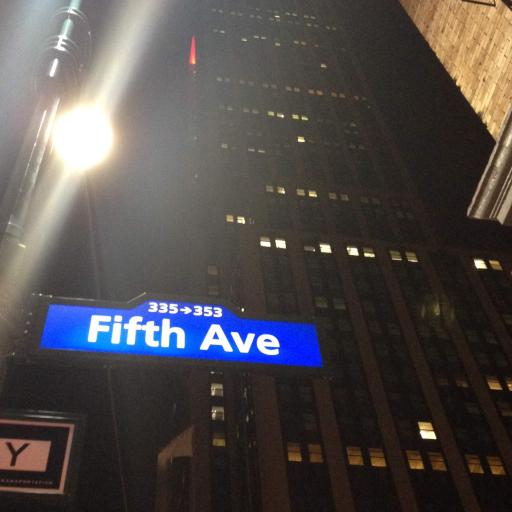}%
    } &
    \imgcell[75pt]{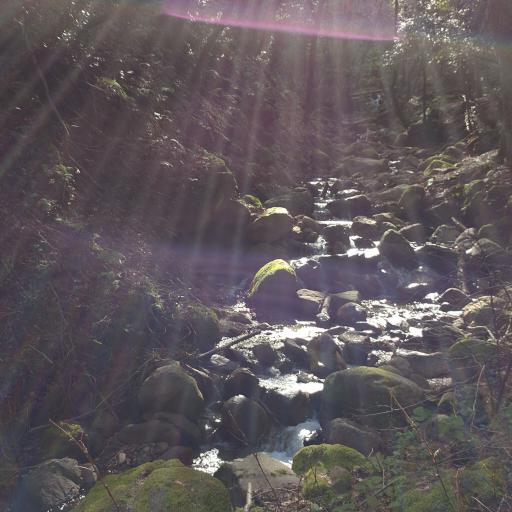} &    \Thickvrule{%
    \imgcell[75pt]{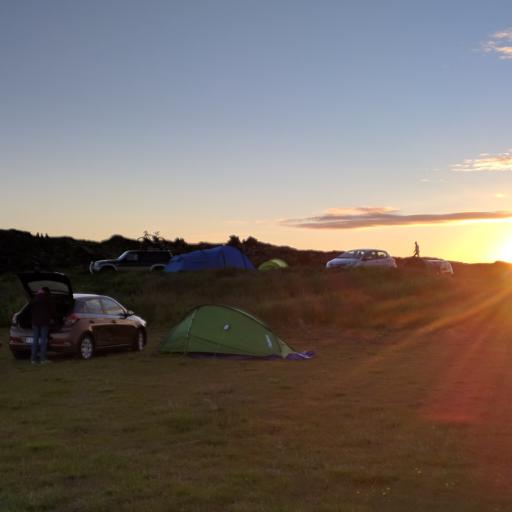}%
    } &
    \imgcell[75pt]{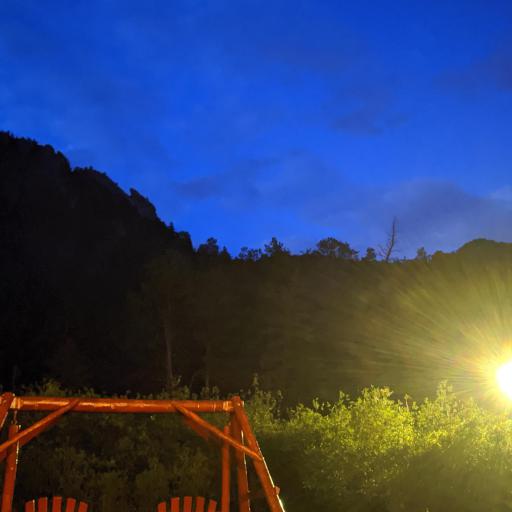} &
    \imgcell[75pt]{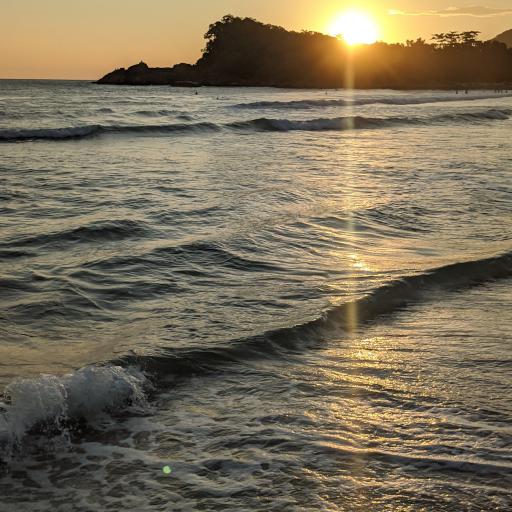} \\[-1pt]
    
    \rotatebox[origin=c]{90}{Output} &
    \imgcell[75pt]{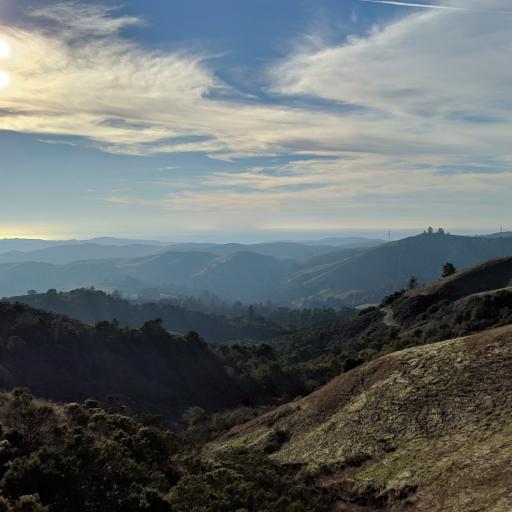} &    \Thickvrule{%
    \imgcell[75pt]{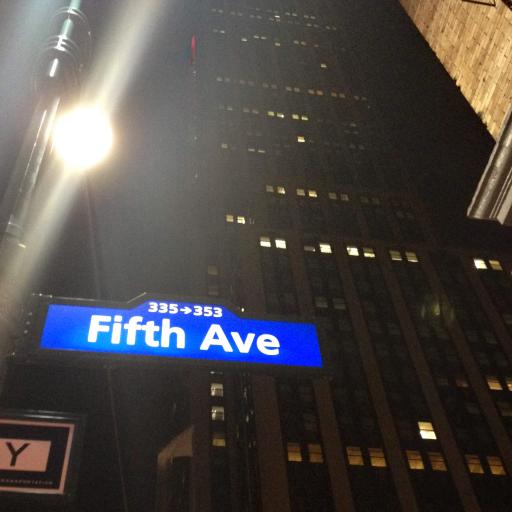}%
    } &
    \imgcell[75pt]{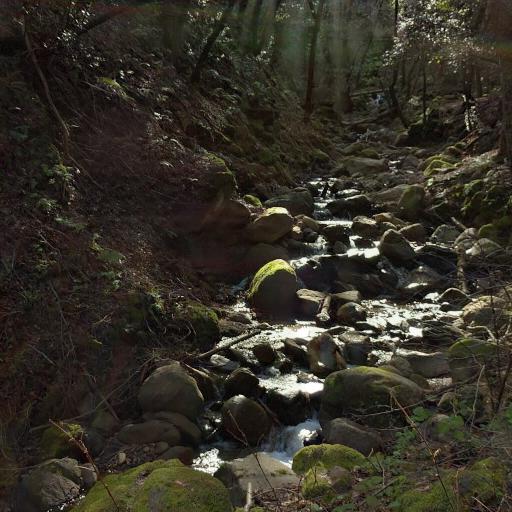} &    \Thickvrule{%
    \imgcell[75pt]{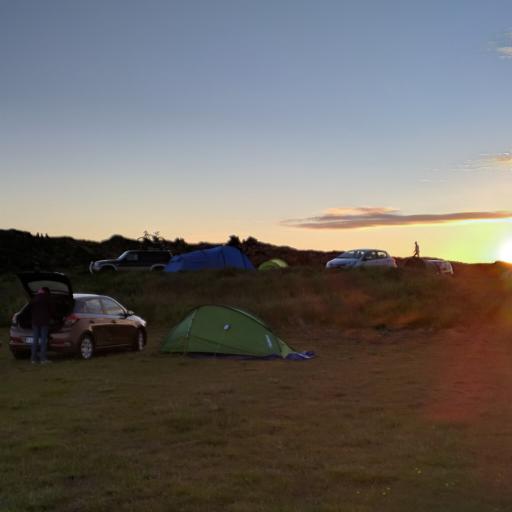}%
    } &
    \imgcell[75pt]{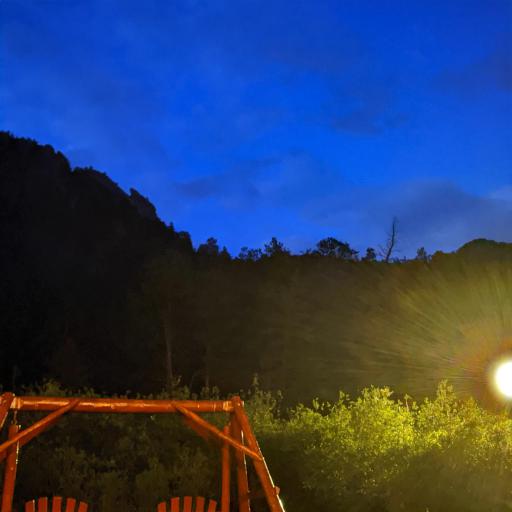} &
    \imgcell[75pt]{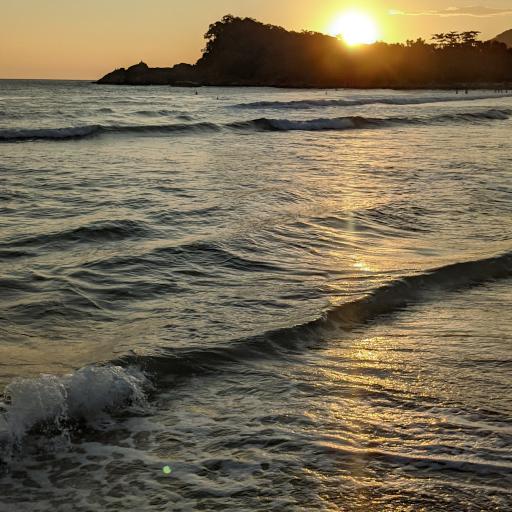} \\[35pt]
    
    \Cline{3-3}\Cline{5-5}
    
    \rotatebox[origin=c]{90}{Input} &
    \imgcell[75pt]{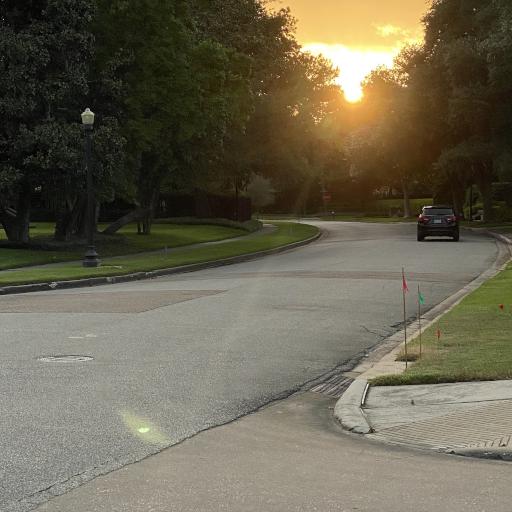} &
    \imgcell[75pt]{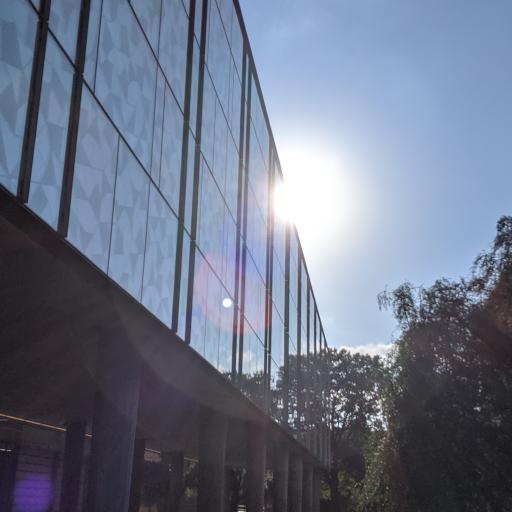} &
    \imgcell[75pt]{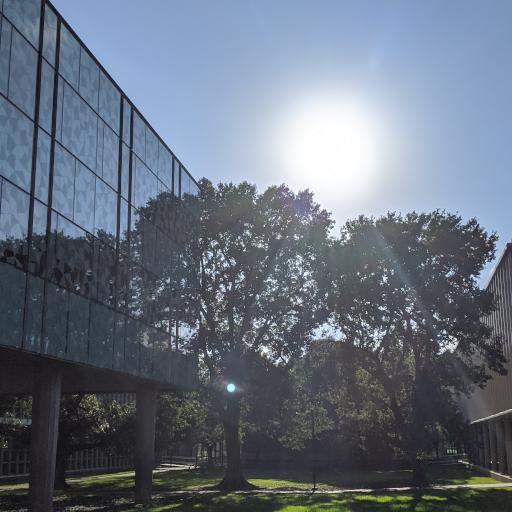} &
    \imgcell[75pt]{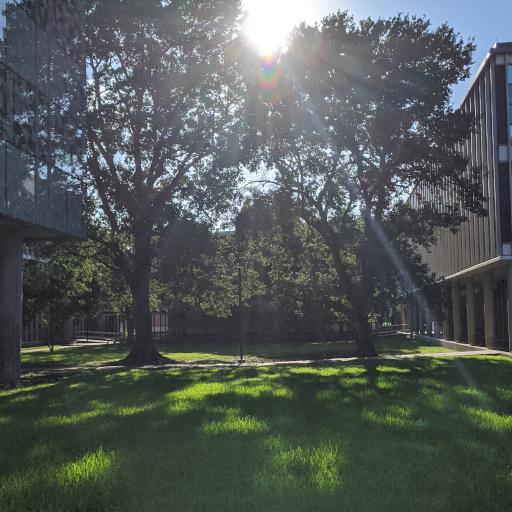} &
    \imgcell[75pt]{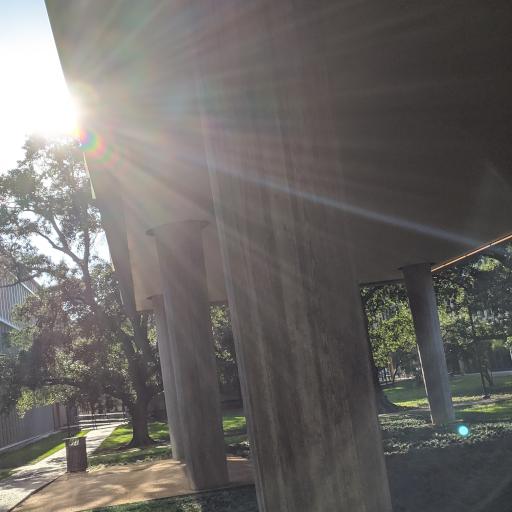} &
    \imgcell[75pt]{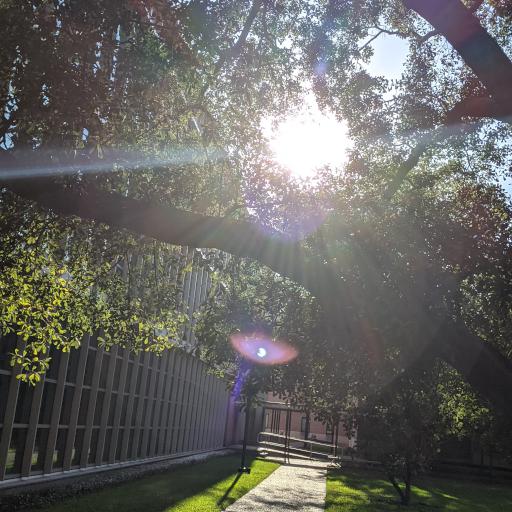} \\[-1pt]
    
    \rotatebox[origin=c]{90}{Output} &
    \imgcell[75pt]{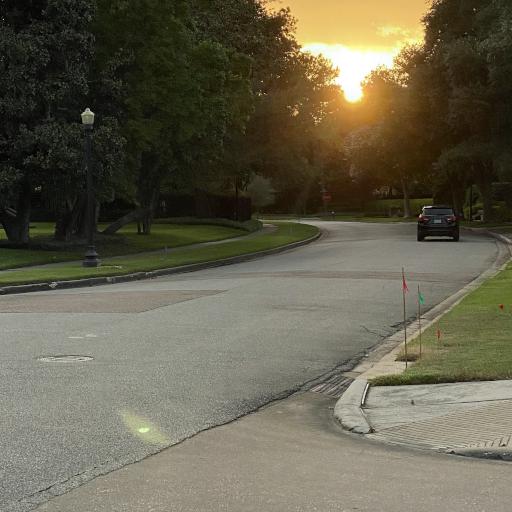} &
    \imgcell[75pt]{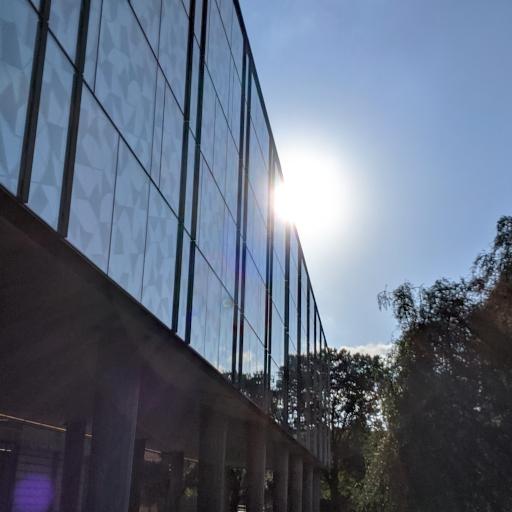} &
    \imgcell[75pt]{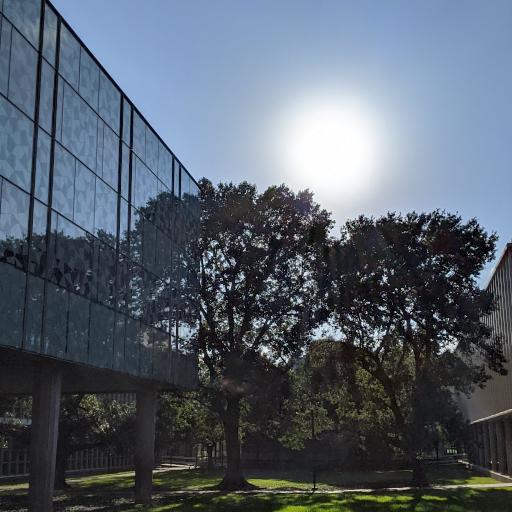} &
    \imgcell[75pt]{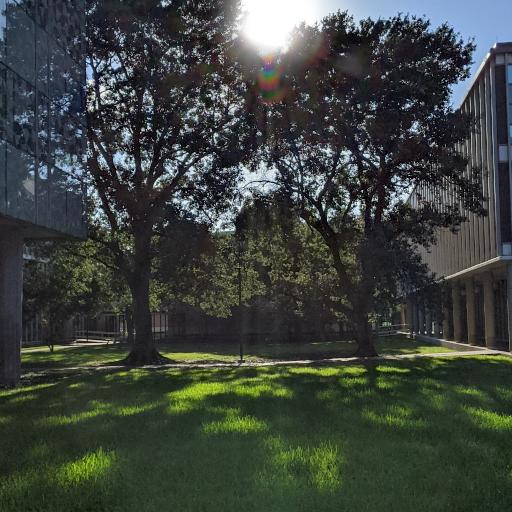} &
    \imgcell[75pt]{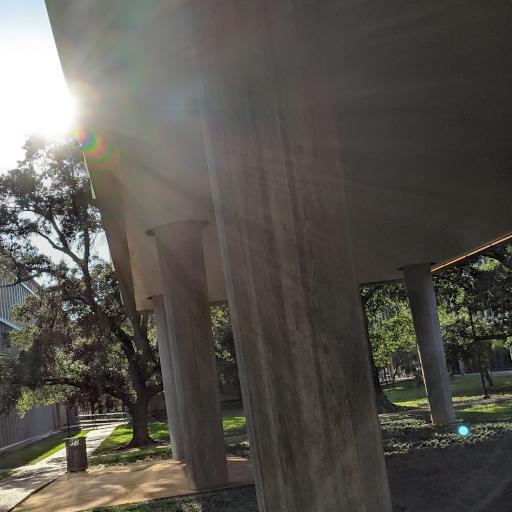} &
    \imgcell[75pt]{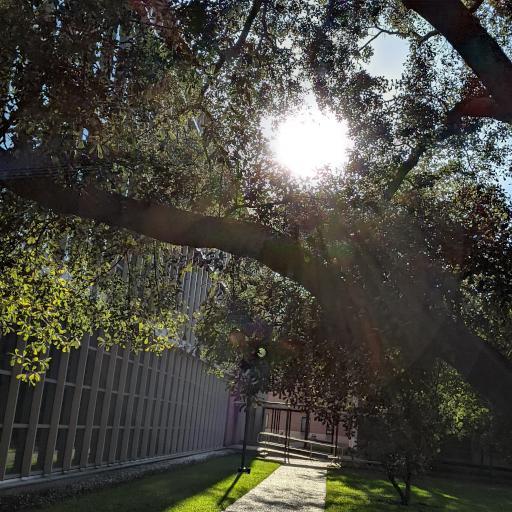} \\[35pt]
    
    \rotatebox[origin=c]{90}{Input} &
    \imgcell[75pt]{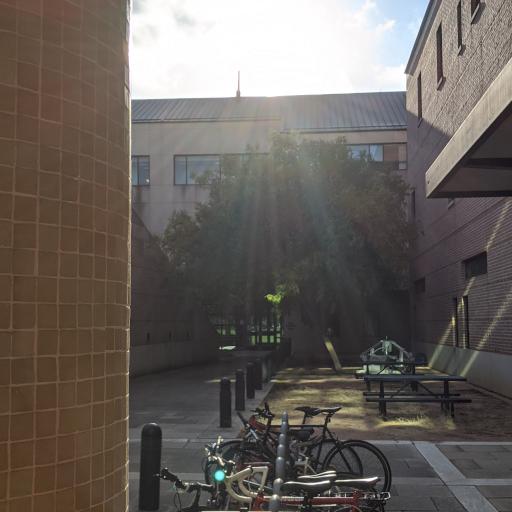} &
    \imgcell[75pt]{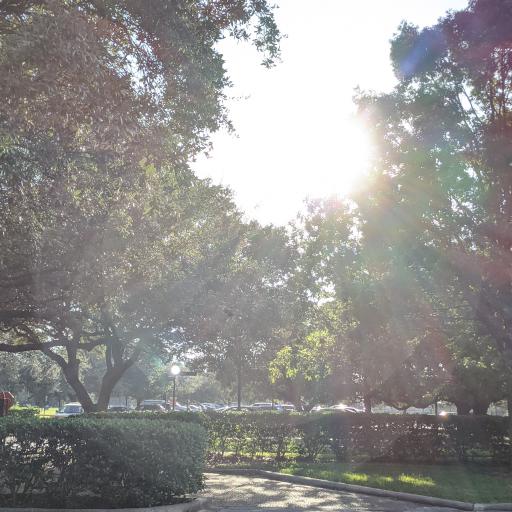} &
    \imgcell[75pt]{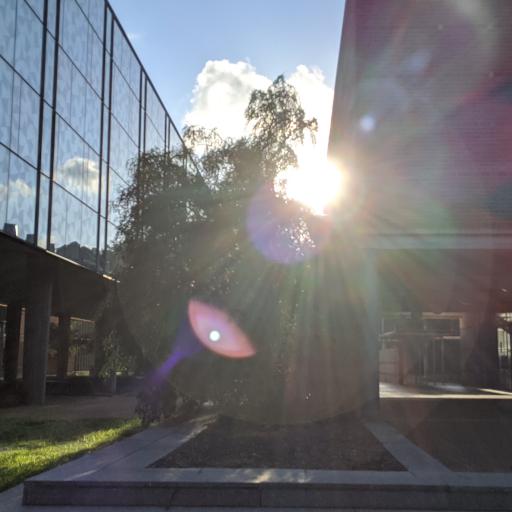} &
    \imgcell[75pt]{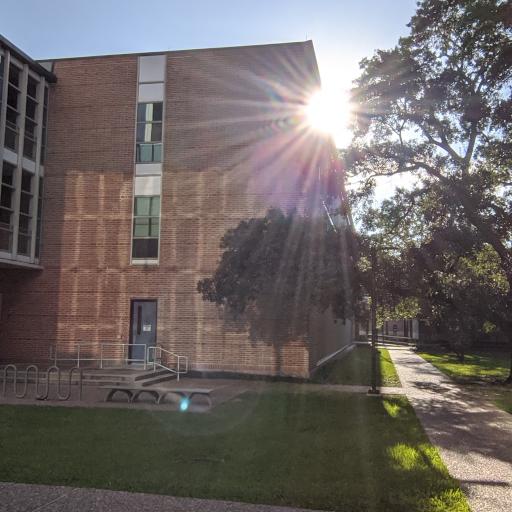} &
    \imgcell[75pt]{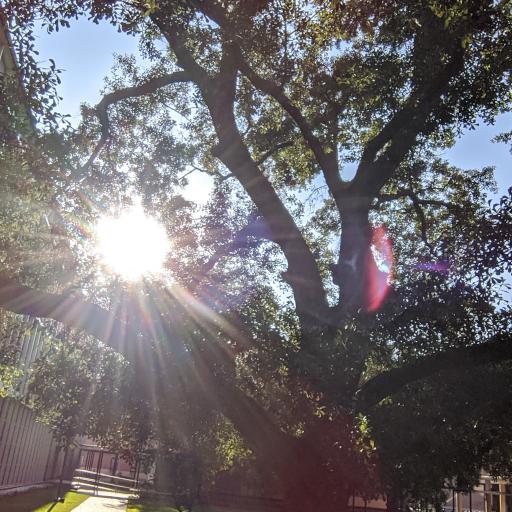} &
    \imgcell[75pt]{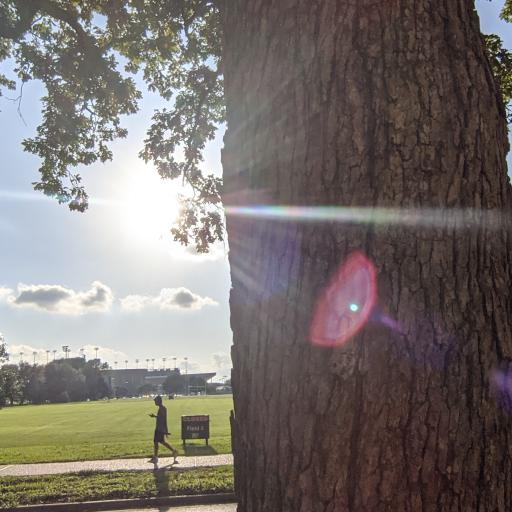} \\[-1pt]
    
    \rotatebox[origin=c]{90}{Output} &
    \imgcell[75pt]{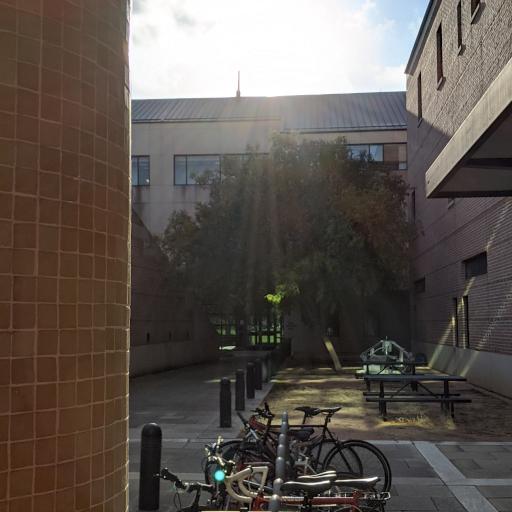} &
    \imgcell[75pt]{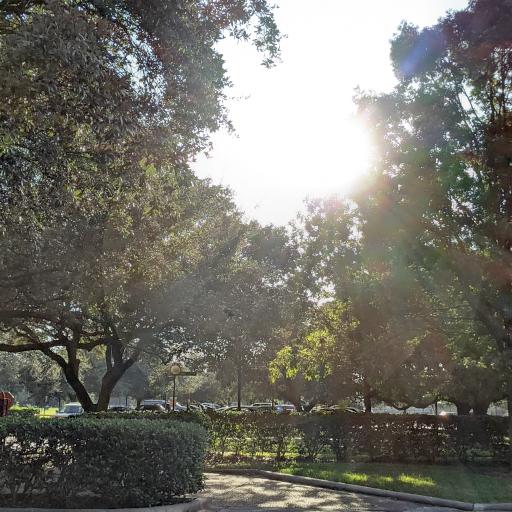} &
    \imgcell[75pt]{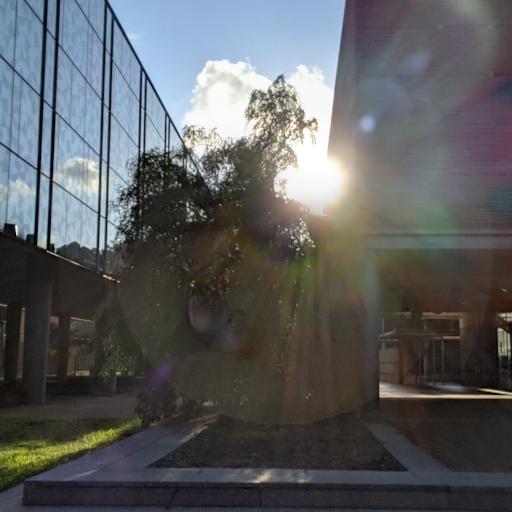} &
    \imgcell[75pt]{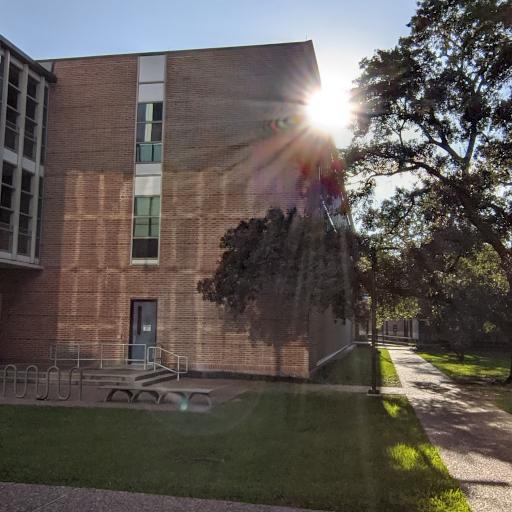} &
    \imgcell[75pt]{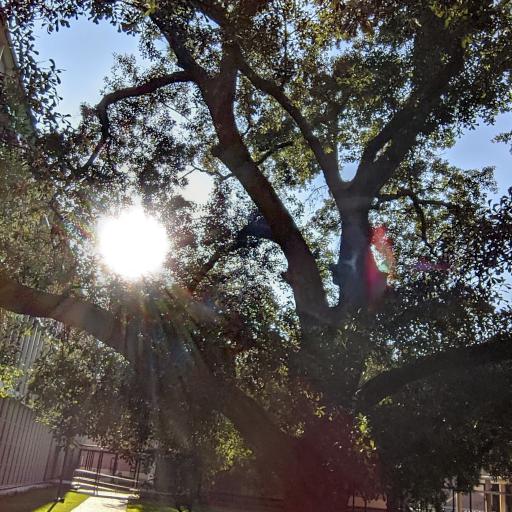} &
    \imgcell[75pt]{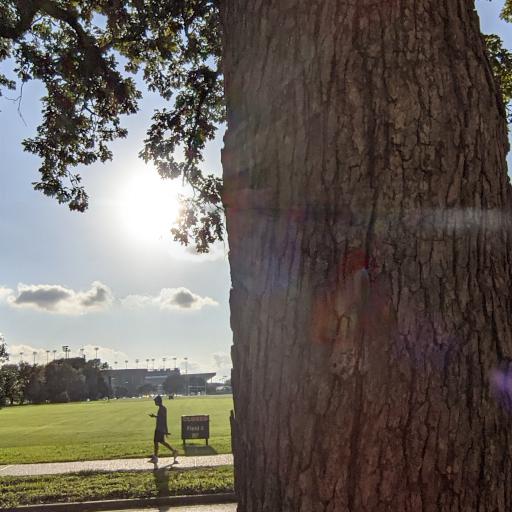} \\[35pt]
    
    \rotatebox[origin=c]{90}{Input} &
    \imgcell[75pt]{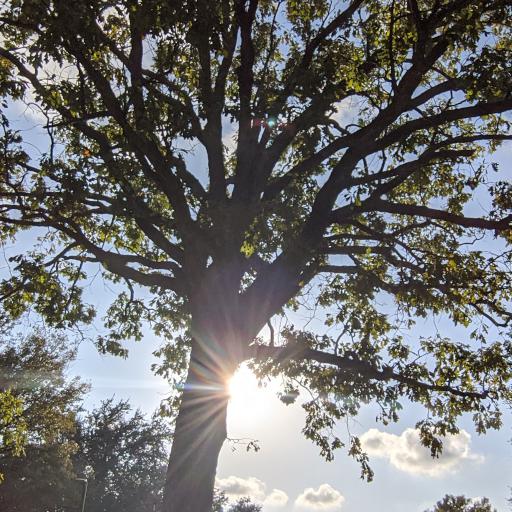} &
    \imgcell[75pt]{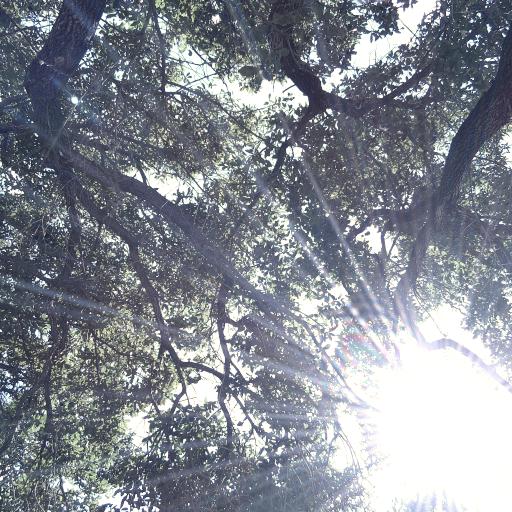} &
    \imgcell[75pt]{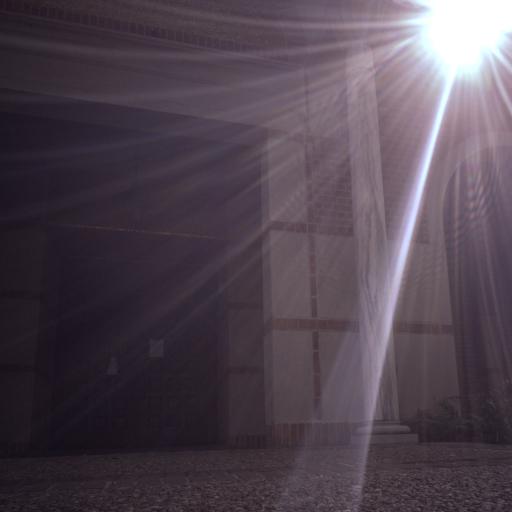} &
    \imgcell[75pt]{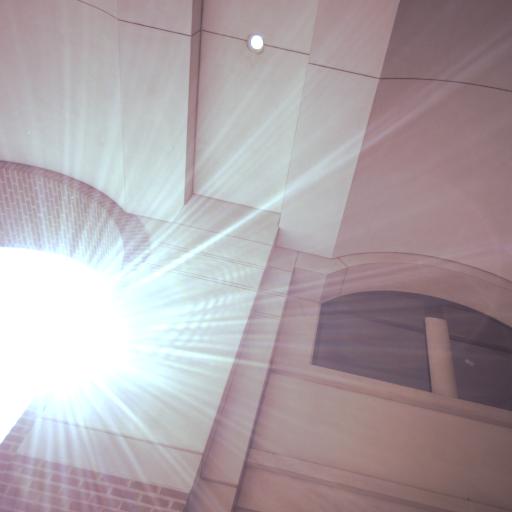} &
    \imgcell[75pt]{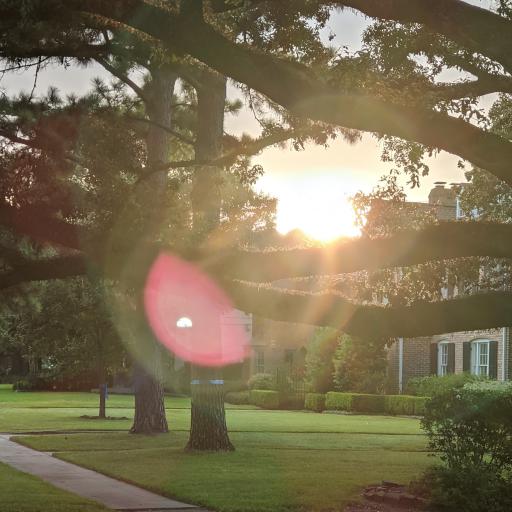} &
    \imgcell[75pt]{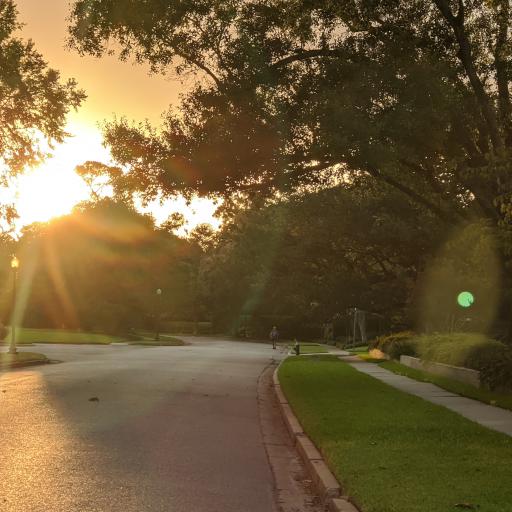} \\[-1pt]
    
    \rotatebox[origin=c]{90}{Output} &
    \imgcell[75pt]{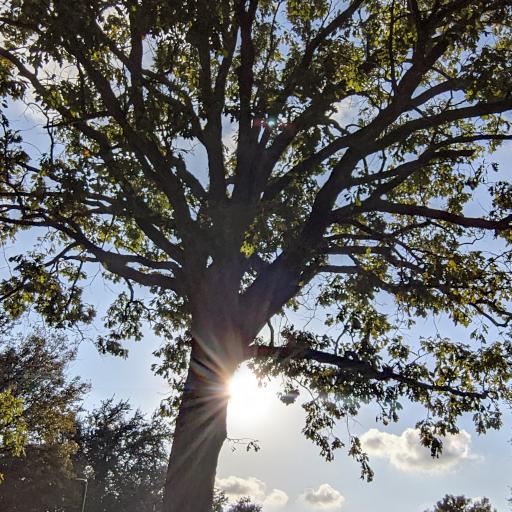} &
    \imgcell[75pt]{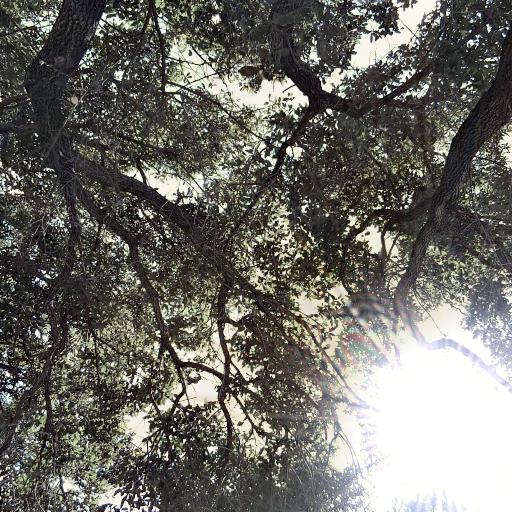} &
    \imgcell[75pt]{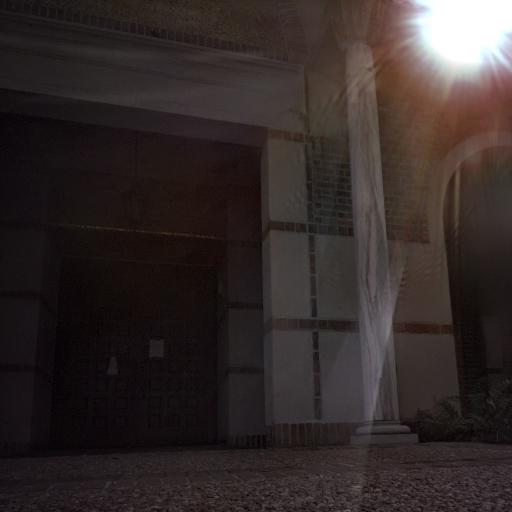} &
    \imgcell[75pt]{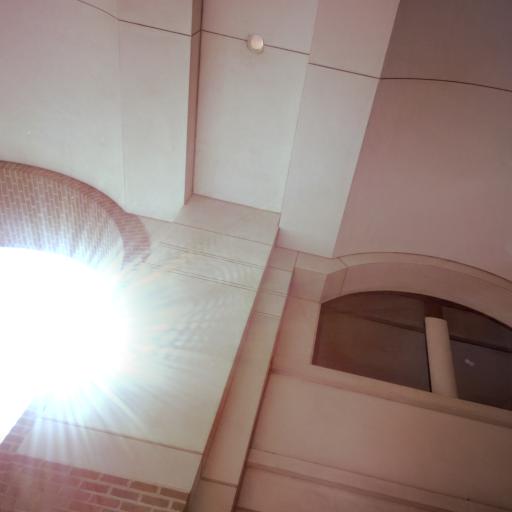} &
    \imgcell[75pt]{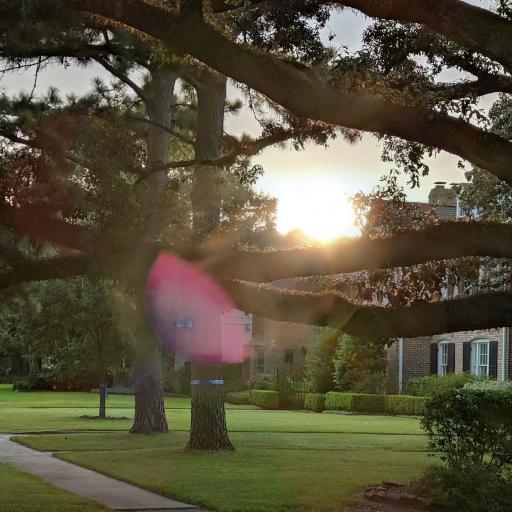} &
    \imgcell[75pt]{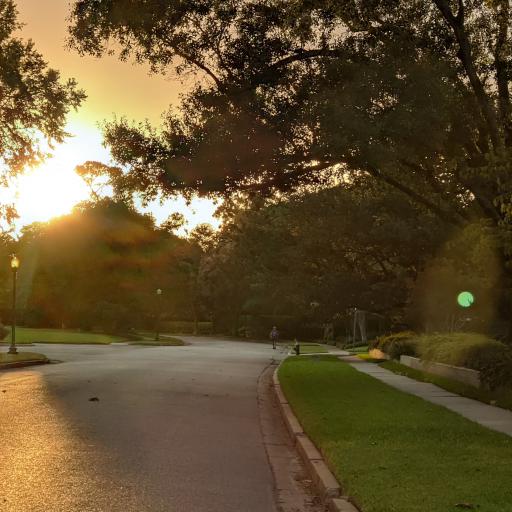}
  \end{tabular}
  \vspace{2pt}
  \caption{
    24 testing images captured by 7 other lens types with different designs and focal lengths. Our method successfully remove most flares, with a few occasional failures that the algorithm either fails to identify flare (e.g., first red box) or incorrectly removes non-flare highlights (e.g., clouds in the second red box).
  }
  \label{fig:supp_gallery_other_cameras}
\end{figure*}

    We also show 24 additional test images captured using the same type of lens as in Sec.~\ref{sec:data:reflective} of the main text (Fig.~\ref{fig:supp_gallery_same_camera}), and 24 test images captured using 7 other lens types with different focal lengths (Fig.~\ref{fig:supp_gallery_other_cameras}).
    
    While our results are mostly satisfactory, there are certainly cases where it does not perform well. They typically have strong flare over the entire image. There is a trade-off between flare removal and scene preservation, and we prefer the latter to reduce artifacts.

{\small
\bibliographystyle{ieee_fullname}
\bibliography{main}
}

\end{document}